\documentclass[longauth]{aa}

\usepackage{euclid}
\usepackage{graphicx}
\usepackage{natbib}
\usepackage{scalerel}

\usepackage[table]{xcolor}

\usepackage{amsmath,amsfonts,amssymb}
\usepackage{txfonts}
\usepackage{yfonts}
\usepackage{longtable}

\usepackage{soul,color,lscape}

\usepackage[pdfencoding=auto,psdextra]{hyperref}
\hypersetup{
    colorlinks=true,
    linkcolor=blue,
    filecolor=magenta,      
    urlcolor=blue,
    citecolor=blue
}
\makeatletter
\renewcommand*\aa@pageof{, page \thepage{} of \pageref*{LastPage}}
\makeatother

\usepackage[utf8]{inputenc}

\usepackage[switch, modulo]{lineno}

\usepackage{physics}
\newcommand{\astroang}[1]{\ang[angle-symbol-over-decimal]{#1}}

\newcommand{\done}[1]{}

\definecolor{brilliantlavender}{rgb}{0.96, 0.73, 1.0}

\newcommand{\MapMapMap}{\expval{\Map^3}}
\newcommand{\Mapn}{\expval{\Map^n}}
\newcommand{\varthetavec}{\vb*{\vartheta}}
\newcommand{\thetavec}{\vb*{\theta}}
\newcommand{\Map}{M_\mathrm{ap}}

\newcommand{\ellvec}{\vb*{\ell}}

\usepackage[normalem]{ulem}

\begin{document}

\title{\Euclid Preparation. XXVIII. Forecasts for ten different higher-order weak lensing statistics}

\titlerunning{\Euclid Preparation. XXVIII. HOWLS Paper I}
\authorrunning{Euclid Collaboration: HOWLS team et al.}

\newcommand{\orcid}[1]{} 
\author{{\normalsize Euclid Collaboration\thanks{Corresponding author: Euclid Collaboration HOWLS team, \email{euclid-howls@lam.fr}}: V.~Ajani\orcid{0000-0001-9442-2527}$^{1,2}$, M.~Baldi\orcid{0000-0003-4145-1943}$^{3,4,5}$, A.~Barthelemy\orcid{0000-0003-1060-3959}$^{6}$, A.~Boyle\orcid{0000-0002-7537-6921}$^{7}$, P.~Burger\orcid{0000-0001-8637-6305}$^{8}$, V.~F.~Cardone$^{9,10}$, S.~Cheng\orcid{0000-0002-9156-7461}$^{11}$, S.~Codis$^{7}$, C.~Giocoli\orcid{0000-0002-9590-7961}$^{4,5}$, J.~Harnois-D\'eraps\orcid{0000-0002-4864-1240}$^{12}$
% ...
}}

\institute{$^{1}$ Universit\'e Paris-Saclay, Universit\'e Paris Cit\'e, CEA, CNRS, Astrophysique, Instrumentation et Mod\'elisation Paris-Saclay, 91191 Gif-sur-Yvette, France\\
$^{2}$ Institute for Particle Physics and Astrophysics, Dept. of Physics, ETH Zurich, Wolfgang-Pauli-Strasse 27, 8093 Zurich, Switzerland\\
$^{3}$ Dipartimento di Fisica e Astronomia "Augusto Righi" - Alma Mater Studiorum Universit\`{a} di Bologna, via Piero Gobetti 93/2, 40129 Bologna, Italy\\
$^{4}$ INAF-Osservatorio di Astrofisica e Scienza dello Spazio di Bologna, Via Piero Gobetti 93/3, 40129 Bologna, Italy\\
$^{5}$ INFN-Sezione di Bologna, Viale Berti Pichat 6/2, 40127 Bologna, Italy\\
$^{6}$ University Observatory, Faculty of Physics, Ludwig-Maximilians-Universit{\"a}t, Scheinerstr. 1, 81679 Munich, Germany\\
$^{7}$ AIM, CEA, CNRS, Universit\'{e} Paris-Saclay, Universit\'{e} de Paris, 91191 Gif-sur-Yvette, France\\
$^{8}$ Argelander-Institut f\"ur Astronomie, Universit\"at Bonn, Auf dem H\"ugel 71, 53121 Bonn, Germany\\
$^{9}$ INFN-Sezione di Roma, Piazzale Aldo Moro, 2 - c/o Dipartimento di Fisica, Edificio G. Marconi, 00185 Roma, Italy\\
$^{10}$ INAF-Osservatorio Astronomico di Roma, Via Frascati 33, 00078 Monteporzio Catone, Italy\\
$^{11}$ Johns Hopkins University 3400 North Charles Street Baltimore, MD 21218, USA\\
$^{12}$ School of Mathematics, Statistics and Physics, Newcastle University, Herschel Building, Newcastle-upon-Tyne, NE1 7RU, UK\\
$^{13}$ Departamento de F\'isica, Faculdade de Ci\^encias, Universidade de Lisboa, Edif\'icio C8, Campo Grande, PT1749-016 Lisboa, Portugal\\
$^{14}$ Instituto de Astrof\'isica e Ci\^encias do Espa\c{c}o, Faculdade de Ci\^encias, Universidade de Lisboa, Tapada da Ajuda, 1349-018 Lisboa, Portugal\\
$^{15}$ Aix-Marseille Universit\'e, CNRS, CNES, LAM, Marseille, France\\
$^{16}$ Institute of Physics, Laboratory of Astrophysics, Ecole Polytechnique F\'{e}d\'{e}rale de Lausanne (EPFL), Observatoire de Sauverny, 1290 Versoix, Switzerland\\
$^{17}$ Dipartimento di Fisica, Sapienza Universit\`a di Roma, Piazzale Aldo Moro 2, 00185 Roma, Italy\\
$^{18}$ Universit\'e Paris-Saclay, CNRS, Institut d'astrophysique spatiale, 91405, Orsay, France\\
$^{19}$ INAF-Osservatorio Astrofisico di Torino, Via Osservatorio 20, 10025 Pino Torinese (TO), Italy\\
$^{20}$ Dipartimento di Fisica, Universit\`{a} di Genova, Via Dodecaneso 33, 16146, Genova, Italy\\
$^{21}$ INFN-Sezione di Roma Tre, Via della Vasca Navale 84, 00146, Roma, Italy\\
$^{22}$ Department of Physics "E. Pancini", University Federico II, Via Cinthia 6, 80126, Napoli, Italy\\
$^{23}$ Instituto de Astrof\'isica e Ci\^encias do Espa\c{c}o, Universidade do Porto, CAUP, Rua das Estrelas, PT4150-762 Porto, Portugal\\
$^{24}$ Dipartimento di Fisica, Universit\'a degli Studi di Torino, Via P. Giuria 1, 10125 Torino, Italy\\
$^{25}$ INFN-Sezione di Torino, Via P. Giuria 1, 10125 Torino, Italy\\
$^{26}$ INAF-IASF Milano, Via Alfonso Corti 12, 20133 Milano, Italy\\
$^{27}$ Institut de F\'{i}sica d'Altes Energies (IFAE), The Barcelona Institute of Science and Technology, Campus UAB, 08193 Bellaterra (Barcelona), Spain\\
$^{28}$ Port d'Informaci\'{o} Cient\'{i}fica, Campus UAB, C. Albareda s/n, 08193 Bellaterra (Barcelona), Spain\\
$^{29}$ Institut d'Estudis Espacials de Catalunya (IEEC), Carrer Gran Capit\'a 2-4, 08034 Barcelona, Spain\\
$^{30}$ Institute of Space Sciences (ICE, CSIC), Campus UAB, Carrer de Can Magrans, s/n, 08193 Barcelona, Spain\\
$^{31}$ INAF-Osservatorio Astronomico di Capodimonte, Via Moiariello 16, 80131 Napoli, Italy\\
$^{32}$ INFN section of Naples, Via Cinthia 6, 80126, Napoli, Italy\\
$^{33}$ Dipartimento di Fisica e Astronomia "Augusto Righi" - Alma Mater Studiorum Universit\'a di Bologna, Viale Berti Pichat 6/2, 40127 Bologna, Italy\\
$^{34}$ Centre National d'Etudes Spatiales, Toulouse, France\\
$^{35}$ Institut national de physique nucl\'eaire et de physique des particules, 3 rue Michel-Ange, 75794 Paris C\'edex 16, France\\
$^{36}$ Institute for Astronomy, University of Edinburgh, Royal Observatory, Blackford Hill, Edinburgh EH9 3HJ, UK\\
$^{37}$ Jodrell Bank Centre for Astrophysics, Department of Physics and Astronomy, University of Manchester, Oxford Road, Manchester M13 9PL, UK\\
$^{38}$ ESAC/ESA, Camino Bajo del Castillo, s/n., Urb. Villafranca del Castillo, 28692 Villanueva de la Ca\~nada, Madrid, Spain\\
$^{39}$ European Space Agency/ESRIN, Largo Galileo Galilei 1, 00044 Frascati, Roma, Italy\\
$^{40}$ Mullard Space Science Laboratory, University College London, Holmbury St Mary, Dorking, Surrey RH5 6NT, UK\\
$^{41}$ Instituto de Astrof\'isica e Ci\^encias do Espa\c{c}o, Faculdade de Ci\^encias, Universidade de Lisboa, Campo Grande, 1749-016 Lisboa, Portugal\\
$^{42}$ Department of Astronomy, University of Geneva, ch. d'Ecogia 16, 1290 Versoix, Switzerland\\
$^{43}$ INAF-Istituto di Astrofisica e Planetologia Spaziali, via del Fosso del Cavaliere, 100, 00100 Roma, Italy\\
$^{44}$ Univ Lyon, Univ Claude Bernard Lyon 1, CNRS/IN2P3, IP2I Lyon, UMR 5822, 69622, Villeurbanne, France\\
$^{45}$ INAF-Osservatorio Astronomico di Trieste, Via G. B. Tiepolo 11, 34143 Trieste, Italy\\
$^{46}$ INAF-Osservatorio Astronomico di Padova, Via dell'Osservatorio 5, 35122 Padova, Italy\\
$^{47}$ Max Planck Institute for Extraterrestrial Physics, Giessenbachstr. 1, 85748 Garching, Germany\\
$^{48}$ Universit\"ats-Sternwarte M\"unchen, Fakult\"at f\"ur Physik, Ludwig-Maximilians-Universit\"at M\"unchen, Scheinerstrasse 1, 81679 M\"unchen, Germany\\
$^{49}$ Leiden Observatory, Leiden University, Niels Bohrweg 2, 2333 CA Leiden, The Netherlands\\
$^{50}$ Jet Propulsion Laboratory, California Institute of Technology, 4800 Oak Grove Drive, Pasadena, CA, 91109, USA\\
$^{51}$ Technical University of Denmark, Elektrovej 327, 2800 Kgs. Lyngby, Denmark\\
$^{52}$ Cosmic Dawn Center (DAWN), Denmark\\
$^{53}$ Institut d'Astrophysique de Paris, 98bis Boulevard Arago, 75014, Paris, France\\
$^{54}$ Max-Planck-Institut f\"ur Astronomie, K\"onigstuhl 17, 69117 Heidelberg, Germany\\
$^{55}$ NASA Goddard Space Flight Center, Greenbelt, MD 20771, USA\\
$^{56}$ Universit\'e de Gen\`eve, D\'epartement de Physique Th\'eorique and Centre for Astroparticle Physics, 24 quai Ernest-Ansermet, CH-1211 Gen\`eve 4, Switzerland\\
$^{57}$ Department of Physics, P.O. Box 64, 00014 University of Helsinki, Finland\\
$^{58}$ Helsinki Institute of Physics, Gustaf H{\"a}llstr{\"o}min katu 2, University of Helsinki, Helsinki, Finland\\
$^{59}$ Institute of Theoretical Astrophysics, University of Oslo, P.O. Box 1029 Blindern, 0315 Oslo, Norway\\
$^{60}$ NOVA optical infrared instrumentation group at ASTRON, Oude Hoogeveensedijk 4, 7991PD, Dwingeloo, The Netherlands\\
$^{61}$ Department of Physics, Institute for Computational Cosmology, Durham University, South Road, DH1 3LE, UK\\
$^{62}$  Universit\'e Paris Cit\'e, CNRS, Astroparticule et Cosmologie, 75013 Paris, France\\
$^{63}$ Institut d'Astrophysique de Paris, UMR 7095, CNRS, and Sorbonne Universit\'e, 98 bis boulevard Arago, 75014 Paris, France\\
$^{64}$ European Space Agency/ESTEC, Keplerlaan 1, 2201 AZ Noordwijk, The Netherlands\\
$^{65}$ Kapteyn Astronomical Institute, University of Groningen, PO Box 800, 9700 AV Groningen, The Netherlands\\
$^{66}$ Department of Physics and Astronomy, University of Aarhus, Ny Munkegade 120, DK-8000 Aarhus C, Denmark\\
$^{67}$ Space Science Data Center, Italian Space Agency, via del Politecnico snc, 00133 Roma, Italy\\
$^{68}$ Institute of Space Science, Bucharest, 077125, Romania\\
$^{69}$ Dipartimento di Fisica e Astronomia "G.Galilei", Universit\'a di Padova, Via Marzolo 8, 35131 Padova, Italy\\
$^{70}$ INFN-Padova, Via Marzolo 8, 35131 Padova, Italy\\
$^{71}$ Dipartimento di Fisica e Astronomia, Universit\'a di Bologna, Via Gobetti 93/2, 40129 Bologna, Italy\\
$^{72}$ Departamento de F\'isica, FCFM, Universidad de Chile, Blanco Encalada 2008, Santiago, Chile\\
$^{73}$ Institut f\"ur Astro- und Teilchenphysik, Universit\"at Innsbruck, Technikerstr. 25/8, 6020 Innsbruck, Austria\\
$^{74}$ Aix-Marseille Universit\'e, CNRS/IN2P3, CPPM, Marseille, France\\
$^{75}$ Institut de Ciencies de l'Espai (IEEC-CSIC), Campus UAB, Carrer de Can Magrans, s/n Cerdanyola del Vall\'es, 08193 Barcelona, Spain\\
$^{76}$ Centro de Investigaciones Energ\'eticas, Medioambientales y Tecnol\'ogicas (CIEMAT), Avenida Complutense 40, 28040 Madrid, Spain\\
$^{77}$ Universidad Polit\'ecnica de Cartagena, Departamento de Electr\'onica y Tecnolog\'ia de Computadoras, 30202 Cartagena, Spain\\
$^{78}$ Institut de Recherche en Astrophysique et Plan\'etologie (IRAP), Universit\'e de Toulouse, CNRS, UPS, CNES, 14 Av. Edouard Belin, 31400 Toulouse, France\\
$^{79}$ Infrared Processing and Analysis Center, California Institute of Technology, Pasadena, CA 91125, USA\\
$^{80}$ INAF-Osservatorio Astronomico di Brera, Via Brera 28, 20122 Milano, Italy\\
$^{81}$ Instituto de Astrof\'isica de Canarias, Calle V\'ia L\'actea s/n, 38204, San Crist\'obal de La Laguna, Tenerife, Spain\\
$^{82}$ Center for Computational Astrophysics, Flatiron Institute, 162 5th Avenue, 10010, New York, NY, USA\\
$^{83}$ School of Physics and Astronomy, Cardiff University, The Parade, Cardiff, CF24 3AA, UK\\
$^{84}$ Department of Physics and Helsinki Institute of Physics, Gustaf H\"allstr\"omin katu 2, 00014 University of Helsinki, Finland\\
$^{85}$ Dipartimento di Fisica "Aldo Pontremoli", Universit\'a degli Studi di Milano, Via Celoria 16, 20133 Milano, Italy\\
$^{86}$ INFN-Sezione di Milano, Via Celoria 16, 20133 Milano, Italy\\
$^{87}$ Junia, EPA department, 59000 Lille, France\\
$^{88}$ Instituto de F\'isica Te\'orica UAM-CSIC, Campus de Cantoblanco, 28049 Madrid, Spain\\
$^{89}$ CERCA/ISO, Department of Physics, Case Western Reserve University, 10900 Euclid Avenue, Cleveland, OH 44106, USA\\
$^{90}$ Laboratoire de Physique de l'\'Ecole Normale Sup\'erieure, ENS, Universit\'e PSL, CNRS, Sorbonne Universit\'e, 75005 Paris, France\\
$^{91}$ Observatoire de Paris, Universit\'e PSL, Sorbonne Universit\'e, LERMA, 750 Paris, France\\
$^{92}$ Astrophysics Group, Blackett Laboratory, Imperial College London, London SW7 2AZ, UK\\
$^{93}$ SISSA, International School for Advanced Studies, Via Bonomea 265, 34136 Trieste TS, Italy\\
$^{94}$ IFPU, Institute for Fundamental Physics of the Universe, via Beirut 2, 34151 Trieste, Italy\\
$^{95}$ INFN, Sezione di Trieste, Via Valerio 2, 34127 Trieste TS, Italy\\
$^{96}$ Departamento de Astrof\'{i}sica, Universidad de La Laguna, 38206, La Laguna, Tenerife, Spain\\
$^{97}$ Dipartimento di Fisica e Scienze della Terra, Universit\'a degli Studi di Ferrara, Via Giuseppe Saragat 1, 44122 Ferrara, Italy\\
$^{98}$ Istituto Nazionale di Fisica Nucleare, Sezione di Ferrara, Via Giuseppe Saragat 1, 44122 Ferrara, Italy\\
$^{99}$ Institut de Physique Th\'eorique, CEA, CNRS, Universit\'e Paris-Saclay 91191 Gif-sur-Yvette Cedex, France\\
$^{100}$ Dipartimento di Fisica - Sezione di Astronomia, Universit\'a di Trieste, Via Tiepolo 11, 34131 Trieste, Italy\\
$^{101}$ NASA Ames Research Center, Moffett Field, CA 94035, USA\\
$^{102}$ INAF, Istituto di Radioastronomia, Via Piero Gobetti 101, 40129 Bologna, Italy\\
$^{103}$ INFN-Bologna, Via Irnerio 46, 40126 Bologna, Italy\\
$^{104}$ Universit\'e C\^{o}te d'Azur, Observatoire de la C\^{o}te d'Azur, CNRS, Laboratoire Lagrange, Bd de l'Observatoire, CS 34229, 06304 Nice cedex 4, France\\
$^{105}$ Institute for Theoretical Particle Physics and Cosmology (TTK), RWTH Aachen University, 52056 Aachen, Germany\\
$^{106}$ Institute for Astronomy, University of Hawaii, 2680 Woodlawn Drive, Honolulu, HI 96822, USA\\
$^{107}$ Department of Physics \& Astronomy, University of California Irvine, Irvine CA 92697, USA\\
$^{108}$ University of Lyon, UCB Lyon 1, CNRS/IN2P3, IUF, IP2I Lyon, France\\
$^{109}$ INFN-Sezione di Genova, Via Dodecaneso 33, 16146, Genova, Italy\\
$^{110}$ Department of Astronomy \& Physics and Institute for Computational Astrophysics, Saint Mary's University, 923 Robie Street, Halifax, Nova Scotia, B3H 3C3, Canada\\
$^{111}$ Ruhr University Bochum, Faculty of Physics and Astronomy, Astronomical Institute (AIRUB), German Centre for Cosmological Lensing (GCCL), 44780 Bochum, Germany\\
$^{112}$ Univ. Grenoble Alpes, CNRS, Grenoble INP, LPSC-IN2P3, 53, Avenue des Martyrs, 38000, Grenoble, France\\
$^{113}$ Department of Physics and Astronomy, University College London, Gower Street, London WC1E 6BT, UK\\
$^{114}$ Department of Physics and Astronomy, Vesilinnantie 5, 20014 University of Turku, Finland\\
$^{115}$ University of Applied Sciences and Arts of Northwestern Switzerland, School of Engineering, 5210 Windisch, Switzerland\\
$^{116}$ Centro de Astrof\'{\i}sica da Universidade do Porto, Rua das Estrelas, 4150-762 Porto, Portugal\\
$^{117}$ Institute of Cosmology and Gravitation, University of Portsmouth, Portsmouth PO1 3FX, UK\\
$^{118}$ Department of Mathematics and Physics E. De Giorgi, University of Salento, Via per Arnesano, CP-I93, 73100, Lecce, Italy\\
$^{119}$ INFN, Sezione di Lecce, Via per Arnesano, CP-193, 73100, Lecce, Italy\\
$^{120}$ INAF-Sezione di Lecce, c/o Dipartimento Matematica e Fisica, Via per Arnesano, 73100, Lecce, Italy\\
$^{121}$ Institute for Computational Science, University of Zurich, Winterthurerstrasse 190, 8057 Zurich, Switzerland\\
$^{122}$ Higgs Centre for Theoretical Physics, School of Physics and Astronomy, The University of Edinburgh, Edinburgh EH9 3FD, UK\\
$^{123}$ Universit\'e St Joseph; Faculty of Sciences, Beirut, Lebanon\\
$^{124}$ Institut f\"ur Theoretische Physik, University of Heidelberg, Philosophenweg 16, 69120 Heidelberg, Germany\\
$^{125}$ Department of Astrophysical Sciences, Peyton Hall, Princeton University, Princeton, NJ 08544, USA}

\date{}

\setcounter{page}{1}

\abstract{

Recent cosmic shear studies have shown that higher-order statistics (HOS) developed by independent teams now outperform standard two-point estimators in terms of statistical precision thanks to their sensitivity to the non-Gaussian features of large-scale structure. The aim of the Higher-Order Weak Lensing Statistics (HOWLS) project is to assess, compare, and combine the constraining power of ten different HOS on a common set of \Euclid-like mocks, derived from N-body simulations. In this first paper of the HOWLS series, we computed the nontomographic ($\Omega_{\rm m}$, $\sigma_8$) Fisher information for the one-point probability distribution function, peak counts, Minkowski functionals, Betti numbers, persistent homology Betti numbers and heatmap, and scattering transform coefficients, and we compare them to the shear and convergence two-point correlation functions in the absence of any systematic bias. We also include forecasts for three implementations of higher-order moments, but these cannot be robustly interpreted as the Gaussian likelihood assumption breaks down for these statistics. Taken individually, we find that each HOS outperforms the two-point statistics by a factor of around two in the precision of the forecasts with some variations across statistics and cosmological parameters. When combining all the HOS, this increases to a $4.5$ times improvement, highlighting the immense potential of HOS for cosmic shear cosmological analyses with \Euclid. The data used in this analysis are publicly released with the paper.}


\maketitle



\section{Introduction \done{Nicolas, Vincenzo}}
\label{sec:intro}

It is a well-established fact that the Universe is undergoing a phase of accelerated expansion \citep[e.g.,][]{Riess+98,Perlmutter+99}. Understanding what is driving this acceleration in the framework of a spatially flat universe is one of the (if not the) greatest challenges of modern-day cosmology. The concordance $\Lambda$ cold dark matter ($\Lambda$CDM) model performs excellently in fitting the available data, yet the cosmological constant $\Lambda$ is far from satisfactory from a theoretical point of view. To make things harder for $\Lambda$CDM, recent tensions have emerged due to an inconsistency between the values of some parameters from independent data measuring the same quantities in radically different ways. The most debated case is the discrepancy between the Hubble constant $H_0$ as measured from local probes and as inferred from cosmological data sets \citep[see, e.g.,][for a review]{DiValentino2021}. Another, albeit less significant, example is the disagreement between the cosmic microwave background (CMB) and lensing estimates of the growth of structure parameter, $S_8 = \sigma_8 \sqrt{\Omega_{\rm m}/0.3}$ \citep[e.g.,][]{Hildebrandt+17,KiDS1000,Amon+22}. Although some unknown systematic effects could have been missed in the analysis, such tensions may also be the first signs that alternative models are needed, relying on either dark energy in a general relativity framework or based on modified gravity \citep[see, e.g.,][and references therein]{JLS16}. Discriminating among the plethora of viable candidates is the aim of Stage IV surveys such as the Dark Energy Spectroscopic Instrument \citep[DESI,][]{DESI2016}, the Prime Focus Spectrograph \citep[PFS,][]{PFS2014}, the \textit{Vera C. Rubin} Observatory Legacy Survey of Space and Time \citep[LSST,][]{LSST2019}, \Euclid \citep{Laureijs+11}, SPHEREx \citep{SPHEREx2014}, and the \textit{Nancy Grace Roman} Space Telescope \citep{Spergel2015}.

In this context, the \Euclid mission will play a pivotal role in cosmology by measuring the dark energy equation of state $w$ and the $S_8$ parameter with exquisite precision and accuracy \citep[see, e.g.,][]{Laureijs+11,EuclidVII}. This will be achieved by exploiting the clustering of galaxies as well as cosmic shear: the Weak gravitational Lensing (WL) signal of galaxies due to the deflection of light rays by the large-scale structure. In this article, we focus on optimizing the extraction of cosmological information from the cosmic shear probe.

The classical cosmic shear analysis involves measuring the correlations between ellipticities of pairs of galaxies as a function of their separation; an estimator called the two-point correlation functions of the shear ($\gamma$-2PCF). Although it benefits from a comprehensive theoretical description, this estimator is only sensitive to the multiscale variance of the lensing field. However, the gravitational collapse of the matter perturbations introduces non-Gaussian features in the shear field. As a consequence, the $\gamma$-2PCF and any estimator probing the field up to second order do not contain all cosmological information. This is seen, for example, in the degeneracies between parameters, such as the one between $\Omega_{\rm m}$ and $\sigma_8$, which means the $\gamma$-2PCF can only efficiently constrain their combination $S_8 =\sigma_8\sqrt{\Omega_{\rm m}/0.3}$. To recover the extra information contained in nonlinear scales, many non-Gaussian estimators -- also referred as higher-order statistics (HOS) in contrast to two-point statistics -- have been introduced in the literature. Such HOS include higher-order moments \citep[e.g.,][]{vanWaerbeke+13,Gatti2021,Porth+21}, peak counts \citep[e.g.,][]{Marian+09,Dietrich+10,Kacprzak+16,Martinet+18,Martinet+21a,Harnois-Deraps+21}, one-point probability distributions \citep[e.g.,][]{Barthelemy20a,Boyle2020,Liu2019,Thiele2020}, Minkowski functionals \citep[e.g.,][]{Kratochvil+12,Petri+15,Vicinanza_2019,Parroni_2020}, Betti numbers \citep[e.g.,][]{Feldbrugge_2019, Parroni_2021}, persistent homology \citep[e.g.,][]{Heydenreich:2021,Heydenreich:2022}, scattering transform coefficients \citep[e.g.,][]{Cheng2020,Cheng2021}, as well as map-level inference \citep{Porqueres2022,Boruah2022}. Despite their increased complexity, which often requires resorting to numerical simulations to model their cosmology dependence, all the references above have demonstrated that these new statistics have superior constraining power compared to the $\gamma$-2PCF. However, each of these new HOS is usually developed and studied by independent teams, which renders a fair comparison between them extremely difficult.

The Higher-Order Weak Lensing Statistics (HOWLS) project has been initiated to remedy this situation. One of its main aims is, indeed, to test HOS probes by relying on the same mock data, here mimicking those that \textit{Euclid} will make available. In contrast to some early \citep[e.g.,][]{Pires+09,Hilbert+12} and recent efforts in the literature \citep[e.g.,][]{Zurcher+22}, HOWLS was designed as a challenge to the community, thus attracting contributions from the largest team of HOS experts ever. Individual teams within the {\it Euclid} community have applied $24$ different algorithms to the same mocks for a total of two second-order statistics (the shear and convergence two-point correlation functions $\gamma$-2PCF and $\kappa$-2PCF) and ten different HOS: convergence one-point probability distribution ($\kappa$-PDF), higher-order convergence moments (HOM), $n$-th order aperture mass moments $\smash{\Mapn}$, aperture mass peak counts (peaks), convergence Minkowski functionals (MFs), convergence Betti numbers (BNs), aperture mass persistent homology Betti numbers (pers. BNs) and heatmap (pers. heat.), and convergence scattering transform coefficients (ST). Such a large number is unprecedented and offers the possibility of investigating which one (or which combination) is best suited to be coupled with the standard $\gamma$-2PCF probe to narrow down the constraints on cosmological parameters (CPs). Different HOS are in fact sensitive to different scales and features in the convergence ($\kappa$) maps and thus they couple to the $\gamma$-2PCF in their own way. Moreover, HOWLS can also check for correlations among the various HOS probes, revealing which ones are sufficiently uncorrelated such that their combination does indeed improve the total constraining power. It is also worth stressing that the present paper is only the first in a series. HOWLS will actually serve as a preparation for the application of WL HOS to the Euclid Survey, defining common tools and pipelines for the consortium.

The HOWLS data set is based on the DUSTGRAIN-\emph{pathfinder} simulations \citep{giocoli18b}, designed to model the cosmological dependence of every statistic, and on the Scinet LIght-Cones Simulations (SLICS, \citealp{SLICS}) for estimating covariances. We have built realistic {\it Euclid} mocks out of these simulations, in particular mimicking the expected galaxy density, intrinsic ellipticities, and redshift distribution. For every mock, we built a convergence map following the \citet{Kaiser+93} implementation described in \citet{Pires+20}. We measured the $\gamma$-2PCF in the ellipticity catalogs; the $\kappa$-2PCF, $\kappa$-PDF, MFs, BNs, ST, and HOM from the convergence maps; $\MapMapMap$, $\Mapn$, and pers. BNs and pers. heat. from the aperture mass calculated from the shear field; and peaks of aperture mass maps calculated from the reconstructed convergence fields. We then developed two independent analysis pipelines to compute the Fisher information and thus forecast the constraining power of HOS compared to two-point statistics.

This first paper in the HOWLS series is intended to introduce the data set (Sect.~\ref{sec:ds}) and HOS (Sect.~\ref{sec:hos}), and to conduct a Fisher analysis (Sect.~\ref{sec:Fisher}) to compare them. Forecasts are presented and discussed in Sect.~\ref{sec:res}. We conclude in Sect.~\ref{sec:ccl} by listing the refinements that we will include in the following HOWLS publications. The HOWLS data set and applied statistics are publicly released with this article\footnote{\url{https://archive.lam.fr/GECO/HOWLS}}.

\section{HOWLS data set}
\label{sec:ds}

To perform a Fisher analysis, one needs to compute data vector (DV) derivatives with respect to individual CPs. We therefore run the DUSTGRAIN-\emph{pathfinder} simulations, varying one parameter at a time among $\Omega_{\rm m}$, $\sigma_8$, and $w$, for four different values around the fiducial ones. We additionally used the DUSTGRAIN-\emph{pathfinder} and SLICS to build the covariance matrix necessary to forecast parameter constraints. The simulations are summarized in Table~\ref{tab:allsim} and described in detail below in Sect.~\ref{sec:ds;subsec:DUST} for DUSTGRAIN-\emph{pathfinder} and Sect.~\ref{sec:ds;subsec:SLICS} for SLICS.

\begin{table} 
\caption[]{Simulations used in HOWLS with CP values for $\Omega_{\rm m}$, $\sigma_8$, and $w$, and the numbers of realizations. The realizations are generated from a single simulation for each of the $13$ DUSTGRAIN-\emph{pathfinder} cosmologies and from $924$ independent simulations for SLICS.}
\centering 
\begin{tabular}{lcccc} 
\hline 
\hline
Name & \# & $\Omega_{\rm m}$ & $\sigma_8$ & $w$ \\
\hline
model &&&&\\
DUSTGRAIN-\emph{path.} Om$--$ & $128$  & $0.2000$ & $0.842$ & $-1.00$ \\
DUSTGRAIN-\emph{path.} Om$-$  & $128$  & $0.3009$ & $0.842$ & $-1.00$ \\
DUSTGRAIN-\emph{path.} Om$+$  & $128$  & $0.3260$ & $0.842$ & $-1.00$ \\
DUSTGRAIN-\emph{path.} Om$++$ & $128$  & $0.4000$ & $0.842$ & $-1.00$ \\
DUSTGRAIN-\emph{path.} s8$--$ & $128$  & $0.3134$ & $0.707$ & $-1.00$ \\
DUSTGRAIN-\emph{path.} s8$-$  & $128$  & $0.3134$ & $0.808$ & $-1.00$ \\
DUSTGRAIN-\emph{path.} s8$+$  & $128$  & $0.3134$ & $0.876$ & $-1.00$ \\
DUSTGRAIN-\emph{path.} s8$++$ & $128$  & $0.3134$ & $0.977$ & $-1.00$ \\
DUSTGRAIN-\emph{path.} w$--$ & $128$  & $0.3134$ & $0.842$ & $-1.16$ \\
DUSTGRAIN-\emph{path.} w$-$  & $128$  & $0.3134$ & $0.842$ & $-1.04$ \\
DUSTGRAIN-\emph{path.} w$+$  & $128$  & $0.3134$ & $0.842$ & $-0.96$ \\
DUSTGRAIN-\emph{path.} w$++$ & $128$ & $0.3134$ & $0.842$ & $-0.84$ \\
\hline
covariance &&&&\\
DUSTGRAIN-\emph{path.} fiducial & $256$  & $0.3134$ & $0.842$ & $-1.00$ \\
SLICS fiducial & $924$  & $0.2905$ & $0.826$ & $-1.00$ \\
\hline
\end{tabular} 
\label{tab:allsim}
\end{table} 

\subsection{DUSTGRAIN-\emph{pathfinder} simulations \done{Carlo, Marco}}
\label{sec:ds;subsec:DUST}
\begin{figure*}
    \centering
    \includegraphics[width=0.99\hsize]{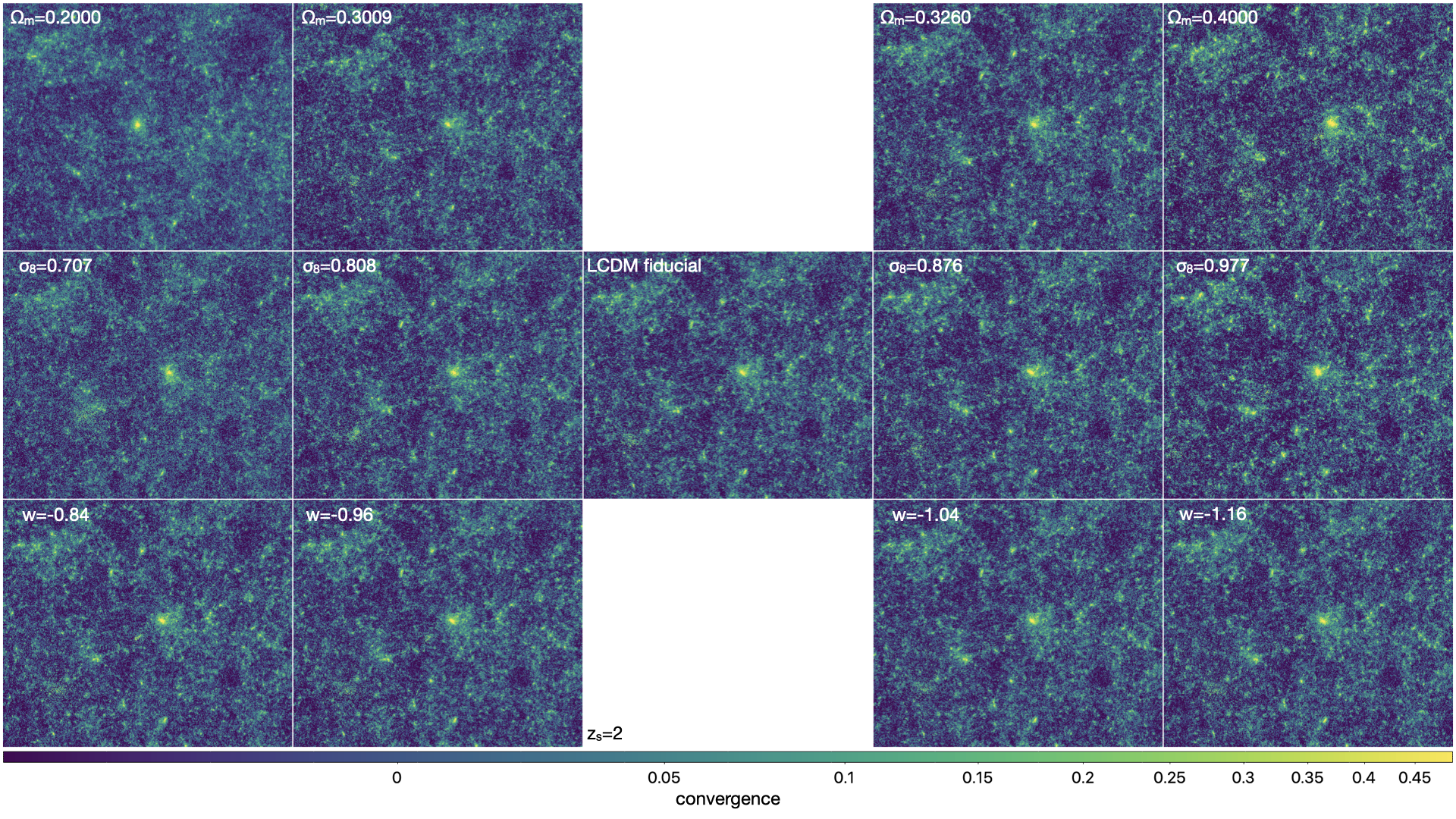}
    \caption{Convergence maps for sources at $z_\mathrm{s}=2$ of the same light-cone random realization constructed from the DUSTGRAIN-\emph{pathfinder} simulations used in this work. The region displayed covers a region of approximately $2.5\times 2.5$ deg$^2$.}
    \label{fig:DPFmaps}
\end{figure*}

\begin{figure*}
    \centering
    \includegraphics[width=0.32\hsize]{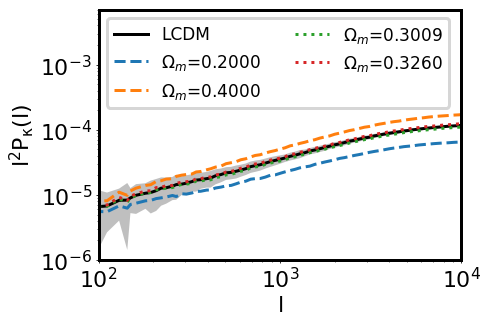}
    \includegraphics[width=0.32\hsize]{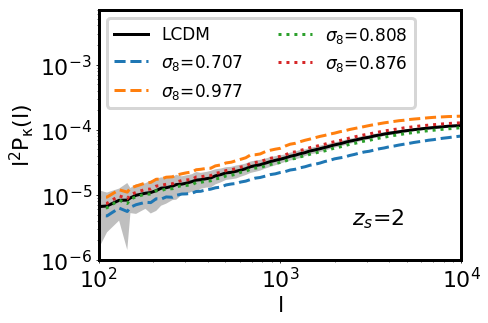}
    \includegraphics[width=0.32\hsize]{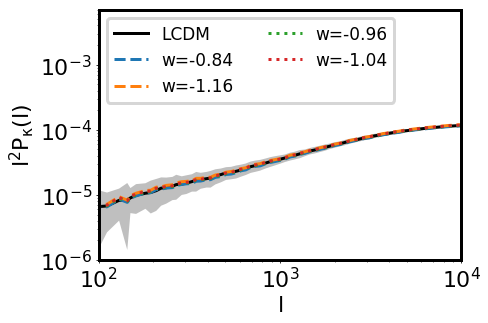}    
    \caption{Convergence power spectra for sources at $z_\mathrm{s}=2$. From left to right, we exhibit the runs that account for $\Omega_\mathrm{m}$, $\sigma_8$ and $w$ variations, respectively. 
    The curves show the average over all $128$ line of sight realizations, and the gray areas show the scatter of the convergence power spectrum around the mean value for the reference $\mathrm{\Lambda}$CDM simulation ($256$ lines of sight) considering a field of view of 5 deg on a side.}
    \label{fig:DPFpowers}
\end{figure*}

The DUSTGRAIN-{\em pathfinder} suite consists of N-body simulations of volume $(750$ $h^{-1}\,$Mpc$)^3$ filled with $N_{\rm p} = 768^3$ particles, corresponding to a particle mass resolution $m_{\rm p}$ of approximately  $8 \times 10^{10} ~ h^{-1}\,M_{\odot}$ \citep{giocoli18b}. The standard reference CPs have been set to values consistent with the results of the Planck-2015 \citep{planck1_15} cosmological data analysis, namely a matter density $\Omega_{\rm m}$ of $0.31345$, a baryon density $\Omega_{\rm b}$ of $0.0491$, Hubble constant $H_0$ of $67.31\,\kmsMpc$, a scalar spectral index $n_{\rm s}$ of $0.9658$, and mean amplitude $\sigma_8$ of the linear density fluctuations on the $8\,h^{-1}\,$Mpc scale of $0.842$. Since the DUSTGRAIN-{\em pathfinder} simulations are part of a cosmological data set that also accounts for modified gravity models, the simulations (including those assuming standard general relativity) were carried out with the MG-Gadget code \citep[][]{Puchwein_etal_2013}.

For the analyses performed for this work, in addition to the reference $\mathrm{\Lambda}$CDM, we used 12 other cosmological runs. In particular, we considered cosmological simulations where only one of the CPs $\Omega_\mathrm{m}$, $\sigma_8$, and $w$ was varied, either by $\pm 4\%$ and $\pm 16\%$ for $\sigma_8$ and $w$, or by ${+28}\%/{-36}\%$ for $\Omega_\mathrm{m}$ to allow for existing data to be reused. When varying $\Omega_\mathrm{m}$ the value of the physical baryon density $\Omega_\mathrm{b}h^{2}$ was kept fixed in the computation of the linear matter power spectrum adopted in the initial conditions, which was performed by means of the Boltzmann code {\small CAMB} \citep[][]{CAMB}.

For each cosmological simulation, we built up mass density planes and then shooting-rays in light cones for $128$ different line-of-sight realizations ($256$ in the case of the fiducial cosmology). The past light cones were built using the MapSim routine \citep{giocoli15} following a pyramidal geometry. This method has been used and tested on a variety of cosmological simulations \citep{tessore15,castro17,giocoli18a} and recently compared with other algorithms \citep{hilbert20} finding only percent-level differences for both cosmic shear two-point and peak statistics. The approach we follow in MapSim is based on using several snapshots from a single realization of an N-body simulation to build a light-cone up to a redshift of 4. For the DUSTGRAIN-{\em pathfinder} suite there are $21$ snapshots available in this redshift range. Given the box length of $750$ $h^{-1}$Mpc, roughly $7$ ($5$) boxes are needed to cover the comoving distance of about $5$ ($3.6$) $h^{-1}\,$Gpc to a source redshift $z_{\rm s}$ of $4$ ($2$). To obtain better redshift sampling, the volume required to construct the light-cone is divided along the line-of-sight into multiple contiguous redshift slices obtained from the individual snapshots. If the redshift slice reaches beyond the boundary of a single box, two lens planes are constructed from a single snapshot. The total number of lens planes up to $z_{\rm s} = 4$ ($z_\mathrm{s} = 2$) is $27$ ($19$). To avoid replicating the same structure along the line of sight, the $7$ boxes needed to cover the light-cone are randomized. This randomization procedure allows us to extract multiple realizations from a single simulation. Randomization is achieved by using seeds that act on the simulation boxes based on: ($i$) changing the location of the observer, typically placed on the center of one of the faces of the box, ($ii$) redefining the center of the box (taking advantage of periodic boundary conditions), and ($iii$) changing the signs of the box axes. From each line of sight realization, we then project, using the Born approximation, to construct convergence and shear maps of $5\times5$ deg$^2$ at various source redshifts. 

In Fig.~\ref{fig:DPFmaps} we exhibit the $\kappa$ maps for our $13$ simulations considering sources at redshift $z_{\rm s}=2$. As can be seen, we display the same line-of-sight realization for the different cosmological simulations: in each subpanel we can recognize by eye the same large-scale structure distribution. More detailed and quantitative statistical information can now be extracted from these maps, for example in Fig.~\ref{fig:DPFpowers} we show the corresponding convergence power spectra, averaged over 128 lines of sight. The left, central and right panels display the $\Omega_{\rm m}$, $\sigma_8$, and $w$ variations, respectively. The black solid line and gray shaded region -- the same in all panels -- exhibit the average convergence power spectra and the corresponding dispersion for the reference $\mathrm{\Lambda}$CDM run.

\subsection{SLICS simulations \done{Joachim}}
\label{sec:ds;subsec:SLICS}

The SLICS are a suite of 924 fully independent N-body simulations specifically designed for the estimation of covariance matrices describing WL observables. They were produced by {\sc cubep$^3$m}, a Poisson solver that computes the nonlinear evolution of $1536^3$ particles starting from initial conditions created at $z=120$ under the Zeldovich approximation \citep{cubep3m}, in boxes of 505 $h^{-1}\,$Mpc on a side. Every run shares the same CPs\footnote{The SLICS cosmology is $\Omega_{\rm m}=0.2905$, $\Omega_{\Lambda}=0.7095$, $\Omega_{\rm b}=0.0447$, $\sigma_8=0.826$, $h=0.6898$, $n_{\rm s}=0.969$, $w=-1.00$.} but embodies a unique noise realization, thereby providing a large ensemble ideally suited for  sample variance estimation. The matter power spectrum agrees to within 2 percent with the {\sc Cosmic Emulator} \citep{CosmicEmu} up to $k = 3.0 ~h^{-1}$ Mpc at $z=0.6$ \citep{SLICS_V1}. 

As detailed in \citet{SLICS}, the particles were collapsed on-the-fly into mass sheets at 18 predetermined redshifts, from which WL light-cones were constructed up to a redshift of $3.0$ following the standard multiple-plane technique. Specifically, convergence and shear maps of 100 deg$^2$ were constructed under the Born approximation at 18 source planes, and subsequently sampled to generate {\it Euclid}-like galaxy mocks with properties listed in Sect. ~\ref{sec:ds;subsec:mock}, similarly to the methods presented in Sect.~\ref{sec:ds;subsec:DUST}. Each plane has a thickness of $257.5 ~ h^{-1}$ Mpc, which, according to \citet{ZorillaMatilla}, results in sub-dominant biases on cosmic shear statistics for upcoming surveys.

The SLICS simulations are publicly available\footnote{SLICS:https://slics.roe.ac.uk} and were used in a number of cosmic shear data analyses including CFHTLenS \citep[e.g.,][]{Joudaki2017}, Kilo Degree Survey \citep{Hildebrandt+17} and Dark Energy Survey \citep{Harnois-Deraps+21} data, as well as clustering data analyses including 2dFLenS \citep{2dFLenS}, GAMA \citep{vanUitert} and BOSS \citep{Xia}. Notably, the covariance matrix estimates of two-point functions have been shown to match well the analytical calculations in \citet{Hildebrandt+17} and \citet{cosmo-SLICS}, leading to comparable constraints on CPs.

Some dissimilarities between the SLICS and the DUSTGRAIN-\emph{pathfinder} simulations are worth noting here, as they might have a small but nonnegligible impact on the results presented in this paper. The two suites are based on distinct N-body codes, which results in residual differences in the nonlinear clustering.  In addition, we note differences in particle count and mass resolution, in the light-cone opening angles, in  pixel sizes\footnote{Although the final convergence maps used in the analysis have the same pixel sizes, the prereconstruction resolution differs: the backbone SLICS pixels are 4.6 arcsec, compared to 8.8 arcsec for the DUSTGRAIN-{\it pathfinder}.} and in the lens randomization procedure. Finally, the shear maps construction pipelines differ in that they are computed directly from the convergence field of view for DUSTGRAIN-\emph{pathfinder}, but over the full periodic box for SLICS, which eliminates residual edge effects and $B$-mode leakage. It is worth mentioning that for the DUSTGRAIN-{\it pathfinder} runs, we assume void boundary condition, with $\kappa=0$ outside the field of view, when computing the shear field. The only scales that are affected are angular modes $l \sim 10^{5}$ and change is below one percent. This last caveat has been also highlighted in \citet{hilbert20}, for a lower resolution map, to have little impact on two-point statistics as also suggested from the good agreement between numerical and theoretical forecasts for $\gamma$-2PCF and $\kappa$-PDF in the present analysis (see Sect.~\ref{sec:Fisher;subsec:theory}). 

\subsection{Mock galaxy catalogs \done{Nicolas}}
\label{sec:ds;subsec:mock}

We generate mock galaxy catalogs by sampling the shear and convergence planes defined earlier from the DUSTGRAIN-\emph{pathfinder} and SLICS simulations. This is to achieve a high degree of realism, allowing us to reproduce \textit{Euclid} survey properties such as the redshift distribution, galaxy density, and shape noise. The procedure closely follows the implementation of \citet{Martinet+21a}, but additionally considers magnitudes in the $\IE$ band of the \textit{Euclid} VIS instrument for sampling the redshift distribution in the case of DUSTGRAIN-\emph{pathfinder}. In contrast to the mentioned article, we do not present results including tomography in the present work. This is delayed to a future HOWLS paper but our mocks already support any tomographic slicing by construction.

\begin{figure}
    \centering
    \includegraphics[width=1.0\hsize]{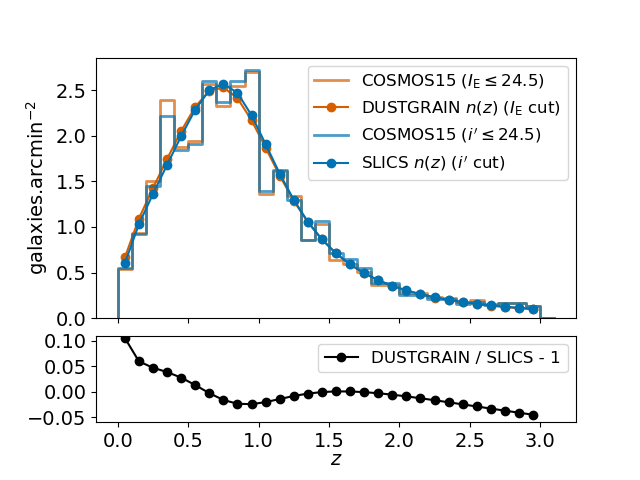}   
    \caption{\textit{Top:} Redshift distributions of the mocks built from the DUSTGRAIN-\emph{pathfinder} (red dots) and SLICS (blue dots) simulations, normalized to $30$ galaxies per arcmin$^2$. These correspond to a \citet{Fu+08} fit to the \citet{Laigle+16} COSMOS2015 catalog after removing galaxies with magnitudes $I_{\scriptscriptstyle\rm E}\geq24.5$ and $i'\geq24.5$ for DUSTGRAIN-\emph{pathfinder} and SLICS, respectively (red and blue histograms). \textit{Bottom:} Fractional difference between the DUSTGRAIN-\emph{pathfinder} and SLICS redshift distributions. The difference is always below $5$\% except for redshifts lower than $0.2$.}
    \label{fig:n(z)}
\end{figure}

The redshift distributions $n(z)$ of the DUSTGRAIN-\emph{pathfinder} and SLICS simulations are shown in Fig.~\ref{fig:n(z)}. They are built from the COSMOS2015 photometric redshift catalogs \citep{Laigle+16} after a cut in magnitudes  $I_{\scriptscriptstyle\rm E}\leq24.5$ in the case of DUSTGRAIN-\emph{pathfinder} and $i'\leq24.5$ for SLICS. This difference is explained by the fact the SLICS mocks were built earlier for \citet{Martinet+21a} than DUSTGRAIN-\emph{pathfinder} and did not yet include the information about the \textit{Euclid} VIS magnitudes (computed as a combination of the $r$, $i'$, and $z'$ magnitudes in the present work). Therefore, the DUSTGRAIN-\emph{pathfinder} $n(z)$ is closer to the expected \textit{Euclid} $n(z)$. The difference is, however, less than $5$\%, except in the redshift range $z\leq0.2$, which contains little information with regard to lensing. Additionally, the model of the dependence on cosmology is built from DUSTGRAIN-\emph{pathfinder} while SLICS enters in the computation of the covariance matrix. This distinction lowers the risk of any impact of the $n(z)$ difference between the two sets of data on the cosmological inference. After the magnitude cut, the COSMOS2015 $n(z)$ is smoothed by fitting the parametrization from \citet{Fu+08} as it was shown in \citet{Martinet+21a} to capture the high-redshift tail better than the standard \citet{Smail+94} fit. The redshift distributions of the DUSTGRAIN-\emph{pathfinder} and SLICS mocks are fully characterized by 
\begin{equation}
\label{eq:Fu}
n(z) = A \frac{z^a+z^{ab}}{z^b+c}\,,
\end{equation}
and the parameters listed in Table~\ref{tab:n(z)}.

\begin{table} 
\caption[]{Parameters of the \citet{Fu+08} redshift distributions of Eq.~(\ref{eq:Fu}) for the DUSTGRAIN-\emph{pathfinder} and SLICS mocks, normalized to $30$ galaxies per arcmin$^{2}$.} 
\centering 
\begin{tabular}{lcc} 
\hline 
\hline
 & DUSTGRAIN-\emph{pathfinder} & SLICS \\
\hline
$A$ (arcmin$^{-2}$) & $1.8048$ & $1.7865$ \\
$a$ & $0.4170$ & $0.4710$ \\
$b$ & $4.8685$ & $5.1843$ \\
$c$ & $0.7841$ & $0.7259$ \\
\hline
\end{tabular} 
\label{tab:n(z)}
\end{table} 

\begin{table*} 
\caption[]{Statistics that have been applied to the HOWLS data set, with their abbreviation in the present article, the filter and smoothing scale employed, the number of independent teams for each statistic and links to the corresponding subsections. Numbers in parentheses refer to additional teams who participated in former HOWLS data sets. When multiple teams exist, we retained only one implementation after coordinating the independent team results. C+02 corresponds to the filter function of \citet{Crittenden2002} and S+07 to that of \citet{Schirmer:2007}.}
\centering 
\begin{tabular}{llllcl} 
\hline 
\hline
Field & Statistics & Abbreviation & Filter \& Scales & Teams & Sections \\
\hline

 $\gamma$ & two-point correlation functions & $\gamma$-2PCF  ($\xi_+/\xi_-$) &  $[\ang{;0,24;},\ang{;8,55;}]/[\ang{;3,51;},\ang{;300;}]$ & $ 1 \, (+1) $ & Sect.~\ref{sec:hos;subsec:2pcf} \\
 $\kappa$ & two-point correlation function & $\kappa$-2PCF ($\xi_\kappa$) &  $[\ang{;0,6;},\ang{;9,23;}]$ & $ 1 \, (+2) $ & Sect.~\ref{sec:hos;subsec:2pcf} \\
 $\kappa$ & one-point probability distribution function& $\kappa$-PDF ($\mathcal P$) & Top-hat, $\ang{;4,69;}$ & $ 1 $ & Sect.~\ref{sec:hos;subsec:PDF} \\
 $\kappa$ & higher-order moments & HOM ($\cal{M}$) & Top-hat,  $\ang{;4,69;}$ & $ 2 \, (+2) $ & Sect.~\ref{sec:hos;subsec:Moments} \\
 $\Map(\gamma)$ & third order moments & $\MapMapMap$  &  C+02,  $[\ang{;1,17;},\ang{;2,34;},\ang{;4,69;},\ang{;9,37;}]$ & $ 1 $ & Sect.~\ref{sec:hos;subsec:Map3} \\
 $\Map(\gamma)$ & $n$-th order moments & $\Mapn$ & Polynomial,  $[\ang{;1,17;},\ang{;2,34;},\ang{;4,69;},\ang{;9,37;}]$ & $ 1 $ & Sect.~\ref{sec:hos;subsec:Map3} \\
 $\Map(\kappa)$ & peak counts & peaks ($N$)& Starlet,  $\ang{;2,34;}$ & $ 1 \, (+2) $ & Sect.~\ref{sec:hos;subsec:peaks} \\
 $\kappa$ & Minkowski functionals & MFs ($V_0$, $V_1$, $V_2$) & Gaussian,  $\ang{;2,34;}$ & $ 1 \, (+2) $ & Sect.~\ref{sec:hos;subsec:MFs} \\
 $\kappa$ & Betti numbers & BNs ($\beta_0$, $\beta_1$) & Gaussian,  $\ang{;2,34;}$ & $ 1 \, (+2) $ & Sect.~\ref{sec:hos;subsec:BNs} \\
 $\Map(\gamma)$ & persistent homology Betti numbers & pers. BNs ($\beta$) & S+07,  $\ang{;2,34;}$ & $ 1 $ & Sect.~\ref{sec:hos;subsec:homology} \\
 $\Map(\gamma)$ & persistent homology heatmap & pers. heat. ($h$) & S+07,  $\ang{;2,34;}$ & $ 1 $ & Sect.~\ref{sec:hos;subsec:homology} \\
 $\kappa$ & scattering transform coefficients &  $s_1, s_2$ & Gaussian,  $\ang{;2,34;}$ & $ 1 $ & Sect.~\ref{sec:hos;subsec:ST} \\

\hline
\end{tabular} 
\label{tab:hos}
\end{table*} 

Shape noise is included by assigning an intrinsic ellipticity to each galaxy. Specifically, we draw each ellipticity component ($\epsilon_{i}$, $i=\{1,2\}$) from a Gaussian random distribution centered on $0$ and with a dispersion $\sigma_{\epsilon_i} = 0.26$. This reference value \citep[e.g.,][]{EuclidIV} has been measured for a sample of galaxies observed with the \textit{Hubble} Space Telescope and with similar photometric properties to the expected VIS sample \citep[magnitudes $I_{814}\sim24.5$, ][]{Schrabback+18}.

The impact of shape noise on the cosmological model is minimized by using the same random realization of galaxy intrinsic ellipticities and positions across all cosmologies for a given mock. This means that we have $128$ independent realizations of shape noise for the DUSTGRAIN-\emph{pathfinder} mocks, but these are identical for the $12$ cosmologies probed. Conversely, the positions and ellipticities are fully random for any realization used in the covariance matrix computation, either with the DUSTGRAIN-\emph{pathfinder} or SLICS simulations. This ensures the shape noise contribution to the error budget is faithfully captured. This process of fixing shape noise in the model and leaving it free in the covariance has become standard practice for simulation-based inference with higher-order mass map estimators \citep[e.g.,][]{Kacprzak+16,Martinet+18,Harnois-Deraps+21}.

These galaxy catalogs are then fed to the \textit{Euclid} convergence map reconstruction pipeline described in \citet{Pires+20} and in Sect.~\ref{sec:ds;subsec:map} to produce the convergence maps on which most HOS will be measured. These catalogs are also used to compute direct statistics from the shear, specifically $\gamma$-2PCFs (Sect.~\ref{sec:hos;subsec:2pcf}) and aperture masses ($\Map$, see Sects.~\ref{sec:hos;subsec:Map3} \&~\ref{sec:hos;subsec:homology}). 

\subsection{Mass mapping \done{Sandrine P.}}
\label{sec:ds;subsec:map}

The statistical properties of the WL field can be assessed by a statistical analysis either of the shear field or of the convergence field. Many HOS are traditionally computed from the $\kappa$ field. This requires solving a mass inversion problem that consists of reconstructing the convergence $\kappa$ from the measured shear field $\gamma$. Using complex notation, the shear field is written as $\gamma = \gamma_1 + {\rm i} \gamma_2 $, and the convergence field as $\kappa = \kappa_{\rm E} + {\rm i} \kappa_{\rm B}$, with $\kappa_{\rm E}$ and $\kappa_{\rm B}$ respectively corresponding to the $E$- and $B$-mode components of the field, by analogy with the electromagnetic field.

We can derive the relation between the shear field $\gamma$ and the convergence field $\kappa$ in the Fourier domain \citep{Kaiser+93} with
\begin{eqnarray}
\label{eq:gamma2kappa}
\hat \kappa =  \hat P^* \, \hat \gamma\;,
\end{eqnarray}
where the hat symbol denotes Fourier transforms, $\hat P^*$ is the complex conjugate, and $\hat P = \hat P_1 + {\rm i} \hat P_2$ with
\begin{eqnarray}
\hat{P_1}(\vec{\ell}) & = & \frac{\ell_1^2 - \ell_2^2}{\ell^2}\,, \nonumber \\
\hat{P_2}(\vec{\ell}) & = & \frac{2 \ell_1 \ell_2}{\ell^2}\,,
\end{eqnarray}
with $\ell^2 \equiv \ell_1^2 + \ell_2^2$ and $\ell_i$ the Fourier counterparts of the angular coordinates $\theta_i$. 
The convergence can only be determined up to an additive constant because there is a degeneracy when $\ell_1$ = $\ell_2$ = 0 \citep[see e.g.,][]{Bartelmann+95}. In practice, we impose that the mean convergence vanishes across the field by setting the reconstructed $\ell = 0$ mode to zero.

Assuming the mass inversion is conducted without noise regularization, the same information is contained in the shear field as in the convergence maps.
However, it is well known that the Kaiser-Squires inversion creates undesirable artifacts at the borders of the reconstructed convergence maps \citep[see e.g.,][]{Seitz+96,Seitz+01,Pires+20}. This is due to the fact that the discrete Fourier transform implicitly assumes periodicity of the image along both dimensions. In a future work, mass mapping systematic effects will be further mitigated and their impact quantified by propagating the errors into CP forecasts using HOS. In addition to border effects and masks, we will test the impact of reduced shear. In this article, however, we assume that the mean ellipticity is an unbiased estimator of the mean shear instead of reduced shear such that we can replace $\gamma$ by $\epsilon$ in Eq.~(\ref{eq:gamma2kappa}), an assumption only correct in the weak regime.

The shear is sampled only at the positions of the galaxies. Therefore, the first step of the mass inversion method is to bin the observed ellipticities of galaxies on a regular pixel grid to create the shear maps. In practice, we bin the galaxies in pixels of size of \ang{;0.59;}, resulting in shear maps of $512 \times 512$ pixels (for DUSTGRAIN-\emph{pathfinder}) and $1024 \times 1024$ pixels (for SLICS) that are then converted into convergence maps using Eq.~\eqref{eq:gamma2kappa}.
The statistical analysis is performed only on the $E$-modes convergence maps because WL only produces $E$-modes. However, the application of the HOS to the $B$-modes map can be used to test for residual systematic effects.

\section{Statistics \done{Vincenzo, Nicolas}}
\label{sec:hos}

Many higher-order probes have been proposed, tested, and measured on present-day Stage III lensing maps with promising preliminary results \citep[e.g.,][]{Martinet+18,Gatti2021,Harnois-Deraps+21,Heydenreich:2022,Zurcher+22,burger+22}. This consideration motivated us to focus our analysis on HOS of scalar fields derived from the shear: convergence and aperture mass. In the following paragraphs, we review the probes we consider, referring the interested reader to the quoted papers for further details. Far from being fully exhaustive, this short review aims at presenting the tools we have used and giving the reader an overview of the many roads that open up when going beyond second-order statistics. A short description of theoretical predictions is given for the 2PCF and the $\kappa$-PDF as they are used to validate the simulated derivatives entering the Fisher analysis, see Section~\ref{sec:Fisher;subsec:theory}. The impatient reader can directly look at Table~\ref{tab:hos}, where we list the HOS we have used together with their abbreviations, the number of independent teams that applied them, and the subsections in which they are described. The fact that each statistic was computed by a different team led to a variety of choices in terms of the filtering of the shear or convergence field. For this publication we do not try to homogenize these choices as we consider it part of each method and list these differences in Table~\ref{tab:hos}.

\subsection{Two-point correlation functions \done{Vincenzo, Nicolas, Simone}}
\label{sec:hos;subsec:2pcf}

Although useful on their own, HOS are at their best when used in combination with standard second-order statistics, coupling the typically larger signal-to-noise ratio (S/N) of lower-order statistics with the degeneracy-breaking power of HOS. In the following, we therefore quantify both the constraints from each HOS probe alone and the improvement of the constraints from joint second- and higher-order statistics with respect to the second-order-only case.  

The most basic cosmic shear observable is the real-space shear two-point correlation functions ($\gamma$-2PCF), since it can be estimated by simply multiplying the ellipticities of galaxy pairs and averaging. The shear can conveniently be decomposed into a tangential, $\gamma_{\rm t}$, and cross\,-\,component, $\gamma_{\cross}$, such that
\begin{equation}
\gamma_{\rm t}(\varthetavec',\varthetavec) = -{\textswab{R}} [ \gamma(\varthetavec') {\rm e}^{-2 {\rm i} \phi} ] \qquad 
\gamma_{\cross}(\varthetavec',\varthetavec) = -{\textswab{I}}[\gamma(\varthetavec') {\rm e}^{-2 {\rm i} \phi}] \;,
\end{equation}
with \textswab{R} and \textswab{I} the real and imaginary parts, $\phi$ the polar angle of the direction vector between the galaxy position $\varthetavec'$ and a reference point $\varthetavec$, and the minus sign a convention to have the tangential shear positive around a mass overdensity. The two shear components can then be combined to get two 2PCFs 
\citep[see, e.g.,][and references therein]{KilbingerReview}
\begin{equation}
\left \{
\begin{array}{l}
\displaystyle{\xi_{+}(\theta) = \langle \gamma \gamma^* \rangle(\theta) 
= \langle \gamma_{\rm t} \gamma_{\rm t} \rangle(\theta) + 
\langle \gamma_{\cross} \gamma_{\cross} \rangle(\theta)} \\ 
 \\ 
\displaystyle{\xi_{-}(\theta) = {\textswab{R}}[\langle \gamma \gamma \rangle(\theta) {\rm e}^{-4 {\rm i} \phi}] =  
\langle \gamma_{\rm t} \gamma_{\rm t} \rangle(\theta) - 
\langle \gamma_{\cross} \gamma_{\cross} \rangle(\theta)} \\ 
\end{array}
\right . ,
\label{eq: 2pcfdef}
\end{equation}
where the dependence is only on the angular separation $\theta$ on the sky because under the Cosmological Principle cosmic fields are statistically invariant under translation and rotation. 

The main virtue of these $\gamma$-2PCF is that they can be straightforwardly estimated from the measured ellipticities $\varepsilon_i$ as 
\begin{align}
\xi_{\pm}(\theta) = 
\frac{\sum_{i,j}{w_i w_j
(\varepsilon_{{\rm t},i} \varepsilon_{{\rm t},j} \pm 
\varepsilon_{\cross,i} \varepsilon_{\cross,j})}}
{\sum_{i,j}{w_i w_j}},
\end{align}
where the sum runs over the galaxy pairs with positions on the sky $(\thetavec_i, \thetavec_j)$ having angular separation $|\thetavec_i - \thetavec_j|$ in a bin centered on $\theta$. The weight $w_i$ of the ellipticity $\varepsilon_i$ can be related to the measurement error, and set to zero if the galaxy is in a masked region. In the present article the lensing weights are set to $1$, as the impact of masks and shear measurement methods are beyond the scope of this analyis.

The $\gamma$-2PCF can be easily computed for a given cosmological model by first going to Fourier space and then converting back into real space. The final result is
\begin{equation}
\left \{
\begin{array}{l}
\displaystyle{\xi_{+}(\theta) = 
\frac{1}{2 \pi} \int{\left[P_{\kappa}^{E}(\ell) + P_{\kappa}^{B}(\ell)\right] \, J_0(\ell \theta) \,\ell\, {\rm d}\ell}} \\ 
\displaystyle{\xi_{-}(\theta) = 
\frac{1}{2 \pi} \int{\left[P_{\kappa}^{E}(\ell) - P_{\kappa}^{B}(\ell) \right] \, J_4(\ell \theta)\, \ell\, {\rm d}\ell}} \\ 
\end{array}
\right. \;,
\label{eq: 2pcfhankel}
\end{equation}
where $P_{\kappa}^{E}(\ell)$ and $P_{\kappa}^{B}(\ell)$ are the power spectra of the convergence $E$ and $B$ modes, while $J_n(x)$ is the $n$-th order spherical Bessel function of the first kind. We note that, in the absence of systematics and neglecting higher-order effects, WL does not produce $B$ modes so that $P_{\kappa}^{B}(\ell) = 0$, and $\xi_{\pm}$ reduce to different Hankel transforms of the same quantity, which can be derived from the matter power spectrum $P_{\delta}(\ell)$.

As done for the shear, we can similarly define the $\kappa$-2PCF for the convergence. In particular, this will inherit the same properties of translation and rotation invariance as in the shear case. The main difference is that, with the convergence being a scalar quantity, there is only one single correlation function. In Fourier space, this can be related to the convergence power spectrum as 
\begin{equation}
\label{eq:kappa2PCF}
    \expval{\hat{\kappa}(\ellvec) \hat{\kappa}^{\ast}(\ellvec')} = (2 \pi)^2 \delta_\mathrm{D} (\ellvec - \ellvec') P_{\kappa}(\ell)\;,
\end{equation}
where $\delta_\mathrm{D}$ is the Dirac-$\delta$ function and the convergence power spectrum depends only on the modulus of $\ellvec$ due to the statistical homogeneity and isotropy. It is then possible to show that $P_{\kappa}(\ell) = P_{\kappa}^{E}(\ell)$, and the $\kappa$-2PCF is given by 
\begin{equation}
\xi_{\kappa}(\theta) = \frac{1}{2 \pi} \int P_{\kappa}(\ell)\, J_0(\ell \theta)\, \ell\; \dd{\ell}\; ,
\label{eq: xikappa}
\end{equation}
which is the same as $\xi_{+}(\theta)$. In the following, we keep this different label in order to distinguish between the $\gamma$-2PCF and $\kappa$-2PCF. 

\begin{figure}
    \centering
    \includegraphics[width=1.0\hsize]{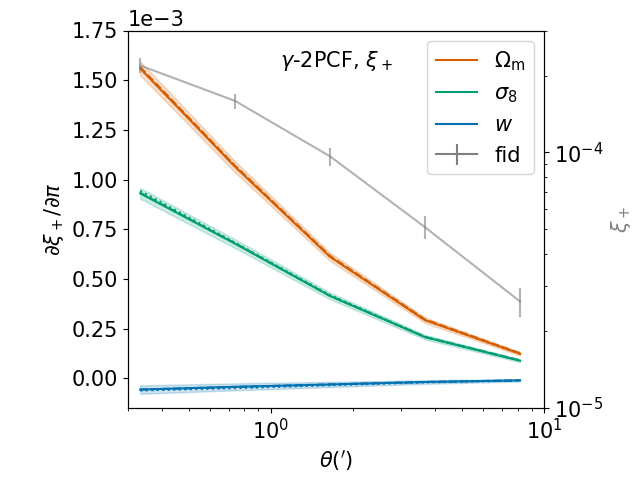}   
    \includegraphics[width=1.0\hsize]{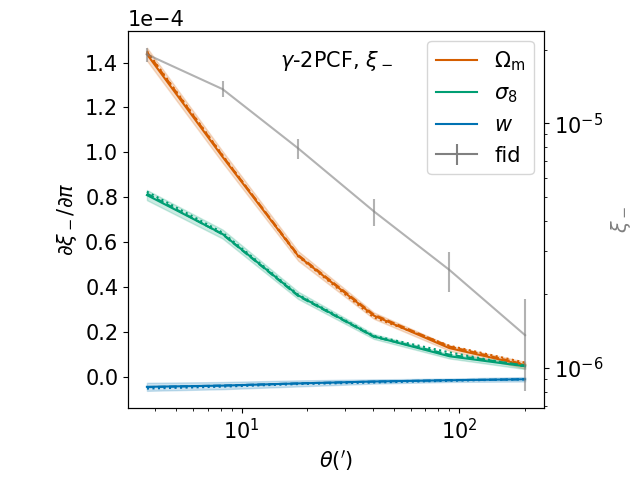}   
    \caption{$\gamma$-2PCF $\xi_+$ (\textit{top}) and $\xi_-$ (\textit{bottom}) DVs and derivatives. The gray lines, whose scale is given by the second axis, correspond to the average DVs computed from the $924$ SLICS realizations. The Fisher derivatives (defined in Eq.~\ref{eq: der3p}) are computed from the DUSTGRAIN-\emph{pathfinder} simulations with large variations of $\Omega_{\rm m}$ (orange), $\sigma_8$ (green), and $w$ (blue). The solid, dashed, and dotted lines respectively correspond to the average over $128$, $64$, and $32$ realizations. The shaded areas represent the uncertainty computed from the $128$ DUSTGRAIN-\emph{pathfinder} realizations, and the gray error bars those of the $924$ SLICS realizations. The inclusion of the dashed and dotted lines within the shaded areas highlights the low numerical noise.}
    \label{fig:DV-g2PCF}
\end{figure}

\begin{figure}
    \centering
    \includegraphics[width=1.0\hsize]{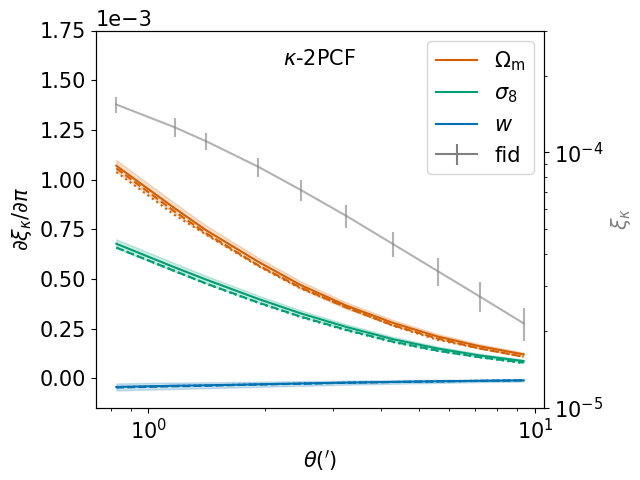}   
    \caption{Fiducial DV for $\kappa$-2PCF (gray) along with its derivatives with respect to $\Omega_{\rm m}$ (orange), $\sigma_8$ (green), and $w$ (blue).}
    \label{fig:DV-k2PCF}
\end{figure}

In this article, the $\gamma$-2PCF has been computed with the \texttt{ATHENA} software \citep{Kilbinger+14} using $10$ logarithmic bins between $\ang{;0.1;}$ and $\ang{;300;}$. The $\kappa$-2PCF has been measured using the public code \texttt{TreeCorr} \citep{Jarvis+04}. The minimum separation considered is $\ang{;0.59;}$ (corresponding to the pixel scale) and the maximum separation is approximately $\ang{;424;}$ (corresponding to the DUSTGRAIN-\emph{pathfinder} map diagonal), with 25 bins. As discussed in Sect.~\ref{sec:Fisher} several bins are later removed to pass our quality criteria, with the final scale range described in Table~\ref{tab:hos}.  We also show these DVs as well as their standard Fisher derivatives with respect to $\Omega_{\rm m}$, $\sigma_8$, and $w$ in Figs.~\ref{fig:DV-g2PCF} \&~\ref{fig:DV-k2PCF} for $\gamma$-2PCF and $\kappa$-2PCF respectively. Finally, we note some difference in the chosen binning by the different teams computing the 2PCFs which leads to some artificial difference between $\xi_+$ and $\xi_\kappa$, for example.

\subsection{Convergence PDF \done{Cora, Alexandre, Sandrine, Aoife}}
\label{sec:hos;subsec:PDF}

The one-point probability distribution function (PDF) of the WL convergence $\kappa$ encodes vital information about the non-Gaussian late-time density field. In the presence of shape noise, which dominates on small scales, this information can be extracted most conveniently from the PDF of the convergence after smoothing on an angular scale big enough that the variance set by the gravitational clustering is larger than the shape noise contribution.
As a one-point statistic, the PDF is straightforward to measure from simulated and real data, as evidenced by recent analyses of density-split statistics and weak lensing moments in the Dark Energy Survey \citep{Friedrich18,Gruen18,Gatti19,Gatti2021}, which carry PDF information in a compressed way and can deal with survey mask effects.

When focusing on mildly nonlinear scales, typically corresponding to smoothing scales of about \ang{;10;} at low redshifts, this one-point PDF also admits accurate theoretical predictions \citep[see for example][]{Barthelemy20a,Boyle2020}. The main idea of this theoretical model relies on the fact that, within the Limber approximation, since the WL convergence probes the matter density in a cone whose opening angle is set by the smoothing scale, the $\kappa$-PDF can be predicted using cylindrical collapse applied to the 2D slices of the density in the circular cross-sections. Hence, for a set of sources regrouped in redshift bins and with a certain source distribution across the survey $n(z)$, we recall that the theoretical cumulant generating function (CGF) $\phi_{\kappa,\theta}$ at a given angular smoothing scale $\theta$ can be expressed as
\begin{equation}
    \phi_{\kappa,\theta}(\lambda) = \int \frac{{\rm d}z \, c}{H(z)} \, \phi_{\rm cyl}[\omega_{n(z)}(z) \lambda,z]\;,
    \label{CGF}
\end{equation}
with $\phi_{\rm cyl}$ the CGF of the density in each individual 2D slice of radius $\chi(z)\theta$, and $\omega_{n(z)}$ the generalized lensing kernel given a wide distribution of sources following the normalized distribution $n(z)$ \begin{equation}
\label{eq:lensing_kernel}
    \omega_{n(z)}(z) = \frac{3\Omega_{\rm m} H_0^2}{2 c^2}\!\!\int\!\! {\rm d}z_\mathrm{s} n(z_\mathrm{s})  \frac{[\chi(z_\mathrm{s})-\chi(z)]\,\chi(z)}{\chi(z_\mathrm{s})}\,\mathcal{H}(z_\mathrm{s}-z)\,(1+z)\;,
\end{equation}
where the Heaviside $\mathcal{H}$ ensures that the integrand vanishes for $z \geq z_\mathrm{s}$.
$\phi_{\rm cyl}$ is obtained from the cylindrical collapse dynamics as a proxy for the whole nonlinear evolution in cylindrically symmetric configurations (top-hat smoothing) and enforces a specific hierarchy of cumulants. The PDF is then recovered through the inverse Laplace transform of the exponential of the $\kappa$ CGF as
\begin{equation}
    \mathcal{P}_{\theta}(\kappa)=\int_{-i \infty}^{+i \infty} \frac{\mathrm{d} \lambda}{2 \pi \mathrm{i}} \exp \left[-\lambda \kappa+\phi_{\kappa, \theta}(\lambda)\right]\;.
    \label{Laplace_PDF}
\end{equation}
Note that the cylindrical collapse model is accurate enough to predict the so-called reduced cumulants of the (2D) density field $\left(S_n = \langle\delta^n\rangle_{\rm c}/\sigma_{\delta}^{2(n-1)}\right)$, where $\langle\delta^n\rangle_{\rm c}$ are the cumulants of the matter density. This means that the model also requires an external input for the prediction of the nonlinear variance of all the 2D slices of the density along the line of sight. Fortunately, the emulation of the matter power spectrum has received lots of attention these past years and can be estimated rather accurately, for example by using the revised Halofit model \citep{Takahashi+12} or the Euclid Emulator \citep{EuclidEmulator}. Galaxy shape noise can be included in the theoretical prediction through a convolution of the noiseless PDF as described in \cite{Boyle2020} or directly at the level of the CGF by simple addition of -- for example -- an associated shape noise variance $\sigma^2_{\rm SN} \lambda^2/2$.

\begin{figure}
    \centering
    \includegraphics[width=1.0\hsize]{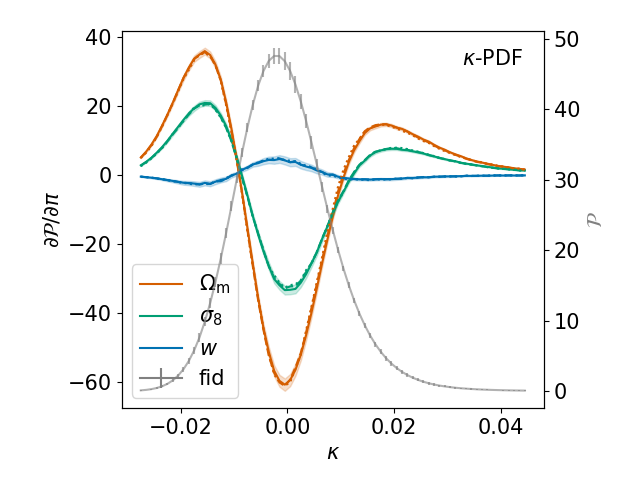}   
    \caption{Fiducial DV for $\kappa$-PDF (gray) along with its Fisher derivatives with respect to $\Omega_{\rm m}$ (orange), $\sigma_8$ (green), and $w$ (blue). The $\kappa$-PDF was computed on $\kappa$ maps smoothed with a top-hat filter with a scale of $\ang{;4.69;}$.}
    \label{fig:DV-kPDF}
\end{figure}

We extracted the $\kappa$-PDF from the simulated maps for two top-hat filters with smoothing scales $\theta_\mathrm{s}\in\{\ang{;4.69;}, \ang{;9.37;}\}$, corresponding to 8/16 pixels in the DUSTGRAIN-\emph{pathfinder} simulations. After smoothing the maps with the appropriate top-hat filter, we excluded all pixels whose smoothing circle of radius $\theta_\mathrm{s}$ would intersect the patch boundary.\footnote{When dealing with masks, this process can be replaced by only removing pixels whose smoothing circle covers less than a given threshold value (like 90\%) in the considered patch and reweighting the $\kappa$ values in the cells according to the coverage, as done in \cite{Friedrich18,Gruen18}.}
The PDF was obtained from a histogram using 201 linearly spaced bins in the range $\kappa\in[-0.1,0.1]$. Cuts were made in the considered range of $\kappa$ to exclude the extreme tails and keep the DV close to Gaussian as discussed in Sect.~\ref{subsec:datasel}. The precise $\kappa$ values are given in Table \ref{tab:hos_gauss_bin} and Fig.~\ref{fig:DV-kPDF}. The cuts exclude 0.05\% and 0.25\% of the cumulative probability from the low-$\kappa$ and high-$\kappa$ tails, respectively, for the smaller smoothing scale. We subsequently compare the theoretical predictions to the measurements in the simulation in Sect.~\ref{sec:Fisher;subsec:theory}.

One alternative to the $\kappa$-PDF is to turn to the (slightly) more involved -- but also more easily accessible through observations -- modeling of the one-point PDF of the aperture mass from Eq.~\eqref{eq:def_aperturemass} discussed in Sect.~\ref{sec:hos;subsec:Map3}. The analogous theoretical model was developed in \cite{2021MNRAS.503.5204B}. The next extension would be to include some tomographic information in the analysis. At the level of the PDF one could rely on the ideas developed in \cite{2022PhRvD.105d3537B} for the joint PDF between -- built to be -- independent lensing kernels that probe structures along the line of sight, but the generalization to any other lensing observables should be straightforward.

\subsection{Higher-order convergence moments \done{Vincenzo}}
\label{sec:hos;subsec:Moments}

The shear and convergence 2PCF and their harmonic space counterparts, the power spectra, are related to the variance of the lensing fields. It is then natural to ask whether additional information is encoded in moments higher than two, given that they are indicators of the non-Gaussianity of the field. We consider here the second, third, and fourth moments of the convergence field.  

Moments of the smoothed WL convergence field $\kappa_\theta$ can  be calculated from weighted averages of the lensing PDF as $\langle\kappa^n\rangle(\theta)=\int \dd{\kappa}\; \kappa^n\, \mathcal P_\theta(\kappa)$,
assuming a lensing convergence field of zero mean. The variance $\langle\kappa^2\rangle=\sigma^2$ fully characterizes a Gaussian one-point distribution, while
the skewness $\langle\kappa^3\rangle/\sigma^3$ and kurtosis $\langle\kappa^4\rangle/\sigma^4-3$ are the most common examples encoding non-Gaussian information. While an analysis of the variance, skewness, and kurtosis promises a significant compression of the $\kappa$-PDF DV, higher-order cumulants are known to be very sensitive to the tails of the distribution, rendering their signal and likelihood hard to predict.

An approximated yet accurate formula for the moments can be obtained under the hierarchical ansatz \citep{BS92,SS93}, using this ansatz to express the bispectrum and trispectrum in terms of the power spectrum and plugging them into the general expression for the convergence moments \citep[see][for details]{MJ01,Vicinanza_2019}.

\begin{figure}
    \centering
    \includegraphics[width=1.0\hsize]{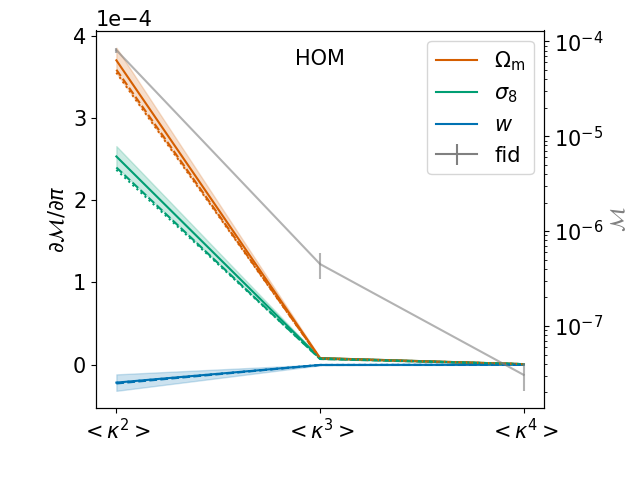}   
    \caption{Fiducial DV (gray) along with its derivatives with respect to $\Omega_{\rm m}$ (orange), $\sigma_8$ (green), and $w$ (blue) for HOM computed on $\kappa$ maps smoothed with a top-hat filter with a scale of $\ang{;4.69;}$.}
    \label{fig:DV-HOM}
\end{figure}

It is worth noting that moments are a function of smoothing radius, so one typically considers as a cosmological probe their scaling with $\theta$. In contrast, here we use them for a fixed $\theta$ since, in a preliminary analysis, we have found moments obtained from different smoothing radii to be highly correlated. Although some information is still present in the correlated moments, we prefer here to first focus on the choice of the filter radius in our two parameter analysis for which combining moments at different radii is not essential. We, therefore, measure them like the $\kappa$-PDF using a top-hat filter with two different values of the aperture radius, namely $\theta\in\{\ang{;4.69;}, \ang{;9.37;}\}$, and take $\left \{ \langle \kappa^2 \rangle(\theta), \langle \kappa^3 \rangle(\theta), \langle \kappa^4 \rangle(\theta) \right \}$ for a fixed $\theta$ as our DV (see Fig.~\ref{fig:DV-HOM}). To this end, we first define

\begin{displaymath}
\kappa_\theta(\varthetavec) = \int \dd[2]{\vartheta'} \cal{H}_\theta(\varthetavec'-\varthetavec)\, \kappa(\varthetavec')
\end{displaymath}
as the smoothed convergence field. We then take it to the second, third, fourth power and compute its average value over the pixels to compute the data vector for every single map. The mean over the different realizations of the simulations makes up our final DV.

\subsection{Higher-order aperture mass moments \done{Sven, Laila, Lucas}}
\label{sec:hos;subsec:Map3}

Higher-order moments as introduced before were extracted for the smoothed convergence maps.
However, it is also possible to compute them for the aperture mass $\Map$ map, at sky location $\varthetavec$ and scale $\theta$ given by \citep{Schneider1998}:

\begin{equation}
\label{eq:def_aperturemass}
    \Map(\varthetavec, \theta)=\int \dd[2]{\vartheta'}\; U_\theta(|\varthetavec - \varthetavec'|)\, \kappa(\varthetavec')\;,
\end{equation}
where $U_\theta$ is a compensated filter function with
\begin{equation}
\label{eq:def_compensated}
    \int \dd[2]{\vartheta}\; U_\theta(|\varthetavec|) = 0\;.
\end{equation}
The higher-order aperture maps statistics $\Mapn$ correlate the aperture masses on $n$ different scales $\theta_1$, $\cdots$,\ $\theta_n$, and are defined as
\begin{equation}
    \Mapn(\theta_1, \cdots, \theta_n) = \expval{\Map(\varthetavec, \theta_1)\, \cdots \, \Map(\varthetavec, \theta_n)}\;.
\end{equation}

The $\MapMapMap$ is closely related to the third-order moment of the convergence maps since both of them are sensitive to the matter bispectrum \citep{Schneider2005}. However, in contrast to other HOS considered in this work, the $\MapMapMap$ can be directly inferred from the shear and are therefore not affected by systematics induced by the convergence reconstruction. Additionally $\MapMapMap$ can be inferred from the shear three-point correlation functions $\Gamma_0=\expval{\gamma\, \gamma\, \gamma}$ and $\Gamma_1=\expval{\gamma^*\, \gamma\, \gamma}$, which can be easily estimated even for irregular survey geometries and in the presence of masks \citep{Schneider2003, Heydenreich2022}. Consequently, $\MapMapMap$ is more straightforward to measure in a realistic survey setting than HOS of $\kappa$.

We infer the $\MapMapMap$ from the tangential shear $\gamma_\mathrm{t}(\varthetavec';\varthetavec)$ at position $\varthetavec'$ with respect to $\varthetavec$. This inference is possible because each filter $U_\theta$ is associated with a function $Q_\theta$ for which
\begin{equation}
    \label{eq: Map3 from gamma_t}
    \Map(\varthetavec, \theta)=\int \dd[2]{\vartheta'}\; Q_\theta(|\varthetavec - \varthetavec'|)\, \gamma_\mathrm{t}(\varthetavec';\varthetavec)\;.
\end{equation}
The function $Q_\theta$ is given by
\begin{equation}
    Q_\theta(\vartheta)=\frac{2}{\vartheta^2}\, \int_0^\vartheta \dd{\vartheta'}\; \vartheta'\, U_\theta(\vartheta') - U_\theta(\vartheta)\;.
\end{equation}
To mimic the application of the $\MapMapMap$ statistics to a survey, we use Eq.~\eqref{eq: Map3 from gamma_t} and work directly with the simulated shear catalogs. We use the filter function $U_\theta$ from \citet{Crittenden2002},
\begin{equation}
    U_\theta(\vartheta)=\frac{1}{2\pi\, \theta^2}\left(1-\frac{\vartheta^2}{2\theta^2}\right)\, \exp(-\frac{\vartheta^2}{2\theta^2})\;,
\end{equation}
and the scale radii $\theta \in \{\astroang{;1.17;}, \astroang{;2.34;}, \astroang{;4.69;}, \astroang{;9.37;}\}$.

\begin{figure}
    \centering
    \includegraphics[width=1.0\hsize]{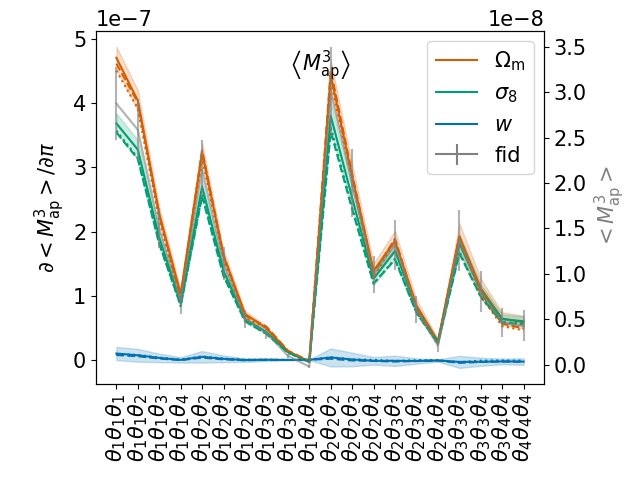}   
    \caption{Fiducial DV (gray) along with its derivatives with respect to $\Omega_{\rm m}$ (orange), $\sigma_8$ (green), and $w$ (blue) for $\MapMapMap$ of aperture mass maps computed from shear fields smoothed with a \citet{Crittenden2002} filter with scales $(\theta_1,\theta_2,\theta_3,\theta_4) = (\ang{;1.17;}, \ang{;2.34;}, \ang{;4.69;}, \ang{;9.37;})$.}
    \label{fig:DV-Map3}
\end{figure}

Our measurement of $\MapMapMap$ proceeds in two steps following the procedure in Section 5.3.1 of \citet{Heydenreich2022}. First, we measure $\Map(\varthetavec, \theta)$. For this, we employ the convolution theorem to solve the convolution of $Q_\theta$ and $\gamma_\mathrm{t}$ in Fourier space. Since the tangential shear $\gamma_\mathrm{t}$ can be written as 
\begin{equation}
    \gamma_\mathrm{t}(\varthetavec',\varthetavec) = - {\textswab{R}} \left[ \gamma(\varthetavec')\frac{(\varthetavec-\varthetavec')^*}{\varthetavec-\varthetavec'} \right] \; ,
\end{equation}
where the vectors are interpreted as complex numbers $\varthetavec = \vartheta_1+\mathrm{i}\vartheta_2$, Eq.~\eqref{eq: Map3 from gamma_t} transforms into
\begin{equation}
    \label{eq:map_from_gamma_via_convolution}
    \Map(\varthetavec, \theta)=-\int \dd[2]{\vartheta'}\; Q_\theta(|\varthetavec - \varthetavec'|)\, \frac{(\varthetavec-\varthetavec')^*}{\varthetavec-\varthetavec'}\,\gamma(\varthetavec')\;.
\end{equation}
Therefore, we can calculate both $Q_\theta(|\varthetavec|)\, \frac{\varthetavec^*}{\varthetavec}$ and $\gamma(\varthetavec)$ on a grid, and solve the convolution in Eq.~\eqref{eq:map_from_gamma_via_convolution} using the Fast Fourier Transform (FFT). To avoid border effects, we cut off a strip of width of $4\theta$ from each border. This large cut-off is needed because the exponential aperture filter is not exactly zero for $\vartheta > \theta$, so at distance $\theta$, we still experience border effects from each side of the $\Map(\varthetavec, \theta)$ map. Our cut-off means we neglect $0.07\%$ of the total filter power, which we deem acceptable. 
Second, we measure $\MapMapMap(\theta_1, \theta_2, \theta_3)$. For this, we multiply $\Map(\varthetavec, \theta_1)$, $\Map(\varthetavec, \theta_2)$, and $\Map(\varthetavec, \theta_3)$ for each line-of-sight and each position $\varthetavec$. Then, we average over $\varthetavec$, which gives $\MapMapMap(\theta_1, \theta_2, \theta_3)$ for each line-of-sight. This DV is shown in Fig.~\ref{fig:DV-Map3}.

\begin{figure}
    \centering
    \includegraphics[width=1.0\hsize]{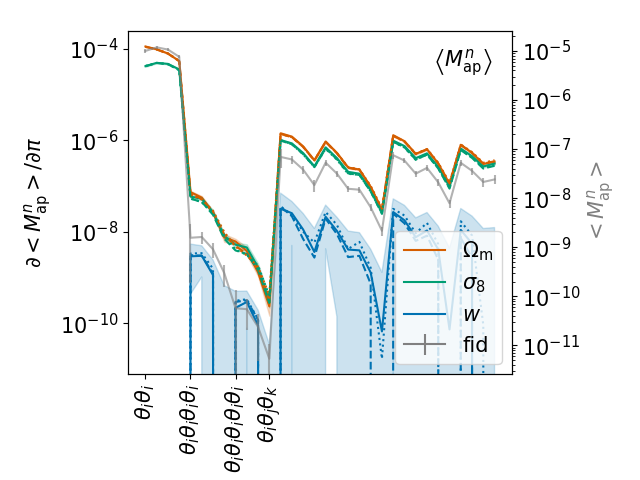}   
    \caption{Fiducial DV (gray) along with its derivatives with respect to $\Omega_{\rm m}$ (orange), $\sigma_8$ (green), and $w$ (blue) for $\Mapn$ of aperture mass maps computed from shear fields smoothed with a polynomial filter with scales $\theta_{i,j,k}\in\{\ang{;1.17;}, \ang{;2.34;}, \ang{;4.69;}, \ang{;9.37;}\}$. We first display second, fourth, and fifth identical scale moments, followed by the third order moments ordered identically to Fig.~\ref{fig:DV-Map3}.}
    \label{fig:DV-Mapn}
\end{figure}

We estimate the $n$-th order moments $\Mapn$ (Fig.~\ref{fig:DV-Mapn}) with a different approach. For this we cover the survey footprint with apertures and estimate $\Mapn$ within each aperture \citep{Schneider1998} as
\begin{align}\label{eq:MapnEstimator}
    \widehat{M_{\mathrm{ap}}^n}(\varthetavec;\theta_1, \cdots, \theta_n) &= \left(\pi \theta_1^2\right) \cdots \left(\pi \theta_n^2\right) 
    \nonumber \\
    &\hspace{-1cm} \times 
    \frac{\sum_{i_1\neq i_2 \neq \cdots \neq i_n} w_{i_1}Q_{\theta_1;i_1}\varepsilon_{\mathrm{t},i_1}\cdots w_{i_n}Q_{\theta_n;i_n}\varepsilon_{\mathrm{t},i_n}}{\sum_{i_1\neq i_2 \neq \cdots \neq i_n}w_{i_1}\cdots w_{i_n}} \ ,
\end{align}
where each $w_i$ is the weight of the ellipticity $\varepsilon_i$ and we abbreviated $Q_{\theta_j}(|\varthetavec_{i_j}-\varthetavec|) \equiv Q_{\theta_j;i_j} $ for an aperture centered at $\varthetavec$. Note that the index $i_k$ in the $k$th sum in \eqref{eq:MapnEstimator} only considers galaxies that lie within the support of $Q_{\theta_k}$. By averaging over all apertures in the footprint with appropriate weights $w_{\mathrm{ap}}$, one then obtains an unbiased estimate for $\Mapn$. As shown in \citet{Porth+20} and \citet{Porth+21}, one can decompose the nested sums appearing in \eqref{eq:MapnEstimator}, so that the full estimation process scales linearly with the number of galaxies. In this work, we estimate the connected parts of the aperture-mass statistics for $n \in \{2,3,4,5\}$, where for $n=3$ we take into account all of the different combinations of aperture radii and in the other cases only consider the components for which  $\theta_1=\dots=\theta_n$. In contrast to the FFT-based estimation procedure described above, we employ a polynomial filter function introduced in \cite{Schneider1998} for the direct estimator method,
\begin{equation}\label{eq:MapnFilter}
    Q_\theta(\vartheta) = \frac{6}{\pi\theta^2} \left(\frac{\vartheta}{\theta}\right)^2\left[1-\left(\frac{\vartheta}{\theta}\right)^2\right]\mathcal{H}(\theta-\vartheta) \ ,
\end{equation}
where $\mathcal{H}(x)$ denotes the Heaviside function. 
For our choice of $w_{\mathrm{ap}}$, we follow \cite{Porth+21} who propose a form that approximates an inverse variance weighting scheme. As the mocks used in this work have $w_i \equiv 1$, it can be shown that in this case the weight of an aperture centered at $\varthetavec$ is equal to the number of multiplet counts within its configuration of aperture radii,
\begin{equation}
    w_{\mathrm{ap}}(\varthetavec;\theta_1, \cdots, \theta_n) = \sum_{i_1\neq i_2 \neq \cdots \neq i_n} 1 \ ,
\end{equation}
where the indices are bound to the same constraints as in \eqref{eq:MapnEstimator}.
Analogous to the FFT-based estimation procedure we need to avoid border effects; with our choice \eqref{eq:MapnFilter} for the filter function this results in cutting off a strip of width of $\mathrm{max}\left(\{\theta_1, \cdots, \theta_n\}\right)$ from each border. In the resulting DV (Fig.~\ref{fig:DV-Mapn}) the first twelve elements correspond to the equal scale statistics of order $\{2,4,5\}$ while the final 20 elements correspond to the third order statistics and are ordered as the DV in Fig.~\ref{fig:DV-Map3}.

\subsection{Aperture mass peak counts \done{Sandrine P.}}
\label{sec:hos;subsec:peaks}

A different way to probe the convergence field at all statistical orders in a single step is to consider peak counts. As the name itself implies, one is now searching for peaks (i.e., local over-densities) on the smoothed convergence maps. 

Some studies only focus on peaks with very large S/N because they are good tracers of massive galaxy clusters \citep{Kruse+99, Marian+06, Gavazzi+06, Hamana+15, Miyazaki+17}. The advantage of this approach is that the dependence on cosmology of the WL cluster abundance can be accurately predicted by theoretical models \citep[see e.g.,][]{Kruse+00, Bartelmann+01,  Hamana+04, Marian+09}.
Another approach consists of also considering low-amplitude peaks.
Since there is no analytical prediction for the full range of S/N peaks, it is necessary to run a large number of N-body simulations to calibrate the dependence of the peak count statistics on cosmology.
However, a significant fraction of these peaks arise from large-scale structure projections, and as such, carry additional cosmological information \citep[see e.g.,][]{Yang+13,Lin+16}. Over the past two decades, many cosmological studies have been performed based on the second approach \citep[see e.g.,][]{Pires+09, Dietrich+10, Martinet+18, Peel+18, Li+19, Ajani+20, Martinet+21a, Martinet+21b, Harnois-Deraps+21} and have shown the strength of peak counts in discriminating between cosmological models. In a preliminary Fisher analysis, we tested the two approaches and, as expected, the best results were obtained when including the full range of S/N peaks, which is therefore what we present in the following.

The computation of the peak count statistics requires filtering the convergence maps because of galaxy shape noise. In practice, the peak count can be evaluated at a given scale by convolving the convergence maps with a filter function of a specific scale (i.e., aperture radius). We choose to filter the maps using compensated filters defined in Eq. (\ref{eq:def_compensated}).
Compensated filters (e.g., aperture mass filters or wavelet filters) have been preferred over low-pass filters (e.g., Gaussian filters)  due to their shapes, which reduce the overlap between different scales and then the correlations between them \citep[see e.g.,][]{Lin+16, Ajani+20}. In \cite{Leonard+12}, it is demonstrated that the aperture mass is formally identical to a wavelet transform at a specific scale.

As such, we use the starlet transform \citep{Starck+07,Leonard+12} to simultaneously compute five aperture mass maps corresponding to scales of $\lbrace\ang{;1.17;}, \ang{;2.34;}, \ang{;4.69;}, \ang{;9.37;}, \ang{;18.74;}\rbrace$. The starlet transform decomposes the convergence as follows:
\begin{equation}
\kappa(\varthetavec) = C_J(\varthetavec) + \sum_{i=1}^J W_\theta^i(\varthetavec)\;,
\end{equation}
where $C_J$ is a smooth version of the convergence $\kappa$ and $W_\theta^i$ are the wavelet maps corresponding to a scale of $\theta = 2^i$ pixels.

The starlet transform is equivalent to applying the following aperture mass filter to the convergence map \citep{Leonard+12}: 
\begin{multline}
  U_\theta(\vartheta) = \frac{31}{3}\, \left \lvert \frac{\vartheta}{\theta}\right\rvert^3\, - \,\frac{64}{9}\, \left(\,\left\lvert\frac{1}{2}\,-\,\frac{\vartheta}{\theta}\,\right\rvert^3\,+\,\left(\,\frac{1}{2}\,+\,\frac{\vartheta}{\theta}\,\right)^3\right) \\
    + 2\,\left(\, \left\lvert\,1\,-\,\frac{\vartheta}{\theta}\,\right\rvert^3\, +\, \left\lvert\,1+ \,\frac{\vartheta}{\theta}\,\right\rvert^3\,\right) - \frac{1}{18}\,\left(\,\left\lvert\,2-\,\frac{\vartheta}{\theta}\,\right\rvert^3+\left\lvert\, 2+\,\frac{\vartheta}{\theta}\,\right\rvert^3 \right) \;. 
 \end{multline}

\begin{figure}
    \centering
    \includegraphics[width=1.0\hsize]{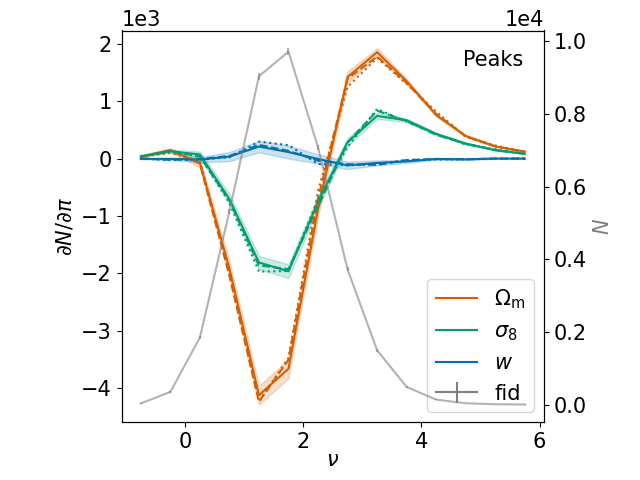}   
    \caption{Fiducial DV (gray) along with its derivatives with respect to $\Omega_{\rm m}$ (orange), $\sigma_8$ (green), and $w$ (blue) for peaks of aperture mass maps, computed from $\kappa$-fields smoothed with a starlet filter with a scale of $\ang{;2.34;}$.}
    \label{fig:DV-peaks}
\end{figure}

In the aperture mass maps, the peaks are identified as individual pixels higher than their eight neighbors. The edges are discarded from the computation.
Once the peaks are detected on each aperture mass map, they are classified depending on their amplitudes with respect to the shape noise. 
Several implementations of the peaks have been tested. The main differences between the implementations are in the range of the amplitudes of the peaks that are considered, the number of equidistant bins and the size of the edges to be discarded. 

The results for the peak counts presented in this study are obtained by sorting the peaks into 14 equidistant bins between $-1$ and $6$ times the dispersion value (see Table~\ref{tab:hos_gauss_bin} and Fig.~\ref{fig:DV-peaks}) to ensure it is Gaussian distributed for the Fisher analysis and by discarding a stripe of 1 pixel on each side of the $\Map$ map before counting peaks.

\subsection{Convergence Minkowski functionals \done{Simone, Vincenzo}}
\label{sec:hos;subsec:MFs}

The HOS probes previously described can be considered extensions of the standard second-order probes since they all aim at probing the convergence field at higher orders to sort out the information contained in its non-Gaussianity. A different way to access this additional constraining power is represented by topological indicators.

Topology is the branch of mathematics that addresses the shapes, boundaries and connectivity of structural features in a field. More precisely, it is concerned with the properties of a geometric object that are preserved under continuous deformations, such as stretching, twisting, crumpling, and bending. Basic examples of topological properties are: the dimension distinguishing between a line and a surface; the compactness differentiating between a straight line and a circle; and connectedness, which separates a circle from two nonintersecting circles.

The first topological statistics we use as a cosmological tool are the Minkowski functionals (MFs). According to Hadwiger's theorem \citep{hadwiger1957vorlesungen}, under a few simple requirements, any morphological descriptor of a $d$-dimensional scalar field is a linear combination of $d + 1$ MFs $V_{n}$ with $n$ ranging from $0$ to $d$. 
The geometrical interpretation of each functional depends on the considered dimension $d$. Therefore, the morphology of the $2$-dimensional convergence scalar field $\kappa(\varthetavec)=\kappa(\vartheta_1, \vartheta_2)$ with variance $\sigma^2$ will be described by the first three MFs\footnote{Note that these integral definitions are always valid for any 2D smooth scalar field, whether it is Gaussian (as for the CMB temperature) or not (as for the lensing convergence).}\,as
\begin{equation} 
\begin{aligned}
    &V_{0}(\nu) = \frac{1}{A} \int_{Q_{\nu}} \dd{a} \ ,   \\
    &V_{1}(\nu) = \frac{1}{4A} \int_{\partial Q_{\nu}} \dd{l} \ ,  \\
    &V_{2}(\nu) = \frac{1}{2 \pi A} \int_{\partial Q_{\nu}} \mathcal{K} \dd{l} \ ,
\end{aligned}
\label{eq: Minkowski functionals definition}
\end{equation}
where $A$ represents the map area, $\partial Q_{\nu}$ is the boundary of the excursion set $Q_{\nu}= \{ (\vartheta_1, \vartheta_2 ) \mid \kappa (\vartheta_1, \vartheta_2 )/\sigma \ge \nu \}$ at a given threshold $\nu$, $\dd{a}$ and $\dd{l}$ are respectively the surface and the line element along $ \partial Q_{\nu} $, and $\mathcal{K}$ is the local geodesic curvature of $ \partial Q_{\nu}$. Qualitatively, one can say that $V_0(\nu)$ and $V_1(\nu)$ quantify, respectively, the area and the perimeter of the excursion set $Q_\nu$, while $V_2(\nu)$ gives the Euler characteristic, which is the topological quantity measuring the connectivity of the field. For an orientable surface, the latter is defined as $\chi = 2 ({\cal{A}} - {\cal{B}})$, that is the difference between the number ${\cal{A}}$ of disconnected regions above the threshold $\nu$ and the number ${\cal{B}}$ of those below $\nu$, that is the number of holes. We note that in Morse theory the Euler characteristics are also given by the alternating sum of critical point counts and hence related to the peak counts described in the previous section.

Equation~\eqref{eq: Minkowski functionals definition}, applied to the convergence field, can be rewritten in an operative form as
\begin{equation} \begin{array}{rl}
\begin{aligned}
    &V_{0}(\nu) = \frac{1}{A} \int_{A} \dd[2]{\varthetavec}\; \mathcal{H}(\kappa - \nu \sigma)\;,     \\
    &V_{1}(\nu) = \frac{1}{4A} \int_{A} \dd[2]{\varthetavec}\; \delta_\mathrm{D}(\kappa - \nu \sigma)\, g(\kappa)\;, \\
    &V_{2}(\nu) = \frac{1}{2 \pi A} \int_{A} \dd[2]{\varthetavec}\; \delta_\mathrm{D}(\kappa - \nu \sigma)\, h(\kappa)\;, 
\end{aligned} \end{array}
\label{Minkowski functionals operative definitions} 
\end{equation}
where 
\begin{equation}
\begin{aligned}
        &g(\kappa) = \sqrt{\kappa_{,1}^{2} + \kappa_{,2}^{2}}  \\
        &h(\kappa) = \left( \frac{ 2 \kappa_{,1} \kappa_{,2} \kappa_{,12} - \kappa_{,1}^{2} \kappa_{,22} - \kappa_{,2}^{2} \kappa_{,11} }{ \kappa_{,1}^{2} + \kappa_{,2}^{2} } \right)
\end{aligned},
\label{MFs: kappa covariant derivatives functions}
\end{equation}
with $ \kappa_{,i}$ and $\kappa_{,ij}$ the first- and second-order partial derivatives of the convergence field with respect to $\vartheta_i$ and $\vartheta_i \vartheta_j$ and for $i,j = 1,2$.

\begin{figure}
    \centering
    \includegraphics[width=1.0\hsize]{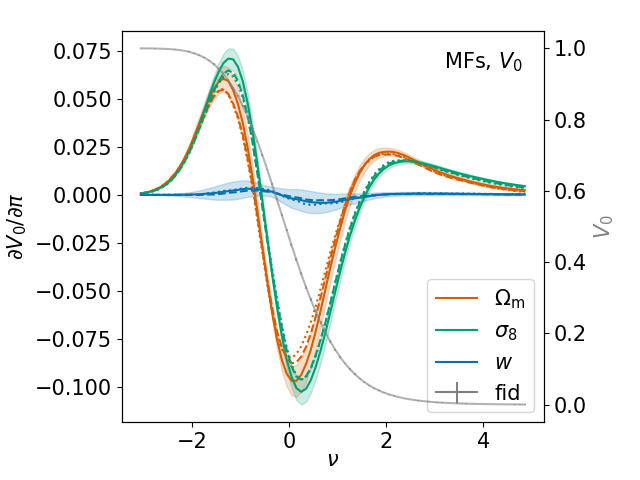}   
    \includegraphics[width=1.0\hsize]{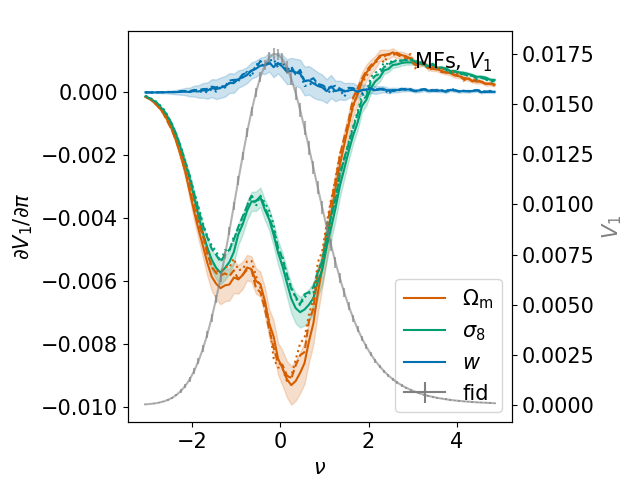}   
    \includegraphics[width=1.0\hsize]{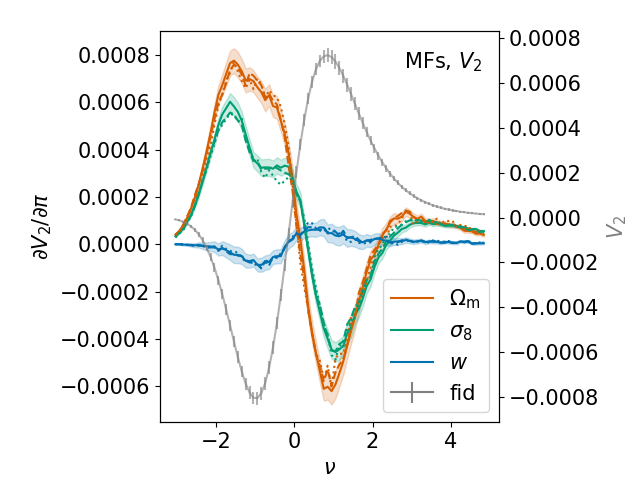}   
    \caption{Fiducial DV (gray) along with its derivatives with respect to $\Omega_{\rm m}$ (orange), $\sigma_8$ (green), and $w$ (blue) for MFs $V_0$ (\textit{top}), $V_1$ (\textit{center}), and $V_2$ (\textit{bottom}) computed on $\kappa$ maps smoothed with a Gaussian filter with a scale of $\ang{;2.34;}$.}
    \label{fig:DV-MFs}
\end{figure}

Equation \eqref{Minkowski functionals operative definitions} suggests that the zeroth MF $V_{0}$ can be evaluated by the integration of the Heaviside step function over the whole excursion set $Q_{\nu}$. The other MFs $V_{1}, V_{2}$ have been transformed from line integrals into surface integrals by inserting a delta function $\delta_\mathrm{D}$ and the appropriate Jacobian. The integrands of the first and second MFs, including the geodesic curvature $\mathcal{K}$, are then given by second-order invariants that depend solely on the threshold $\nu$, the field value $\kappa$ and functions of its first- and second-order derivatives $\kappa_{,i}, \kappa_{,ij}$ defined in equation \eqref{MFs: kappa covariant derivatives functions} \citep[see Sect. 2.3 of][for a detailed calculation]{Schmalzing_1998}\footnote{Here, we replaced the covariant derivatives with the partial derivatives since they coincide when evaluated for a scalar field in Cartesian coordinates, formally $ \kappa_{;i} \equiv \kappa_{,i}$ and $\kappa_{;ij} \equiv \kappa_{,ij}$.}.

Equation~\eqref{Minkowski functionals operative definitions} is the basis for developing an algorithm able to numerically measure MFs for a given threshold on a given convergence map. For our analysis, we make use of the code proposed and implemented in \citet{Vicinanza_2019,Parroni_2020}. We apply it to the convergence map after performing a Gaussian smoothing with four different apertures, namely $(\ang{;1.17;}, \ang{;2.34;}, \ang{;4,69;}, \ang{;9.37;})$. Figure~\ref{fig:DV-MFs} shows the corresponding DV for the $\ang{;2.34;}$ smoothing. The code has been validated by measuring MFs on simulated Gaussian maps since, in this case, it is possible to analytically predict the expected values of $V_{n}(\nu)$. This is no longer the case for non-Gaussian fields. A perturbative approach can be used when deviations from Gaussianity are very small \citep[see e.g.,][]{2016IAUS..308...61P}, but it gives approximate results that require the evaluation of high-order polyspectra to be more accurate. Moreover, this formulation does not account for the presence of noise and systematics from the convergence reconstruction from biased shear data. We refer the reader to \citet{Parroni_2020} and references therein for the derivation and testing of an approximate formulation dealing with these issues. However, in that same paper, it has been shown that a large number of nuisance parameters must be introduced, so it is desirable to rely on direct measurements from simulated convergence maps as we do here, which can be used to develop future emulators.

\subsection{Convergence Betti numbers \done{Simone, Vincenzo}}
\label{sec:hos;subsec:BNs}

\begin{figure}
    \centering
    \includegraphics[width=1.0\hsize]{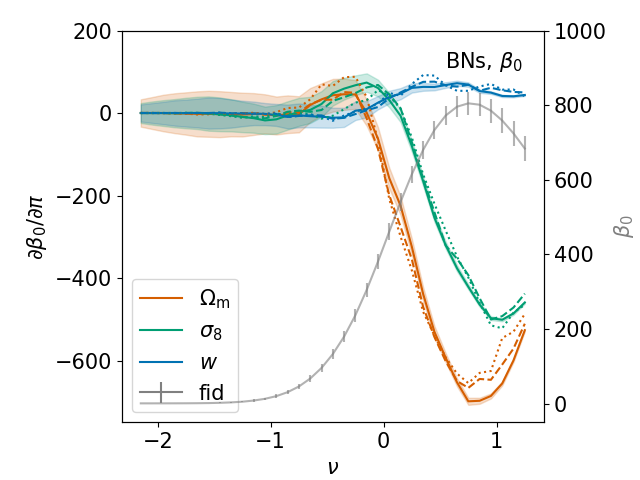}   
    \includegraphics[width=1.0\hsize]{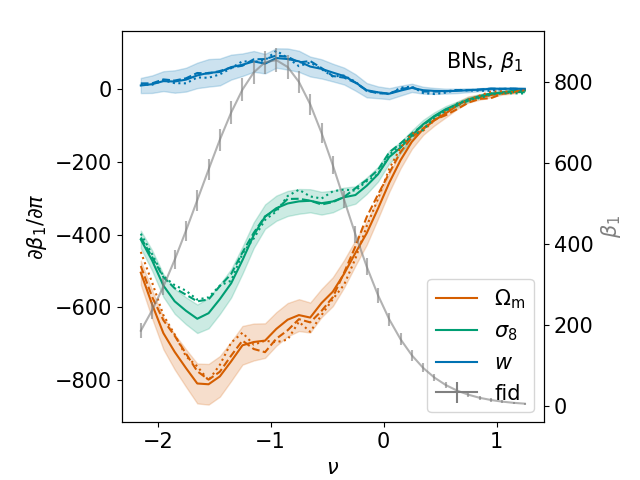}   
    \caption{Fiducial DV (gray) along with its derivatives with respect to $\Omega_{\rm m}$ (orange), $\sigma_8$ (green), and $w$ (blue) for BNs $\beta_0$ (\textit{top}) and $\beta_1$ (\textit{bottom}) computed on $\kappa$ maps smoothed with a Gaussian filter with a scale of $\ang{;2.34;}$.}
    \label{fig:DV-BNs}
\end{figure}

In order to individually study the different topological features in the maps (holes, islands),
one can rely on the Betti Numbers (BNs) $\beta_{i}(\nu)$ (whose alternating sum gives back the Euler characteristics). BNs consist of topological invariants that formalize the pure topological information in a given field through the relationship between singularity structure and connectivity. They represent homology measures since they describe the spatial distribution of the critical points of the field and their connectivity. The homology of a $3$-dimensional manifold $M$ of a $2$-dimensional scalar field such as the convergence $\kappa(\vec{\vartheta}) = \kappa(\vartheta_1, \vartheta_2)$ is characterized by the first three BNs\footnote{In general, the pure topology of a $d$-dimensional scalar field (and thus the homology of its $(d+1)$-dimensional manifold $M$) is described by $d+1$ BNs.},  $\left. \beta_{i} : \mathbb{R} \mapsto \mathbb{Z} \right. \mid i = 0, 1, 2 $. The manifold $M$ of $\kappa$ is a subset of $\mathbb{R}^{3}$ defined by the relation
\begin{equation}
    M = \{ (\vec{\vartheta}, \kappa(\vec{\vartheta}) ) \mid \vec{\vartheta} \in \mathbb{R}^{2} \} \subset \mathbb{R}^{3}.
\label{manifold_definition}
\end{equation}
In particular, similarly to the excursion set $Q_{\nu}$ introduced in Sect. \ref{sec:hos;subsec:MFs}, it is useful to construct a superlevel set filtration at a threshold $\nu$ formally defined as
\begin{equation}
    M(\nu) = \{ (\vec{\vartheta}, \kappa(\vec{\vartheta}) ) \in M \mid \kappa(\vec{\vartheta}) / \sigma \geq \nu \}\,,
\label{superlevel-set-filtration}
\end{equation}
with $\sigma^2$ the variance of the convergence field. For all the thresholds $\nu_{1} \geq \nu_{2}$ made inline $
    M(\nu_{1}) \subseteq M(\nu_{2})\,.$
Indeed, lowering the threshold $\nu$ from $\infty$ to $-\infty$, new points are included into the manifold $M(\nu)$. It is trivial to show that $M(\infty) = \emptyset $ and $M(-\infty) = M$. 

Following the Morse--Smale theoretical approach introduced in \citet{Feldbrugge_2019}, it is possible to infer integral relations for BNs valid for an arbitrary $2$-dimensional random field (not necessarily Gaussian):
\begin{equation} \begin{array}{rl}
\begin{aligned}
    \beta_{0}(\nu)  &= \int_{\nu}^{\infty} \left\lbrace \mathcal{N}_{\rm max}(\nu) - [ 1 - g(\nu) ]\, \mathcal{N}_{\rm saddle}(\nu) \right\rbrace\, \dd{\nu},  \\
    \beta_{1}(\nu)  &= \int_{\nu}^{\infty} \left[ g(\nu)\, \mathcal{N}_{\rm saddle}(\nu) - \mathcal{N}_{\rm min}(\nu) \right]\, \dd{\nu}, \\
    \beta_{2}(\nu)  &= 0, 
\end{aligned} \end{array}
\label{eq:BNs}
\end{equation}
with $\left. \mathcal{N}_{\rm min}, \mathcal{N}_{\rm saddle}, \mathcal{N}_{\rm max} : \mathbb{R} \mapsto \mathbb{R} \right.$ respectively the minima, saddle points and maxima density at threshold $\nu$. The function $\left. g(\nu): \mathbb{R} \mapsto [0,1] \right.$ and the opposite $1-g(\nu)$ respectively represent the probability of two different topological transitions caused by the introduction of a saddle point of the field at threshold $\nu$. These latter affect the value of $\beta_1$ or $\beta_0$ as described in the \textit{incremental algorithm} introduced in \citet{Feldbrugge_2019}. As is evident in Eq.~\eqref{eq:BNs}, BNs depend solely on the diagnostic parameter $\nu$, similarly to MFs. Furthermore, in the $2$\,-\,dimensional case, only the first two BNs $\beta_0(\nu),\beta_1(\nu)$ are relevant since $\beta_2(\nu)$ will increase only when the lowest minimum exceeds the threshold $\nu$. Unfortunately, there is no analytical expression for $g(\nu)$ in either the Gaussian or non-Gaussian case. In fact, only an approximated form has been found by \citet{Feldbrugge_2019} for Gaussian $2$\,-\,dimensional fields. On the other hand, by relying on the \textit{incremental algorithm} presented in \citet{Feldbrugge_2019}, codes can be developed in order to numerically count BNs. For our analysis, we measure BNs through a code developed in \cite{Parroni_2021} and tested against Gaussian maps. As done for the MFs, the convergence field is smoothed with the same Gaussian filter and with identical apertures $\{\ang{;1.17;}, \ang{;2.34;}, \ang{;4,69;}, \ang{;9.37;}\}$. The $\ang{;2.34;}$ scale DV used for the Fisher forecasts is displayed in Fig.~\ref{fig:DV-BNs}.

\subsection{Persistent homology of aperture mass \done{Sven}}
\label{sec:hos;subsec:homology}

\begin{figure}
    \centering
    \includegraphics[width=1.0\hsize]{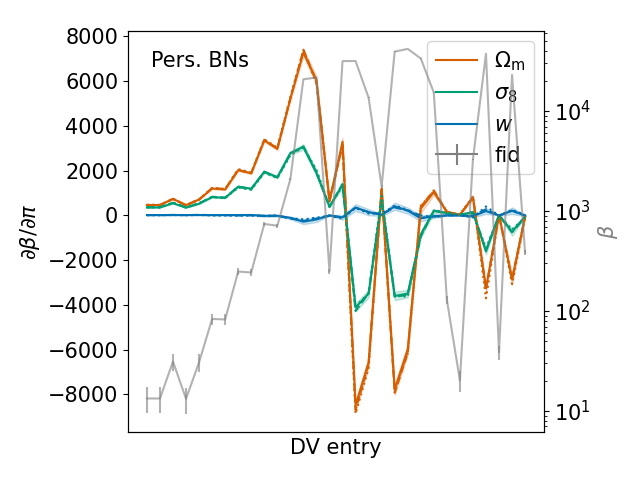}   
    \includegraphics[width=1.0\hsize]{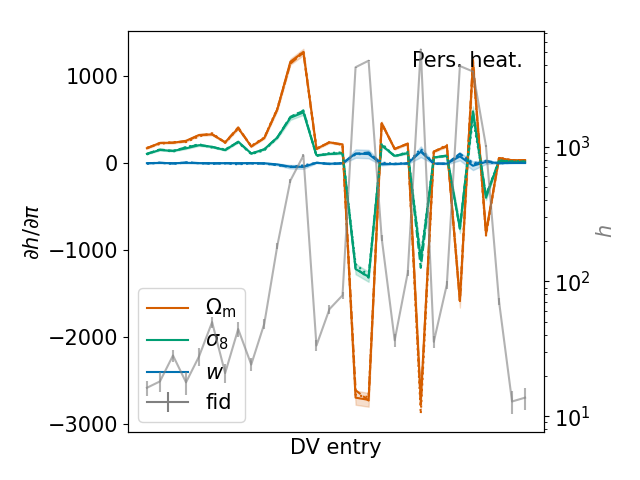}   
    \caption{Fiducial DV (gray) along with its derivatives with respect to $\Omega_{\rm m}$ (orange), $\sigma_8$ (green), and $w$ (blue) for pers. BNs (\textit{top}) and pers. heat. (\textit{bottom}) computed on aperture mass maps from $\gamma$, smoothed with a $\ang{;2.34;}$ \citet{Schirmer:2007} filter. The x-axis corresponds to compression coefficients with no physical meaning.}
    \label{fig:DV-pers}
\end{figure}

Persistent homology is a tool from topological data analysis to quantify the topological structure of data in the presence of noise. We briefly introduce the method here. Further details can be found in \cite{Heydenreich:2021,Heydenreich:2022}.

As in Sect.~\ref{sec:hos;subsec:BNs}, we construct excursion sets on a manifold $M$; here, $M$ is represented by the S/N map of aperture masses $\Map/\sigma\left(\Map\right)$ on a square patch on the sky, seen as a subset of $\mathbb{R}^2$. The filter function of the aperture mass map was optimized to detect dark matter halos and reads \citep{Schirmer:2007}
\begin{align}
  Q_\theta(\vartheta) = & \left[1 + \exp \left(6 -150 \frac{\vartheta}{\theta}\right) + \exp \left(-47 +50 \frac{\vartheta}{\theta}\right)\right]^{-1}
             \nonumber\\
             & \quad \times \left(\frac{\vartheta}{x_{\rm c}\theta}\right)^{-1} \tanh \left(\frac{\vartheta}{x_{\rm c}\theta}\right)\, .
             \label{eq:filterfunction_PH}
\end{align}
Here we use the filter scale of $\theta=\ang{;2,34;}$ and a concentration index of $x_\mathrm{c}=0.15$ \citep{Hetterscheidt:2005}. By varying the threshold, topological features such as connected components and holes emerge. The number of these features at varying threshold levels depends on the topological structure of the aperture mass map. The topological features also have a physical interpretation: Connected components correspond to local minima of the aperture mass map, and thus to underdensities in the integrated matter distribution. On the other hand, holes correspond to local maxima of the aperture mass map, or overdensities in the integrated matter distribution.

In contrast to BNs, which just count the number of features at each filtration level, persistent homology tracks the birth $b$ and death $d$ of each topological feature. The "persistence" of a feature, defined as its "lifetime" $d-b$, is a measure of how far a topological feature protrudes from its surroundings. In other words, features with low persistence are more likely to be attributed to noise \citep{2011MNRAS.414..350S}.

To visualize a persistent homology statistic, the features $(b,d)$ are shown in a scatter plot, which is called a "persistence diagram." Here, we distinguish between two summary statistics that we generate from the persistence diagram. The first are the "persistent Betti numbers" (pers. BNs) $\beta_{i}(\nu_\mathrm{b},\nu_\mathrm{d})$. In contrast to regular Betti numbers $\beta_{i}(\nu)$, which just count the number of features alive at threshold $\nu$, persistent Betti numbers $\beta_{i}(\nu_\mathrm{b},\nu_\mathrm{d})$ count all features that were born before $\nu_\mathrm{b}$ and die after $\nu_\mathrm{d}$. This allows us to take into account the persistence of topological features and disregard the features that are only short-lived \citep[for more detail, see][]{Heydenreich:2021}. As a more robust summary statistic, we generate "persistent heatmaps" (pers. heat.) by replacing each point in the persistence diagram by a Gaussian of width $\sigma=0.2$.\footnote{In general, $\sigma$ can be chosen freely, and we found in previous works that a value of $0.2$ provides stable results.} For both summary statistics, we then construct a data vector by employing the $\chi^2$-maximiser method discussed in \citet{Heydenreich:2022}. This method evaluates how much a pers. BNs or pers. heat. varies between different cosmologies with respect to the expected standard deviation across realizations, and then picks a sample of evaluation points that maximize the cosmological information content. For both pers. BNs and pers. heat., we generate a DV containing 30 entries each and show them in Fig.~\ref{fig:DV-pers}. The jagged features in these DVs are due to the ordering of the x-axis which is dictated by the cosmological information content of the respective point in the heatmap, and not by its value.


\subsection{Scattering transform coefficients \done{Sihao, Vincenzo}}
\label{sec:hos;subsec:ST}

\begin{figure}
    \centering
    \includegraphics[width=1.0\hsize]{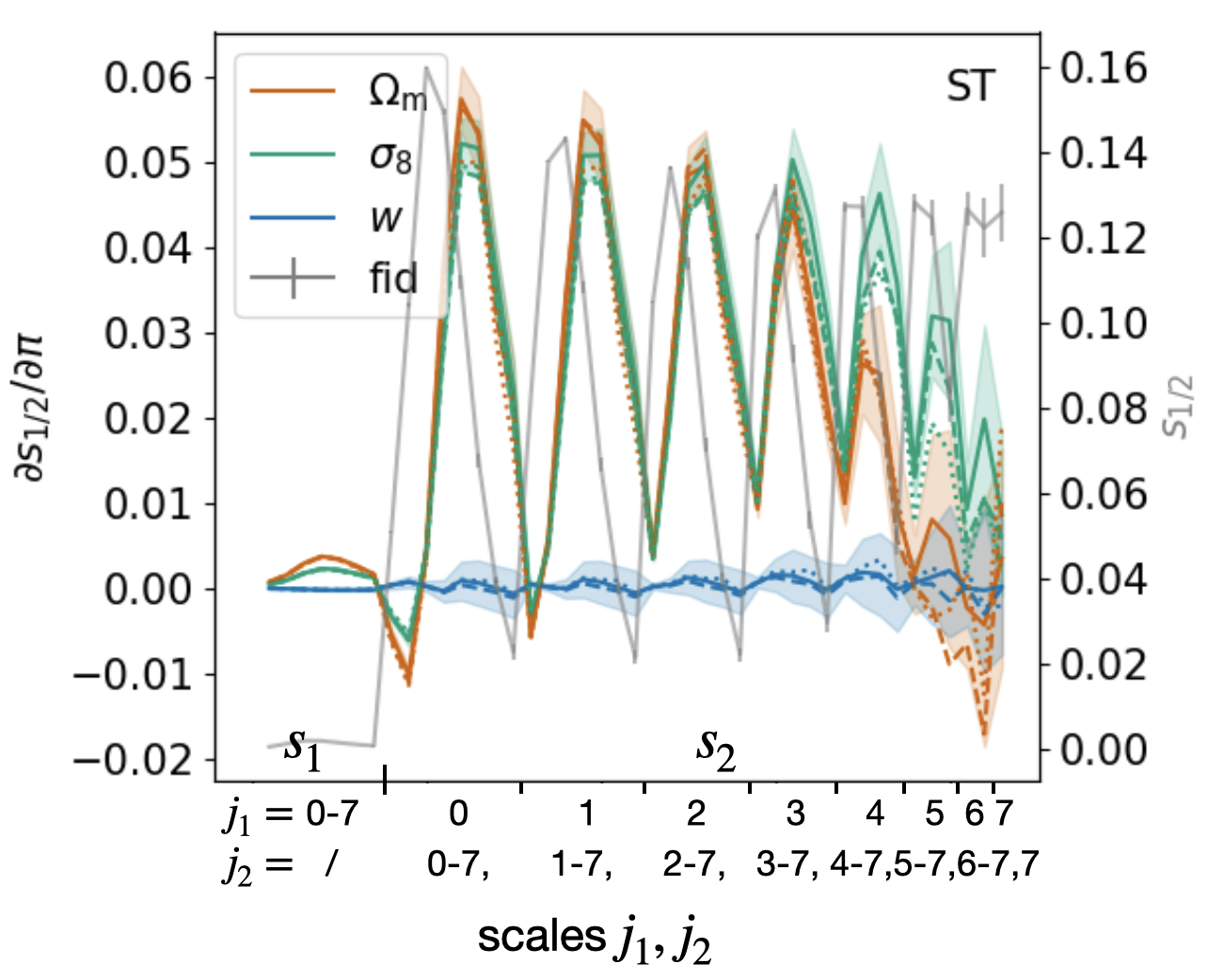}   
    \caption{
    Fiducial data vector (gray) along with its derivatives with respect to $\Omega_{\rm m}$ (red), $\sigma_8$ (green), and $w$ (blue) for ST computed on $\kappa$ maps smoothed with a Gaussian filter with a scale of $\ang{;2.34;}$. The first $8$ points correspond to the first-order scattering coefficients, followed by $8$ series of second-order scattering coefficients, one per scale number $j_1$ and spanning all $j_2$ values for each.}
    \label{fig:DV-ST}
\end{figure}

The scattering transform \citep{Mallat2012} was invented as a novel summary statistic which borrows ideas from convolutional neural networks but does not need training. It was first applied in computer vision, and shortly after introduced to cosmology \citet{Cheng2020}, it has gained an increasing attention for various applications \citep[e.g.,][]{Cheng2021, Valogiannis2022, Chung2022, Delouis2022, Valogiannis2022b, Greig_2022}. Below, we present a brief description of the ST. A nontechnical guide to this new statistic and more intuitive understandings and details can be found in \citet{Cheng2021b} and references therein. 

The ST statistics is similar to a lower order and compressed version of the N-point polyspectra. It keeps the ability to describe different morphological configurations without raising the power of variables by multiplication, thus numerically much more stable. In addition, it also uses logarithmic binning of frequencies, which significantly reduces the size without losing much information for fields with multiscale structures in comparison to polyspectra. 

In this study, we use up to second order ST, which resembles up to some binned tri-spectrum and is defined in the following way:
\begin{equation}
\left \{
\begin{array}{l}
\displaystyle{S_0 = \langle \kappa \rangle} =0 \\
 \\
\displaystyle{S_{1}^{j_1, l_1} = \langle |\kappa \star \psi^{j_1, l_1}| \rangle} \\ 
 \\ 
\displaystyle{S_2^{j_1, l_1, j_2, l_2} = \langle ||\kappa \star \psi^{j_1, l_1}| \star \psi^{j_2, l_2} |\rangle}
\end{array}
\right .  \ ,
\label{eq: stcoeff}
\end{equation}
where we apply the ST to the lensing convergence field $\kappa$. In the above equations, $\kappa \star \psi$ is a convolution with wavelets $\psi$ labeled by a size index $j$ and orientation index $l$. Wavelets are localized band-pass filters, so convolution with a wavelet selects Fourier modes in a wide frequency bin. The wavelets chosen here are qualitatively similar to the aperture mass filters, but they are oriented instead of isotropic and thus also complex-valued, and we use wavelets with a wide range of sizes. 
$|\cdot|$ denotes the modulus of the complex-valued fields, which is the key operation in the design of the ST. Finally, $S_{0,1,2}$ are the zeroth, first, and second order scattering coefficients, which are translation invariant.

Following \citet{Cheng2020}, we use Morlet wavelets and sample $L=4$ different orientations. Morlet wavelets is a sinusoidal wave multiplied by a Gaussian profile and modified to be strictly band-pass:
\begin{equation}
    \psi_0(x,y) = \frac{1}{\sigma\sqrt{2}}{\rm e}^{-(x^2+y^2)/2\sigma^2}\left({\rm e}^{i k_0 x}-\beta\right)\,,
\end{equation}
where $\beta={\rm e}^{-k_0^2\sigma/2}$, $\sigma=0.8$ pixel, $k_0=\frac{3\pi}{4}$ pixel$^{-1}$. The whole library of wavelet $\psi$ is obtained by dilating this prototype wavelet by factors of $2$ for $j$ times and rotating it by $l\times 45$ degrees.

Now we explain the meaning of the ST in some detail. Those coefficients are obtained by first nonlinearly transforming the input field in a nested way and then taking the spatial average. In Fourier space, wavelets are wide window functions with different sizes (labeled by $j$) and orientations (labeled by $l$), so wavelet convolution will select Fourier modes around a frequency. The nonlinear operation, modulus, is not common in traditional statistics but proves useful in computer vision. Note that the modulus is applied in real space to each pixel (not to Fourier coefficients), similar to the pointwise nonlinearity used in convolutional neural networks. One intuitive way to understand the role of modulus is by noticing that for any field $|f(x,y)|=\sqrt{f(x,y)f^*(x,y)}$ is the square root of a quadratic form of $f$. \citet{Cheng2021b} explained that the field $|\kappa\star\psi|$ is similar to a map of locally measured square root of the power spectrum averaged over the pass-band of filter $\psi$. So $S_1$ resembles a binned power spectrum, and $S_2$ resembles the spatial variance of power spectrum (4-point information), but all in a low-order form.

The total number of scattering coefficients depends on the number of distinct wavelets used. We find that the orientation dependence and the coefficients with $j_1\leq j_2$ contain little cosmological information for weak lensing, so the scattering coefficients can be reduced to a smaller set by averaging over all orientations:
\begin{equation}
s_1^{j_1} = \expval{ S_1^{j_1,l_1} }_{l_1}\\
s_2^{j_2} = \expval{ S_2^{j_1,l_1,j_2,l_2} }_{l_1,l_2}\,,
\label{eq: defsnaverage}
\end{equation}
with $j_1 \leq j_2$. If in total $J$ dyadic scales (spaced by power of 2 in size) are used, the first and second-order scattering coefficients have sizes of $J$ and $J(J+1)/2$, respectively. In our study, we explore $J=8$ dyadic scales, resulting in a compact set of 44 scattering coefficients used as the summary statistic. In Fig.~\ref{fig:DV-ST}, we show the mean and dispersion of the scattering transform coefficients in fiducial cosmology and its cosmological sensitivity. The jagged features are again due to the ordering of the x-axis. ST is a function of two scales, and would be a smooth function on a two-dimension space; however to display it in one dimension, we need to slice it along one direction, creating these wiggles.
Besides compactness, the ST statistics also have the desirable properties of being informative and numerically stable, which is confirmed in the following cosmological forecast analysis and the reliability test in Fig.~\ref{fig:Fishderiv}.

\section{Fisher analysis}
\label{sec:Fisher}

The aim of this paper is to estimate the constraints on (a subset of) CPs for each of the HOS probes described in Sect.~\ref{sec:hos}.
To this end, one could perform a Monte Carlo Markov Chain (MCMC) analysis fitting the data as measured on, for example, the SLICS simulated maps. For this approach to be feasible, a mandatory ingredient would be a way to estimate the expected HOS value for any given set of CPs -- in other words, a theoretical model and/or an emulator-like method. For most of the probes of interest here, these are not available, and so we need to look for an alternative approach that can be implemented for all the statistics. This is the Fisher matrix method, which is routinely used to forecast the accuracy that a probe can achieve on constraining CPs given the specifics of the survey at hand. In the following, we first give a brief introduction to the method, highlighting the key quantities, and then discuss in some more detail how we estimate them and the necessary caveats.  

\subsection{Fisher formalism \done{Vincenzo}}

According to Bayes' theorem, the posterior distribution $P({\bf p}\,|\,{\bf D}_{\rm obs})$ of the model parameters ${\bf p}$ given the observed data ${\bf D}_{\rm obs}$ is given by
\begin{equation}
P({\bf p}\,|\,{\bf D}_{\rm obs}) = \frac{{\cal{L}}({\bf D}_{\rm obs}\,|\,{\bf p})\, \mathrm{Pr}({\bf p})}{{\cal{E}}({\bf D}_{\rm obs})}\;,
\label{eq: bayes}
\end{equation}
where ${\cal{L}}({\bf D}_{\rm obs}\,|\, {\bf p})$ is the likelihood of the data ${\bf D}_{\rm obs}$ for the parameters ${\bf p}$, $\mathrm{Pr}({\bf p})$ is the prior on the model parameters, and ${\cal{E}}({\bf D}_{\rm obs})$ is the evidence. The Fisher matrix $\tens{F}$ is defined as the expectation value of the second derivative of the likelihood function, that is its elements are \citep{Bunn95,VS96,TTH97}
\begin{equation}
F_{\alpha \beta} = - \left \langle \frac{\partial^2 \ln{{\cal{L}}({\bf D}_{\rm obs}\,|\,{\bf p})}}{\partial p_{\alpha} \,\partial p_{\beta}} \right \rangle\,,
\label{eq: fabdef}
\end{equation}
where $(\alpha, \beta)$ label the parameters of interest, and the derivatives are evaluated at a fiducial point ${\bf p}_{\rm fid}$ assumed to be the maximum of the likelihood function. The Fisher matrix is then the Hessian of the logarithmic likelihood, thus estimating how fast it decreases from the maximum. 

Under the usual assumption of a Gaussian likelihood function, the Fisher matrix elements reduce to
\begin{eqnarray}
F_{\alpha \beta} & = & \frac{1}{2} 
{\rm tr}{\left [ \frac{\partial \tens{C}_\mathrm{d}}{\partial p_{\alpha}} 
\tens{C}_\mathrm{d}^{-1} \frac{\partial \tens{C}_\mathrm{d}}{\partial p_{\beta}} 
\tens{C}_\mathrm{d}^{-1} \right ]} \nonumber \\
 & + & 
\frac{\partial {\bf D}_{\rm fid}({\bf p})}{\partial p_{\alpha}}
\ \tens{C}_\mathrm{d}^{-1} \ 
\frac{\partial {\bf D}_{\rm fid}({\bf p})}{\partial p_{\beta}}\;,
\label{eq: fabgauss}
\end{eqnarray}
where 
\begin{equation}
\tens{C}_\mathrm{d} = \left \langle ({\bf D}_{\rm obs} - \langle {\bf D} \rangle) ({\bf D}_{\rm obs} - \langle {\bf D} \rangle)^{T} \right \rangle
\label{eq: covmatdata}
\end{equation} 
is the data covariance matrix, and we have assumed that the mean of the data $\langle \mathbf{D} \rangle$ coincides with the theoretical DV ${\bf D}_{\rm fid}$ computed for the fiducial parameters. As in most lensing studies, we assume that the covariance matrix is model independent so that the Fisher matrix reduces to the second term only in Eq.~(\ref{eq: fabgauss}). Strictly speaking, the Fisher matrix is the expectation of the likelihood Hessian, while in Eq.~(\ref{eq: fabgauss}), we have implicitly assumed that this is the same as computing the derivatives at the fiducial point. Because of the Cramer--Rao inequality, this means that the inverse of the Fisher matrix gives a lower limit to the full expected parameter covariance matrix of the CPs. That is to say, the marginalized errors (i.e., having included all of the degeneracies with respect to other parameters) estimated from the diagonal elements $C_{\alpha \alpha}$ of $\tens{C}$ as 
\begin{equation}
\sigma(p_{\alpha}) = C_{\alpha \alpha}^{1/2} = \left [ (\tens{F}^{-1})_{\alpha \alpha}
\right ]^{1/2},
\label{eq: margerr}
\end{equation}
are a lower limit to the actual errors on the CPs. 

The Fisher matrix is also useful for estimating the degeneracy directions in parameter space. The correlation coefficient among the parameters $(p_{\alpha}, p_{\beta})$ is given by
\begin{equation}
\rho_{\alpha \beta} = C_{\alpha \beta}/\sqrt{C_{\alpha \alpha} C_{\beta \beta}}
\label{eq: rhoab} \ , 
\end{equation}
while the angle $\phi_{\alpha \beta}$ defining the degeneracy direction in the 2D plane $(p_{\alpha}, p_{\beta})$ may be estimated as
\begin{equation}
\phi_{\alpha \beta} = \frac{1}{2} \arctan{\left ( \frac{C_{\alpha \beta}}
{C_{\alpha \alpha} - C_{\beta \beta}} \right )} \ . 
\label{eq: phiab}
\end{equation}
Both Eqs.~(\ref{eq: rhoab}) and (\ref{eq: phiab}) show that two parameters are completely independent if $C_{\alpha \beta} = 0$, that is the corresponding off-diagonal term of the Fisher matrix vanishes.

The above description of the Fisher matrix formalism makes it clear that three assumptions are implicitly used when computing the Fisher matrix: i. the likelihood may be approximated as Gaussian; ii. a reliable estimate of the covariance matrix is available; iii. one knows how to compute the derivatives of the DV with respect to the CPs. We subsequently discuss in the next few paragraphs the caveats associated with each of these points and how we deal with them.

\subsection{Gaussian likelihood and data selection \done{Vincenzo}}
\label{subsec:datasel}

Equation (\ref{eq: fabgauss}) has been obtained under the assumption that the likelihood is Gaussian. However, such an assumption must be tested for each particular probe at hand since it is manifestly violated for some of them under certain conditions. As an example, one can think of the number of peaks at very large $\kappa$, which is expected to be zero in the absence of noise. Similarly, the value of the zeroth order MF $V_0(\kappa)$ is identically 1 for very small $\kappa$. As a consequence, every measurement of these quantities will give the same value so that the distribution of repeated measurements would definitely be non-Gaussian. Should we include them in the DV, the Gaussian assumption of the likelihood would break down, introducing a systematic error that is hard to quantify. We therefore need to perform data selection to be confident that the likelihood can be well approximated by a Gaussian. 

To this end, we note that if the likelihood is indeed Gaussian, the quantity
\begin{equation}
y_i = \left ( {\bf D}_{i} - \langle {\bf D} \rangle \right ) 
\tens{C}_\mathrm{d}^{-1} 
\left ( {\bf D}_{i} - \langle {\bf D} \rangle \right )^\mathrm{T}\;,
\label{eq: defy}
\end{equation}
with ${\bf D}_{i}$ the DV measured on the $i$\,-\,th SLICS map, and $\langle {\bf D} \rangle$ the mean over the 924 realizations, must be distributed as a $\chi^2$ distribution with $N_{\rm d}$ degrees of freedom (${{N}}_{\rm d}$ being the length of the DV). For each HOS probe, we therefore compare the $y_i$ distribution to the $\chi^2$ one, and quantify the agreement by computing the weighted average SMAPE (Symmetrized Mean Absolute Percentage Deviation) defined as \citep{RS22}
\begin{equation}
{\cal{S}} = \frac{\sum{\omega_i {\cal{S}}_i}}{\sum{\omega_i}}\,,
\label{eq: smapemean}
\end{equation}
with 
\begin{equation}
{\cal{S}}_i = \frac{\left | P_\mathrm{obs}(y_i) - P_{\chi^2}(y_i) \right |}
{\left | P_\mathrm{obs}(y_i) \right | + \left | P_{\chi^2}(y_i) \right |} \ ,
\label{eq: smapesing}
\end{equation}
and
\begin{equation}
\omega_i = P_{\chi^2}(y_i) \ ,
\label{eq: defweights}
\end{equation}
where $P_\mathrm{obs}(y_i)$ and $P_{\chi^2}(y_i)$ are the measured and $\chi^2$ distribution in $y_i$, and the sum is over the points used to sample the $y$ distribution.\footnote{For a DV with ${{N}}_{\rm d}$ elements, we sample $P(y_i)$ over the range $(1, 3 {{N}}_{\rm d})$ in steps of 1, except for HOM, for which we use a coarser binning because of the low number of degrees of freedom (${{N}}_{\rm d}=3$).} Note that we have defined the weights $\omega_i$ in such a way that we ask for a better match in the central regions of the distributions, reducing the sensitivity to outliers. There is no general rule to define a limiting value for ${\cal{S}}$, but it is clear that the closer it is to zero, the more the two distributions match. For each probe and configuration, we therefore implement a four-step procedure to decide whether the distribution is compatible with the one expected for a Gaussian probe according to the following scheme:
i. We first remove all the elements in the DV that are consistent with zero within $2 \sigma$, with $\sigma$ estimated from the corresponding diagonal element of the covariance matrix. Note that this step is equivalent to saying that we are removing points with ${\rm S/N} < 2$, so that we are actually deleting those elements carrying no valuable information;
ii. We recompute the weighted average SMAPE defined above using only the surviving elements and comparing them with a $\chi^2$ distribution with ${{N}}_\mathrm{obs}$ degrees of freedom, and with ${{N}}_\mathrm{obs} \le {{N}}_\mathrm{d}$ the number of surviving elements of the full DV;
iii. We generate 924 mock realizations of the DV sampling from a Gaussian distribution with a mean and covariance matrix equal to the observed one, and compute the corresponding ${\cal{S}}_\mathrm{mock}$ value;
iv. We repeat the above step 500 times, and estimate the $95\%$\,confidence limit (CL) range of the distribution of ${\cal{S}}_\mathrm{mock}$, finally considering the observed DV Gaussian if its ${\cal{S}}_\mathrm{obs}$ is within this range.\footnote{We have decided to use the $95$ rather than the narrower $68\%$\,CL range although this choice makes it easier to pass the cut. The use of $95\%$ range is, however, a more conservative choice to account for the noise in the estimate of the ${\cal{S}}_\mathrm{mock}$ distribution inherited from the finite number of values, and the use of a covariance matrix inferred from a finite number of simulations.} This was empirically found to roughly correspond to a limit of ${\cal{S}}_\mathrm{obs} \leq 0.15$.

\begin{figure*}
    \centering
    \includegraphics[width=1.0\hsize]{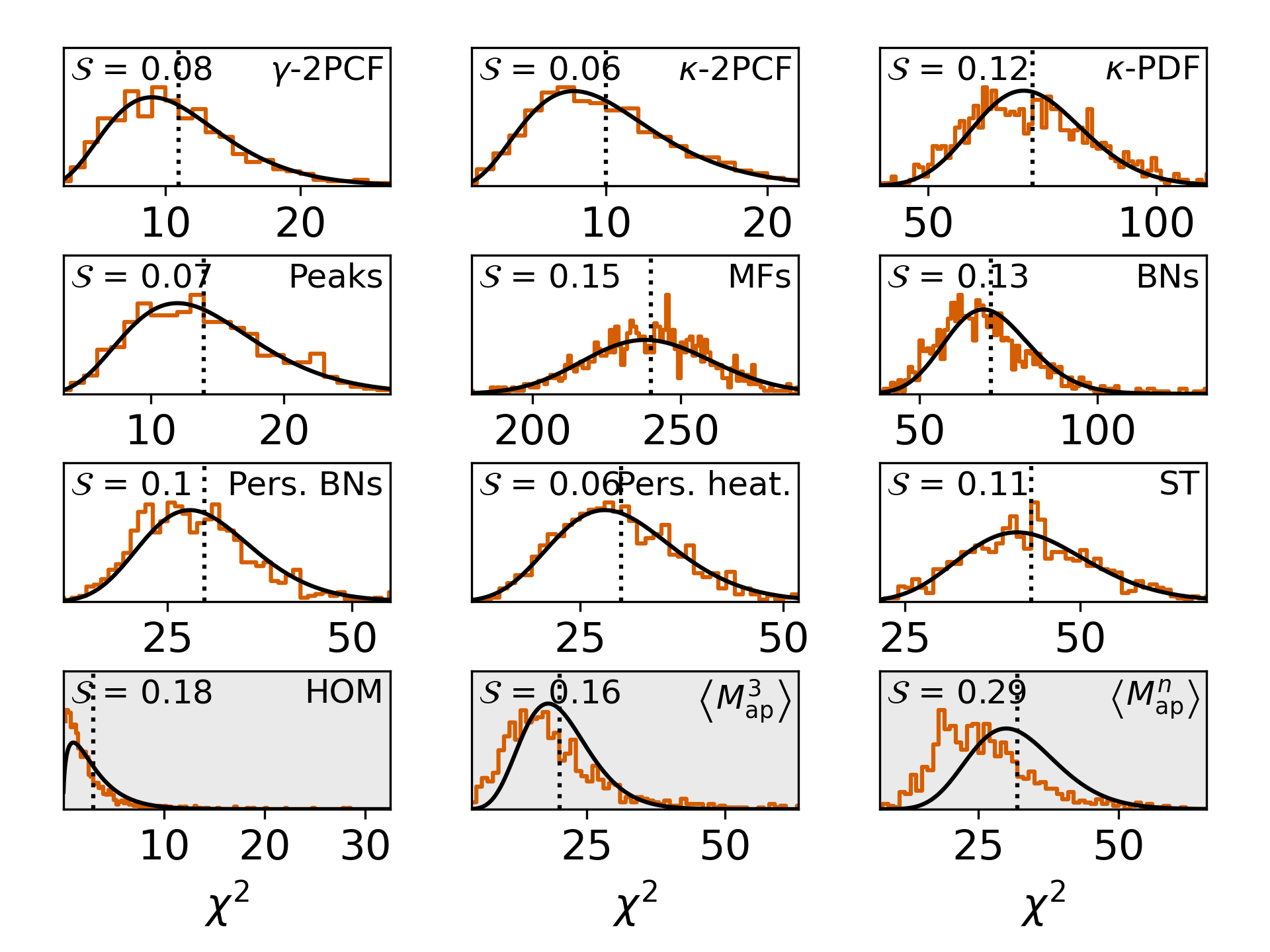}   
    \caption{Verification of the Gaussian hypothesis for all statistics computed in this analysis. For a Gaussian-distributed DV, the histogram of $\chi^2$ values from the SLICS simulations (red) should match the theoretical prediction (black) modulo the sampling noise across the finite set of $924$ realizations. The corresponding SMAPE value (Eq.~\ref{eq: smapesing}) is given in the top left quadrant of each panel. ${\cal{S}} \leq 0.15$ indicates compatibility with the Gaussian hypothesis, which is verified for all statistics but the three bottom panels with gray shaded background. The y-axes correspond to the frequency of each $\chi^2$ bin value across the SLICS and are different for each panel.}
    \label{fig:SMAPE}
\end{figure*}

Applying this procedure to the full list of probes is, however, nontrivial given the different peculiarities. Roughly speaking, we can divide them into two classes. First, we have the $\kappa$-PDF as a function of $\kappa$, the number of peaks as a function of S/N $\nu$, the BNs and the MFs versus the threshold $\nu$. All are measured by first choosing a range $(x_\mathrm{min}, x_\mathrm{max})$, and splitting it into ${{N}}_x$ bins (with $x = \kappa, \nu$). Second, we have other probes that are measured for a given fixed configuration so that they are set once for all by the measurement pipeline. In the first case, we can vary $(x_\mathrm{min}, x_\mathrm{max}, {{N}}_x)$ until the Gaussianity criterion has been successfully passed. We investigate, in particular, configurations with $(\kappa_\mathrm{min}, \kappa_\mathrm{max})$ in the range $(-0.1, 0.1)$ and $(\nu_\mathrm{min}, \nu_\mathrm{max})$ in the range $(-5, 10)$, varying the number of bins from 10 to 70 in steps of 5. We then retain the configurations passing the SMAPE test according to the above procedure. This is illustrated in Fig.~\ref{fig:SMAPE}, where we can appreciate the correspondence between the measured and theoretical $\chi^2$ distribution for all probes but moments in configurations where ${\cal{S}} \leq 0.15$. 

For the other family of HOS probes, we could only compute the value of the average SMAPE, and checked whether it passed our Gaussianity criterion. We  computed the Fisher matrix forecasts for all the probes, but left a warning if the SMAPE value is outside the expected $95\%$\,CL range, that is if the $y_i$ distribution is too different from a $\chi^2$ one. This is the case for all moments-based statistics: HOM, $\MapMapMap$, $\Mapn$ (gray shaded panels in Fig.~\ref{fig:SMAPE}), which have significant non-Gaussian distributions across the different realizations, rendering them unsuitable for a Fisher analysis. This issue, which seems partially related to the small patch sizes considered here, could be mitigated by considering "bulk" moments obtained from integrating the $\kappa$-PDF over a limited range using the tail cuts employed for the PDF.
We nonetheless prefer to report the forecasts for these probes too since the fact that their likelihood is non Gaussian does not prevent one from using them in the future. Indeed, should the corresponding data be available, one can simply run an MCMC analysis based on a non-Gaussian likelihood. The classification of Gaussian and non-Gaussian DVs is given in Table~\ref{tab:hos_gauss_bin}, together with the selected range of scales, convergence, or S/N that allows each probe to pass the Gaussianity test. 

As a final remark, we note that the implemented criterion looks at the overall properties of the DV. However, one could also apply a more stringent criterion looking at the distribution of each single element of the DV. A way to quantitatively implement this approach is based on the D'Agostino---Pearson test. It turns out that this approach is, however, overly restrictive, cutting a large part of the DV for most probes. This is because this method looks for a perfect match with a Gaussian likelihood even in the tail of the distributions, which are actually not of interest for the computation of the Fisher matrix. Indeed, the Fisher matrix requires that the likelihood can be approximated as Gaussian in the neighborhood of its peak, so one is not really interested in what happens in the tails. This is why we finally prefer to rely on the $y_i$ distribution rather than the D'Agostino--Pearson test. This choice is also supported by the results of \citet{Lin+20}, who found that the skewness and kurtosis of the $\gamma$-2PCF in LSST-like mocks have little impact on forecasts when computed with a Gaussian or an Edgeworth non-Gaussian likelihood.

\begin{table} 
\caption[]{Verification of the Gaussian hypothesis. The second, third, and fourth columns respectively correspond to the SMAPE value, the length of the DV, and the range of scales $\theta$, convergence $\kappa$, or SN $\nu$ that is used for each DV.}
\centering 
\begin{tabular}{lccc} 
\hline 
\hline
Statistics & $\cal{S}$ & $N_{\rm d}$ & range \\
 
 \hline
 Gaussian & $\leq 0.15$ & & \\
 $\gamma$-2PCF & $ 0.08 $ & 5 & $\ang{;0,24;}<\theta<\ang{;8,55;}$ ($\xi_+$) \\
  &  & 6 & $\ang{;3,51;}<\theta<\ang{;300;}$ ($\xi_-$) \\
 $\kappa$-2PCF & $ 0.06 $ & 10 & $\ang{;0,6;}<\theta<\ang{;9,23;}$ \\
 $\kappa$-PDF & $ 0.12 $ & 73 & $-0.028<\kappa<0.045$ \\
 Peaks & $ 0.07 $ & 14 & $-1.0<\nu<6.0$ \\
 MFs & $ 0.15 $ & 240 & $-3.1<\nu<4.9$ \\
 BNs & $ 0.13 $ & 35 & $-2.2<\nu<1.3$ \\
 Pers. BNs & $ 0.10 $ & 30 & $-$ \\
 Pers. heat & $ 0.06 $ & 30 & $-$ \\
 ST & $ 0.11 $ & 44 & $-$ \\

\hline
 Non-Gaussian & $> 0.15$ & & \\
 HOM & $ 0.18 $ & 3 & $-$ \\
 $\MapMapMap$ & $ 0.16 $ & 20 & $-$ \\
 $\Mapn$ & $ 0.29 $ & 32 & $-$ \\

\hline
\end{tabular} 
\label{tab:hos_gauss_bin}
\end{table} 

\subsection{Covariance matrices \done{Vincenzo, Nicolas}}

The covariance matrix for each HOS probe can be estimated using the SLICS ellipticity mocks and convergence maps. To this end, we simply measure each probe on all realizations, take the mean as our fiducial DV, and use the definition (\ref{eq: covmatdata}) to compute the data covariance matrix $\tens{C}_\mathrm{d}$. We also use this simple method when investigating the joint use of second and HOS, or of more HOS probes at the same time. We then just concatenate the individual DVs into a single DV, which we then measure on the different mock realizations for the estimate of the covariance matrix. Unless otherwise stated, however, we focus first on single HOS probes and later combine them with the $\gamma$-2PCF. Moreover, to reduce the number of possible combinations, we choose a reference smoothing angle of 4 pixels for all probes, except for the convergence PDF and moments, which we evaluate from maps smoothed with an 8 pixel radius.\footnote{Note that, although the smoothing angle is the same, the smoothing filter can change from one probe to another so that a straightforward comparison among different probes should be avoided. We consider the filter choice as part of the probe itself, so we do not ask different contributors to use the same filter.} We have checked that, for each probe, there is indeed a strong correlation between values estimated from the same map with different smoothing radii, so that using one smoothing angle per statistics captures most of the cosmological information. 

Using a finite number of realizations alters the accuracy of the inverse covariance matrix \citep[e.g., ][]{Hartlap+07}. This arises from the fact that the DV cannot be considered as Gaussian distributed (this is only the case for an infinite number of realizations according to the central limit theorem) but rather follows a student-t distribution. \citet{Sellentin+17} propagated the impact of using a student-t likelihood in the Fisher formalism instead of the Gaussian likelihood, deriving a simple correction factor to this effect to be applied to the inverse Fisher matrix, which depends on the number of realizations used in the computation of the covariance matrix $N_{\rm f}$, the number of degrees of freedom, that is the DV size $N_{\rm d}$, and the number of CPs to be determined $N_{\rm p}$: $c = (N_{\rm f} - 1) / (N_{\rm f} - N_{\rm d} + N_{\rm p} - 1)$ \citep[see Eq. 28 of][]{Sellentin+17}.

Although we use the $N_{\rm f} = 924$ SLICS maps to compute the covariance in our analysis, we also use the $256$ DUSTGRAIN-\emph{pathfinder} realizations as a cross-check. However, the fact that the DUSTGRAIN-\emph{pathfinder} realizations are  pseudo-independent (i.e., derived from a single simulation) renders the comparison with the truly independent SLICS difficult. This nonetheless allows us to quantify the impact of the choice of a simulation set on the covariance computation and to a lesser extent on the possible cosmology dependence of the covariance as the SLICS and DUSTGRAIN-\emph{pathfinder} have slightly different CPs. This comparison is done at the level of the Fisher forecasts, keeping the derivatives fixed and using three different covariance matrices: computed from the $256$ pseudo-independent DUSTGRAIN-\emph{pathfinder} simulations, and $256$ and $924$ SLICS simulations. We find perfect agreement between the SLICS for the different number of realizations when factoring in the \citet{Sellentin+17} correction factor, highlighting the robustness of our covariance estimate in terms of the number of realizations used. However, we see some variations between SLICS and DUSTGRAIN-\emph{pathfinder}. Overall, the order of magnitude of the forecasts remains the same and it is difficult to attribute the difference we measure to a particular effect. This could stem from the simulations themselves, in particular the differences in the computation of the lensing quantities discussed in Sect.~\ref{sec:ds}, the differing fiducial cosmologies, the different area of the mocks ($25$ deg$^2$ for DUSTGRAIN-\emph{pathfinder} and $100$ deg$^2$ for SLICS), or to the level of independence of the realizations in the DUSTGRAIN-\emph{pathfinder} simulations. To specifically assess whether this difference could be due to the different field of view of the SLICS and DUSTGRAIN-\emph{pathfinder} mocks, we also tried the following experiment. We computed the covariance from the central $5\times5$ deg$^2$ region of each of the $924$ SLICS mocks, scaled the covariance matrix by four, and compared it with the $10\times10$ deg$^2$ result. We found excellent agreement showing that the difference with the DUSTGRAIN-\emph{pathfinder} $5\times5$ deg$^2$ covariance is not due to the change in area. The good agreement between the SLICS covariance for both areas also supports our decision to scale up the survey area to the nominal $15\,000$ deg$^2$ expected for \textit{Euclid}. In the case of the $\kappa$-PDF, it has also been shown that the numerical DUSTGRAIN-\emph{pathfinder} covariances agree with those obtained from shifted lognormal maps created with FLASK \citep{Xavier2016} for a fixed source redshift, which can be created in large quantities and tuned to replicate the desired skewness thus exquisitely matching the more expensive simulated covariances \citep{Boyle2020}.

Ultimately we chose to use the SLICS covariance because of the larger field-of-view and number of realizations, and the independence of each realization of this simulation set. We also rescale the covariance to $15\,000$ deg$^2$. We note that our simulated covariances from small patches with vanishing mean density neglect the super-sample covariance effect. As its impact depends on the specific summary statistic, including it in the analysis is beyond the scope of this work.
In the future, those effects can be estimated by treating the full data covariance as a sum of the simulated covariance and a super-sample term built from the response of the summary to a background density, as done for angular power spectra \citep{Lacasa2019ssc} and the $\kappa$-PDF \citep{Uhlemann2020pdfcov}. Alternatively, an analytic covariance model including super-sample effects can be derived from the estimator, as done for the $\kappa$-PDF \citep{Uhlemann2020pdfcov} and $\MapMapMap$ \citep{Linke+2022}.

\subsection{Derivatives and numerical noise \done{Vincenzo, Nicolas}}

\begin{figure*}
    \centering
    \includegraphics[width=1.0\hsize]{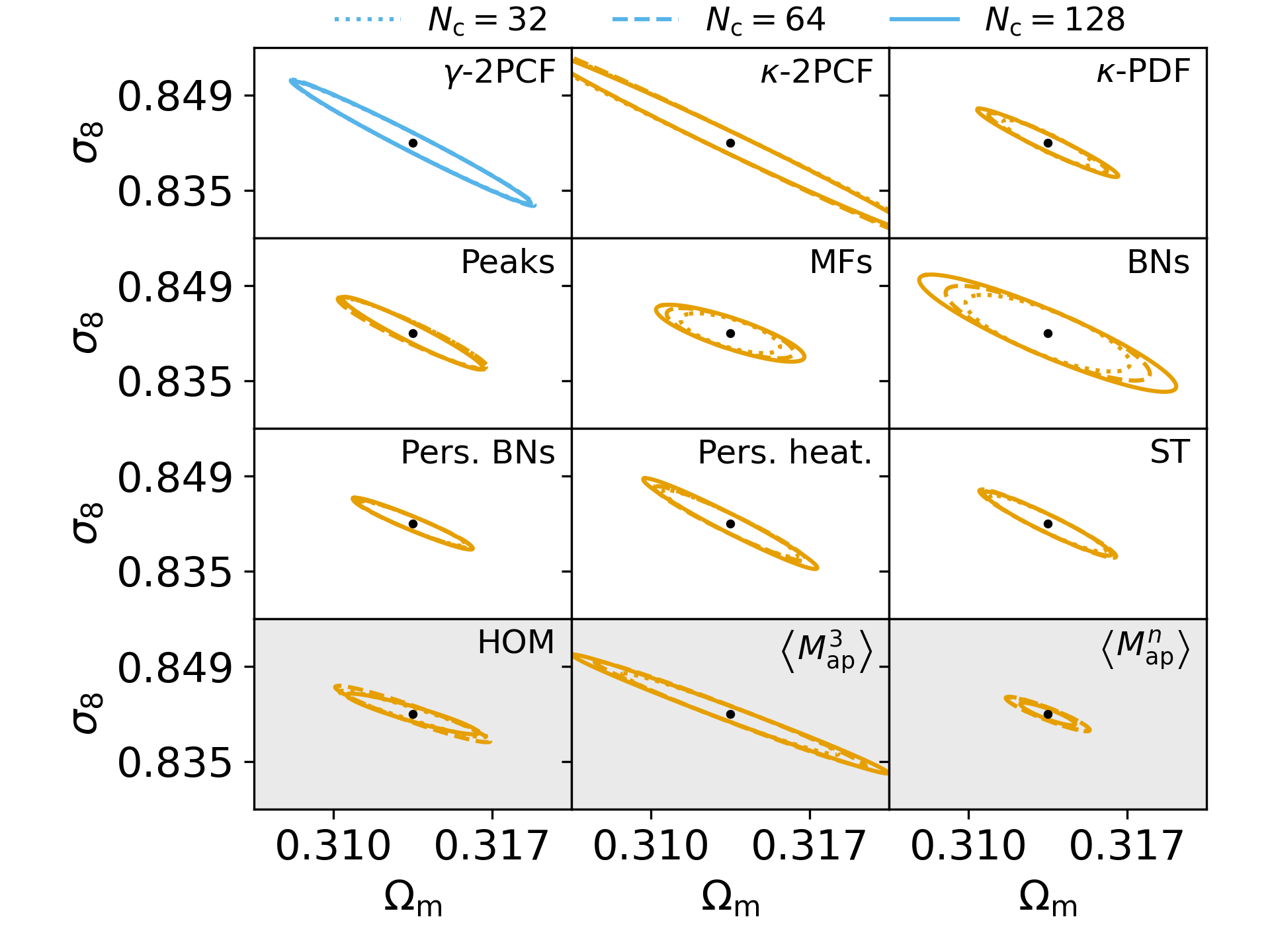}   
    \caption{Fisher forecasts for all probes computed for a \textit{Euclid}-like survey without tomography. The different ellipses refer to different numbers of DUSTGRAIN-\emph{pathfinder} realizations $N_{\rm c}$ used to compute the DV derivatives: dotted, dashed, and solid lines, respectively for $32$, $64$, and $128$ realizations. The covariance matrix is held fixed to SLICS with $N_{\rm f} = 924$ realizations. For $\gamma$-2PCF and $\kappa$-PDF we separately assess the convergence of derivatives by a comparison with theoretical predictions in Figs.~\ref{fig:biasnoise_gamma2pcf} \&~\ref{fig:PDF_dustgrain_theory}.}
    \label{fig:Fishderiv}
\end{figure*}

\begin{figure*}
    \centering
    \includegraphics[width=0.45\hsize]{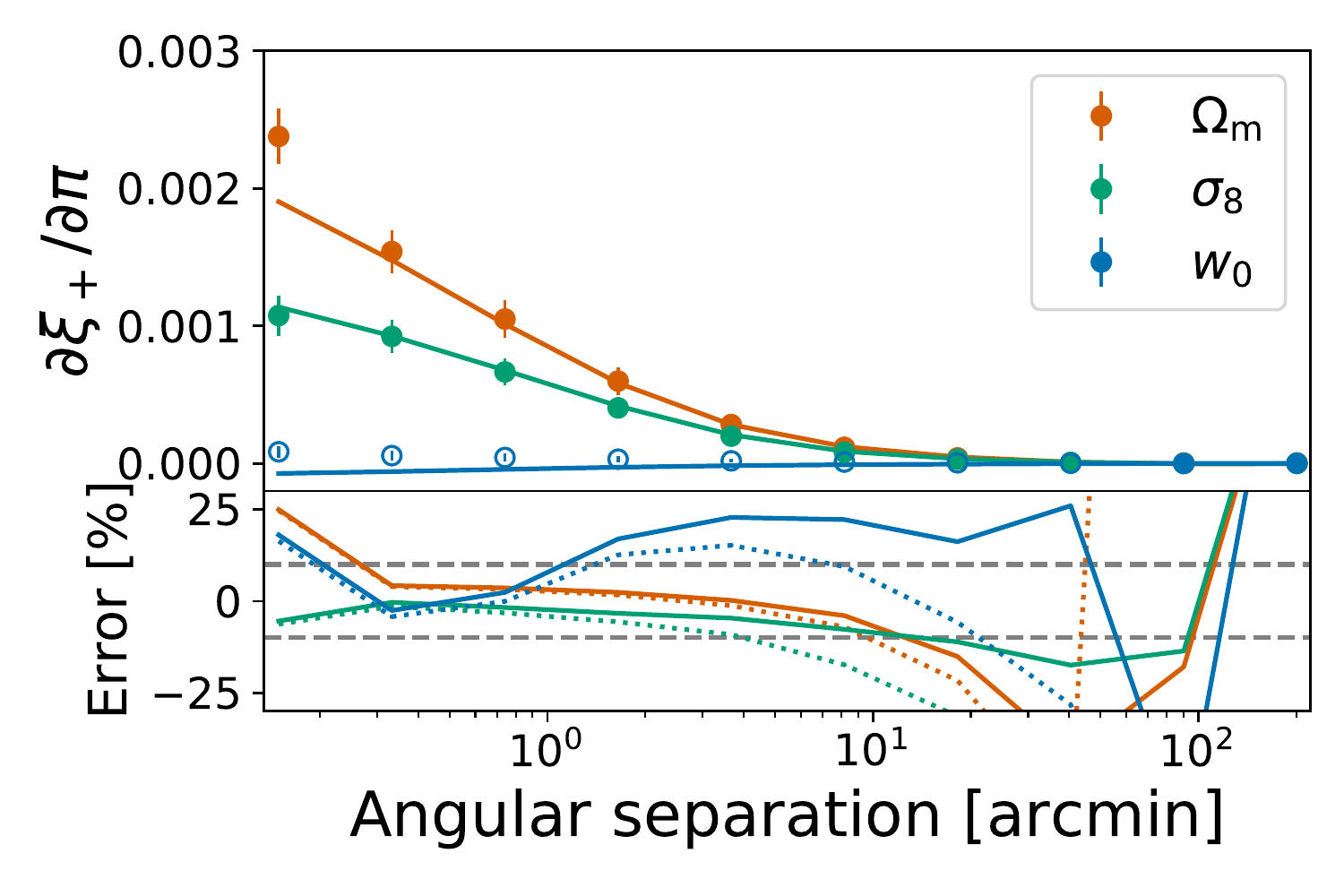} \ \ \ 
    \includegraphics[width=0.45\hsize]{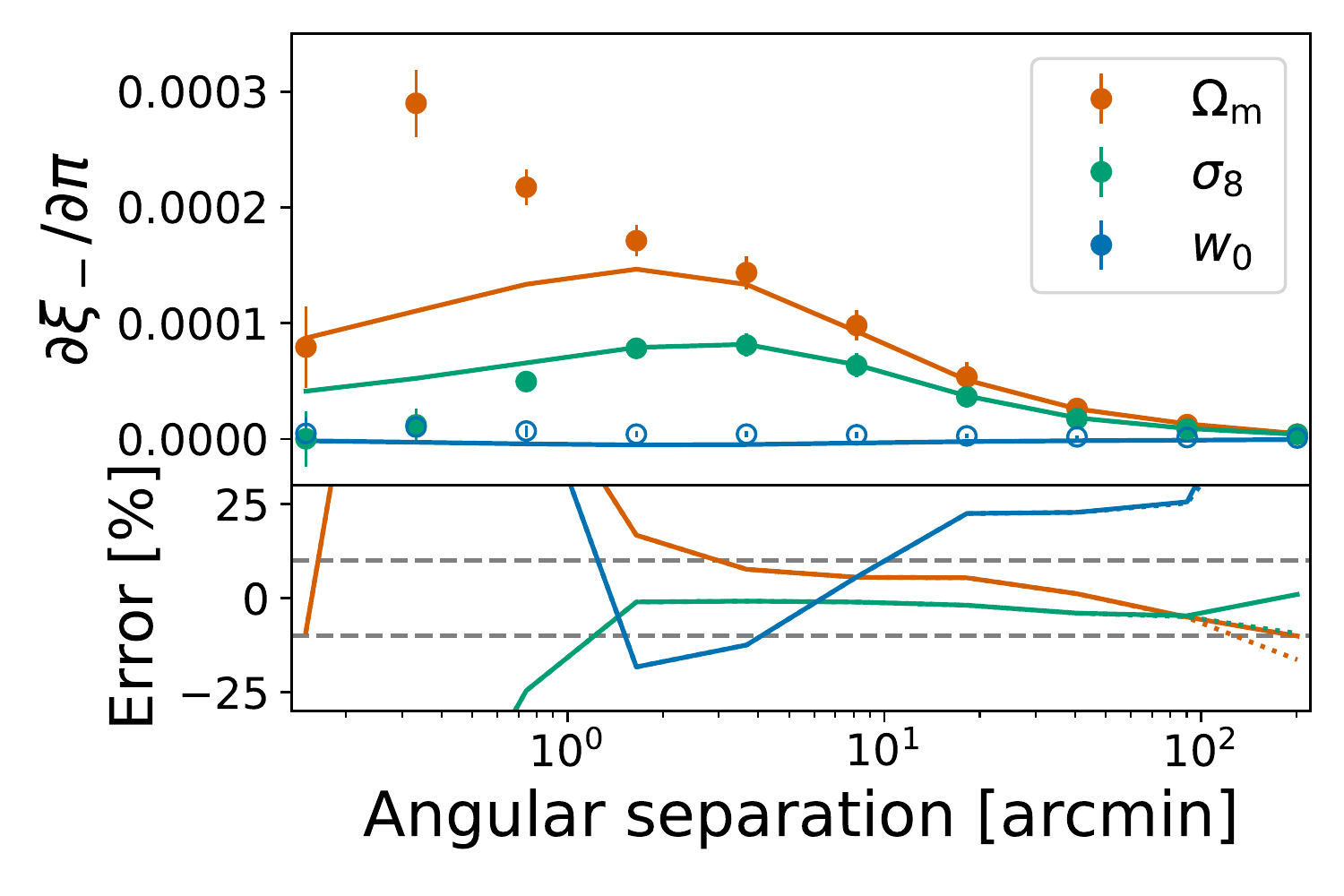}\\ 
    \includegraphics[width=0.45\hsize]{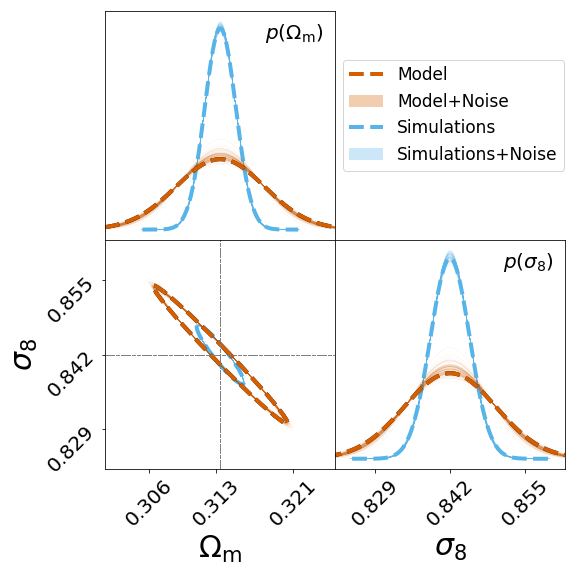} \ \ \ 
    \includegraphics[width=0.45\hsize]{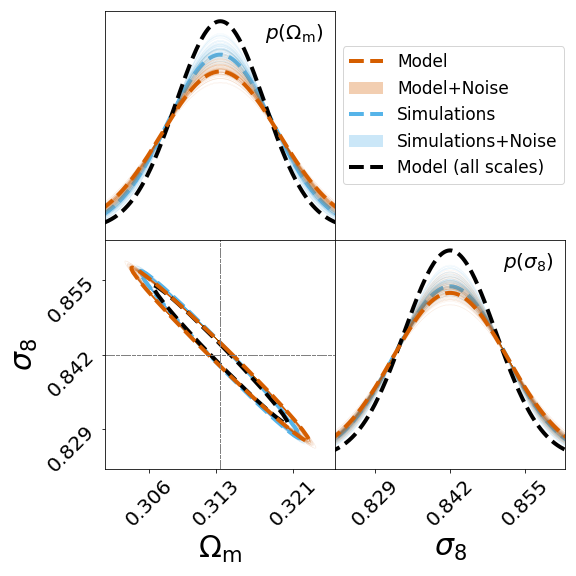} \\ 
    \caption{\textit{Upper row:} Comparison of the derivatives of the $\gamma$-2PCF as predicted from theory (solid lines) and as measured from the DUSTGRAIN-\emph{pathfinder} simulations (dots). The errorbars are scaled to a single line-of-sight in the DUSTGRAIN-\emph{pathfinder} ensemble. The shaded regions display the scales that were discarded from the Fisher analysis presented in this work. In the lower panel we plot the relative deviation between the measured quantities and the theoretical predictions when accounting for (solid lines) or neglecting (dotted lines) finite field effects. The gray dashed lines display the 10 percent errorband. \textit{Lower row:} Comparison of the Fisher forecast using the theoretical model (orange dashed) or the simulation measurements (blue dashed). The thin solid lines correspond to simulated analyses that are used to test the stability of the ellipses given the numerical noise in the derivatives. For the panel on the left we use all available scales on the $\gamma$-2PCF while for the figure on the right only the scales with realistic numerical derivatives (i.e., removing the gray shaded area of the upper panel) are included in the forecast. In the figure on the right we show for reference again the constraints of the theoretical model with all scales included.}
    \label{fig:biasnoise_gamma2pcf}
\end{figure*}

\begin{figure}
    \centering
    \includegraphics[width=\columnwidth]{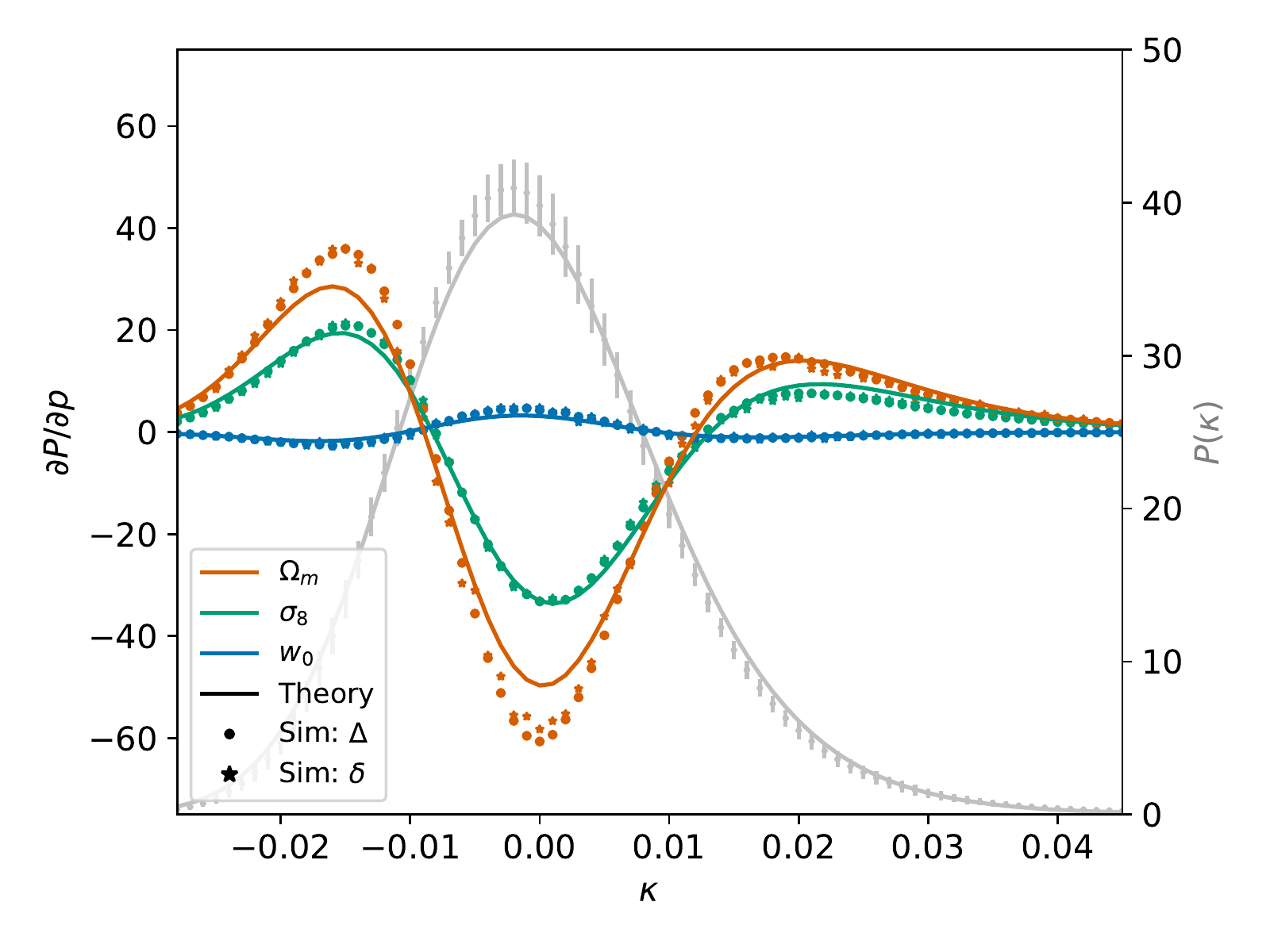}
    \includegraphics[width=\columnwidth]{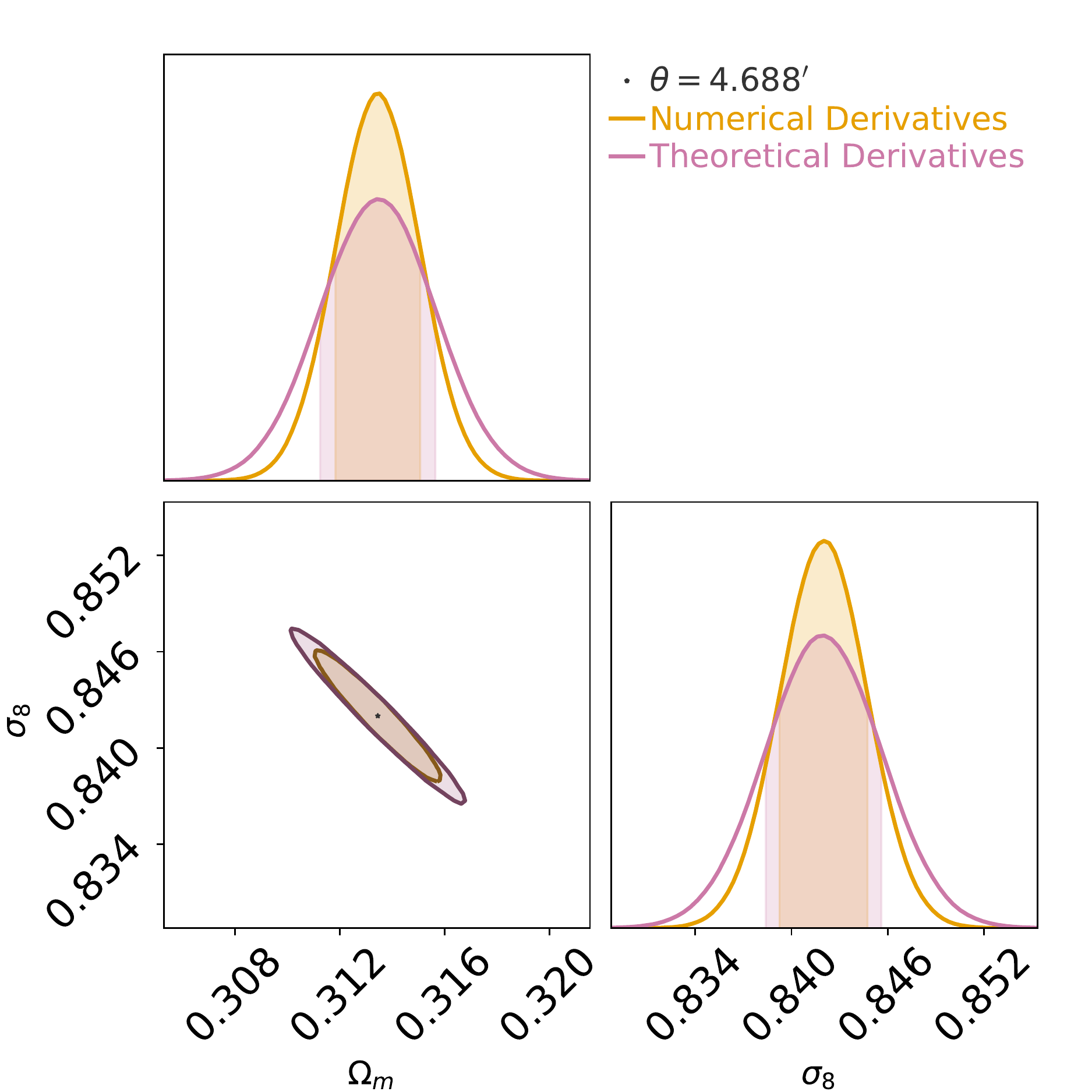}
    \caption{\textit{Top:} Overall shape of the lensing $\kappa$-PDF, smoothed with a top-hat filter of radius \ang{;4.69;} and including shape noise. The fiducial cosmology is shown in gray as predicted from large deviation theory (line) and measured in the DUSTGRAIN-\emph{pathfinder} simulations (data points with error bars indicating the standard deviation across the 256 realizations). Derivatives of the $\kappa$-PDF with respect to the CPs are shown in color, with theoretical predictions using nonlinear variances predicted by Halofit shown as solid lines. The markers represent the derivatives obtained from the simulations using finite differences based on larger and smaller increments (points and stars, respectively). {\textit{Bottom:}} Fisher forecast constraints on $\Omega_{\rm m}$ and $\sigma_8$ for a \textit{Euclid}-like survey from the $\kappa$-PDF shown above. Numerically measured (orange) and theoretically predicted (pink) derivatives find good agreement, with differences in the final $1\sigma$ parameter constraints of approximately $20\%$.} 
    \label{fig:PDF_dustgrain_theory}
\end{figure}

Computing the derivatives of the observable is a key step in the estimate of the Fisher matrix. This would be a trivial task should a theoretical procedure be available to compute the HOS corresponding to a given set of CPs. Unfortunately, this is not the case for most of the probes of interest here. For some of them, a theoretical formulation has been developed, but this typically does not account for noise in the data, and assumes that the $\kappa$ map is perfectly reconstructed from the $\gamma$ data.

Here, we must therefore compute derivatives numerically using the DUSTGRAIN-\emph{pathfinder} simulations for shifted values of $(\Omega_{\rm m}, \sigma_8, w)$. For each given probe, we first measure the DV as the mean over the $N_{\rm c} = 128$ mocks available for each model. We then use four different methods to estimate numerical derivatives. First, we consider the simple 3\,-\,point stencil derivatives defined as
\begin{equation}
\frac{\partial {\bf D}}{\partial p_{\alpha}} = 
\frac{{\bf D}(p_{\alpha} + \varepsilon p_{\alpha}^{\rm fid})
- {\bf D}(p_{\alpha} - \varepsilon p_{\alpha}^{\rm fid})}
{2 \varepsilon |p_{\alpha}^{\rm fid}|}\;,
\label{eq: der3p}
\end{equation}
with $D(p_{\alpha})$ the DV measured on the maps generated from simulations with all CPs but $p_{\alpha}$ set to the fiducial values. Two choices are possible for the shift $\varepsilon$, the small and large increments, respectively corresponding to $4$ and $16\%$ except for the large increment of $\Omega_\mathrm{m}$ for which $\varepsilon$ takes the values ${+28}\%/{-36}\%$. Alternatively, we can use all four values as entries in a linear fit, taking the best fit slope as an estimate of the first derivative. We perform both an unweighted fit and a weighted fit with weights given by
\begin{equation}
w^{\rm fit}_{i} = \left [ 1 - \left ( \frac{| \varepsilon_i - \varepsilon_{\rm ref} |}
{1.1 \varepsilon_{\rm ref}} \right )^3 \right ]^3 \;,
\end{equation}
with $\varepsilon_{\rm ref} = 0.12$, and $\varepsilon_i = (-0.16, -0.04, 0.04, 0.16)$, again with adapted values for the large increment of $\Omega_{\rm m}$. In practice, we find that the weighted and unweighted fits are both equal to the large increment derivatives so we only discuss the large and small increment cases in the following.

As an example, we show in Figs.~\ref{fig:DV-g2PCF} to \ref{fig:DV-ST} the numerical derivatives for the large CP variation of all probes with respect to $\Omega_{\rm m}$, $\sigma_8$, and $w$ in red, green, and blue, respectively. Shaded areas correspond to the noise computed as the dispersion over the $N_{\rm c} = 128$ DUSTGRAIN-\emph{pathfinder} realizations at each cosmology. We also show in gray the mean DV over the $N_{\rm f} = 924$ SLICS realizations with error bars corresponding to the diagonal elements of the SLICS covariance matrix to better visualize the trend of the derivatives on the DV. We note from these figures that the derivatives with respect to $w$ are close to $0$, suggesting that we actually measure very little information on this parameter. This could be alleviated in the future for instance by increasing the number of realizations in the simulation, by adding simulations with a larger step in $w$ in order to strengthen its effect, or by including a tomographic decomposition. On the other hand, there is a strong signal for the other two parameters and we therefore focus the forecasts on $\Omega_{\rm m}$ and $\sigma_8$ in the remainder of the paper.

As a further check on the reliability of the numerical derivatives, one can investigate whether they are stable against the number of DUSTGRAIN-\emph{pathfinder} maps used to compute them. In Figs.~\ref{fig:DV-g2PCF} to \ref{fig:DV-ST}, we therefore add the derivatives computed for $N_{\rm c} = 32$ and $64$ realizations as dotted and dashed lines, respectively. For the large variation derivatives, these lines are within the error bars computed from the $128$ realizations, highlighting the robustness against numerical noise. This is, however, not always the case at the level of the Fisher forecasts. In Fig.~\ref{fig:Fishderiv}, the covariance matrix is held fixed to the $924$ SLICS realizations and the dotted, dashed, and solid ellipses correspond to the forecasts computed with Fisher derivatives of $32$, $64$, and $128$ DUSTGRAIN-\emph{pathfinder} realizations, respectively. We note that several probes suffer from a slow convergence, for example $\kappa$-PDF, MFs, BNs, and pers. heat.. The fact that the forecasts are not yet fully converged is due to the complex interplay between the errors on the derivatives and that of the covariance matrix, which can amplify some noise fluctuations in the derivatives. We demonstrate the trustworthiness of the not fully converged $\kappa$-PDF derivatives through a comparison with theoretical predictions that lead to  marginally wider and slightly tilted Fisher contours as shown in Fig.~\ref{fig:PDF_dustgrain_theory}. Although one could try to develop some numerical methods to correct for this bias, we did not do so because of the three following reasons. First, the convergence of the numerical derivatives is different for each probe with only half of them being affected with the current number of realizations of the DUSTGRAIN-\emph{pathfinder} simulations, suggesting a nontrivial correction scheme that would introduce individual differences in the analysis of each probe. Second, the effect remains small as shown by the smaller difference between the $64$ and $128$ realizations than between the $32$ and $64$. Finally, the comparison with theoretical predictions in the case of the $\kappa$-PDF strongly suggests that convergence has been reached in spite of the small variation still being seen between the $64$ and $128$ realizations in Fig.~\ref{fig:Fishderiv}. Although we are confident in the forecasts when using the large CP increment to compute the derivatives, when considering the small increment derivatives, these are not even converged at the level of the Fisher derivatives. This is due to the CP step being too small, so that the change in the observable due to the shift from the fiducial CPs gets partially lost into the numerical noise. Apart from a few additional tests when comparing numerical and theoretical derivatives in Sect.~\ref{sec:Fisher;subsec:theory}, we therefore only consider the $\pm16\%$ derivatives in the rest of this paper.

\subsection{Theory versus numerical forecasts \done{Lucas, Laila, Sven, Vincenzo}\done{Aoife, Cora, Alexandre}\done{Nicolas}}
\label{sec:Fisher;subsec:theory}

For assessing the robustness of our method in creating reliable Fisher forecasts from the simulated maps, we compare them to the theoretical results for a set of statistics for which such a description is available, namely the $\gamma$-2PCF, $\kappa$-2PCF, $\kappa$-PDF, MFs, HOM, and $\MapMapMap$. We here only report on two of them: the $\gamma$-2PCF in Sect.~\ref{sec:Fisher;subsec:theory;2pcf} and $\kappa$-PDF in Sect.~\ref{sec:Fisher;subsec:theory;pdf}, but the general conclusions are the following. Numerical noise in the Fisher derivatives can generate some artificial degeneracy breaking in the forecasts from the simulation-based model compared to the theoretical ones. Except for moments where we could not reconcile the theory and simulations due to the small field-of-view of our mocks, we find a fair agreement for other probes when applying some restrictions to our analysis:
\begin{itemize}
    \item we only consider Fisher forecasts for the two parameters $\Omega_{\rm m}$ and $\sigma_8$ as the derivatives with respect to $w$ are too noisy in our nontomographic setup;
    \item we use only the large variations of the CPs ($\pm 16\%$) as the small variation ($\pm 4\%$) is too small to capture the signal compared to the numerical noise;
    \item we further restrict the range of scales used in the 2PCFs because of the resolution and field-of-view of the DUSTGRAIN-\emph{pathfinder} mocks.
\end{itemize}

\subsubsection{Shear two-point correlation functions \done{Lucas, Laila, Sven}}
\label{sec:Fisher;subsec:theory;2pcf}

As a first example, we choose the $\gamma$-2PCF introduced in Sect.~\ref{sec:hos;subsec:2pcf}. For computing the theoretical prediction, we use a modified version of the public software \texttt{NICAEA} \citep{Kilbinger+09} and choose the transfer function from \cite{EisensteinHu98} and the revised Halofit prescription of \cite{Takahashi+12} for the modeling of the nonlinear power spectrum. We also account for finite field effects in the ray-tracing of the DUSTGRAIN-\emph{pathfinder} mocks by restricting the support of the convergence power spectrum in Eq.~(\ref{eq: 2pcfhankel}) to the accessible field-of-view; in particular we reweight $P_\kappa$ by the fraction of inverse $\ell$-modes that can be confined to a $5\times5$ deg$^2$ region. This modification introduces a strong ringing effect in the $\gamma$-2PCF which becomes significant at around $5'/150'$ for $\xi_-$/$\xi_+$. In the upper panels of Fig.~\ref{fig:biasnoise_gamma2pcf}, we compare the theoretical and numerical derivatives and we see that there is a nonnegligible discrepancy on small scales for both $\xi_+$ and $\xi_-$ and an additional discrepancy on large scales for $\xi_+$.

In the lower-left panel of Fig.~\ref{fig:biasnoise_gamma2pcf}, we show how these biases propagate in the forecasts with fixed $w$. In particular, when including a broad range of scales, we find that the simulation-based constraints seem much stronger as the different level of bias in the derivatives artificially breaks parameter degeneracies. When using only the scales for which the numerical derivatives are in reasonable agreement with their theoretical counterparts, this effect disappears at the cost of overall reduced constraining power. From the theoretical point of view this reduction remains small as shown by comparing the orange and black ellipses in Fig.~\ref{fig:biasnoise_gamma2pcf}. Besides the bias, noise in the simulation-based derivatives can also lead to artificial degeneracy breaking. To assess the magnitude of the latter effect, we first estimate the covariance matrix of the simulation-based derivatives and, from there, generate $100$ noise realizations. When adding these realizations to the theoretical derivatives, we can assess the stability of our forecast in the ideal case, that is when no bias is present in the simulations. On the other hand, when adding the noise on top of the numerical derivatives, we get an upper bound on the instability of the actual forecast, even without a theoretical description of the observables underlying the analysis. For the $\gamma$-2PCF, we find that the ellipses are stable in all cases. For our subsequent analysis we limit ourselves to those scales in which the mean bias on the derivatives in $(\Omega_\mathrm{m}, \sigma_8)$ is less than $10\,\%$. This range corresponds to the scales displayed in Table~\ref{tab:hos_gauss_bin}. We note that it is more restrictive than the range used in \citet{Martinet+21a,Martinet+21b} based on the same SLICS mocks for the covariance because of the better resolution and larger field-of-view of the cosmo-SLICS used for the model compared to the DUSTGRAIN-\emph{pathfinder} simulations here.

\subsubsection{Convergence PDF \done{Aoife, Cora, Alexandre}}
\label{sec:Fisher;subsec:theory;pdf}

We can use the theoretical model for the $\kappa$-PDF discussed in Sect.~\ref{sec:hos;subsec:PDF} to obtain the PDF at different cosmologies and cross-validate it with the PDFs measured from the simulations, which are also used to obtain the PDF covariance matrix entering the Fisher forecast. 

We compare the mean of the measured $\kappa$-PDF for the fiducial cosmology with the theoretical predictions (including shape noise) in the top part of Fig.~\ref{fig:PDF_dustgrain_theory} (gray points and solid line) for a smoothing scale of $\theta=\ang{;4.69;}$. We see that for the chosen cuts in the tails of the PDF and the measured error bars (standard deviation across all realizations) the theory does perform pretty well, which is reinforced by the fact that the derivatives with respect to the CPs are also well captured. Measured derivatives for both smaller increment sizes (shown as stars) and larger increment sizes (shown as points) in the CP variation are compared with our theoretical model for the $\kappa$-PDF, where the nonlinear variances along the line of sight are set externally. For $\sigma_8$ (green), we see very good agreement between the theoretical and simulated derivatives. We also show the results for $\Omega_{\rm m}$ (red). In this case, a more substantial degree of discrepancy is seen between the theoretical and measured derivatives. The underprediction of the amplitude of the derivative stems from an underprediction of the difference in variances between the two cosmologies as predicted by Halofit. This is consistent with the simulations predicting a stronger response to $\Omega_{\rm m}$ than the theory in Fig. \ref{fig:biasnoise_gamma2pcf}. We found a similar discrepancy in the $\Omega_{\rm m}$ derivatives when cross-checking with a larger smoothing scale of $\ang{;9.37;}$. We observe a similar underprediction for the $w$ derivative (blue), which is also consistent with results for the $\gamma$-2PCF in Fig. \ref{fig:biasnoise_gamma2pcf}. We also checked that replacing the input Halofit nonlinear variance with that from the \textit{Euclid} emulator in each case resulted in minor changes of the predictions.

\begin{table*} 
\caption[]{Fisher forecasts for a \textit{Euclid}-like nontomographic analysis. The precision on $\sigma_8$ and $\Omega_{\rm m}$ is given as a percentage of the fiducial values for the probes taken individually and when combined with the $\gamma$-2PCF. We also report the figure of merit (FoM) as defined in the Dark Energy Task Force \citep{Albrecht+06}. The last columns display the expected gain over the $\gamma$-2PCF analysis for the scales available in the DUSTGRAIN-\emph{pathfinder} simulations. Results for non-Gaussian statistics are also presented as part of the HOWLS project; we stress, however, that a non-Gaussian likelihood would be necessary to assess the robustness of these three probes.}
\centering 
\begin{tabular}{lccccccccc} 
\hline 
\hline
Statistics & \multicolumn{3}{c}{individual
} & \multicolumn{3}{c}{added $\gamma$-2PCF} & \multicolumn{3}{c}{Gain over $\gamma$-2PCF} \\
& $\delta \sigma_8 / \sigma_8$ & $\delta \Omega_{\rm m} / \Omega_{\rm m}$ & FoM & $\delta \sigma_8 / \sigma_8$ & $\delta \Omega_{\rm m} / \Omega_{\rm m}$ & FoM & $\delta \sigma_8$ & $\delta \Omega_{\rm m}$ & FoM \\

\hline
2nd order statistics &  &  &  &  &  &  &  & \\
 $\gamma$-2PCF & $ 0.75 \% $ & $ 1.17 \% $ & $ 2.56 \times 10^{5} $ & $ - $ & $ - $ & $ - $ & $ - $ & $ - $ & $ - $ \\
 $\kappa$-2PCF & $ 1.13 \% $ & $ 1.88 \% $ & $ 1.40 \times 10^{5} $ & $ 0.71 \% $ & $ 1.10 \% $ & $ 2.74 \times 10^{5} $ & $ \times 1.06 $ & $ \times 1.07 $ & $ \times 1.07 $ \\

\hline
HOS (Gaussian) &  &  &  &  &  &  &  &  \\
 $\kappa$-PDF & $ 0.42 \% $ & $ 0.70 \% $ & $ 4.96 \times 10^{5} $ & $ 0.37 \% $ & $ 0.60 \% $ & $ 7.36 \times 10^{5} $ & $ \times 2.01 $ & $ \times 1.95 $ & $ \times 2.87 $ \\
 Peaks & $ 0.45 \% $ & $ 0.70 \% $ & $ 4.28 \times 10^{5} $ & $ 0.40 \% $ & $ 0.64 \% $ & $ 5.28 \times 10^{5} $ & $ \times 1.87 $ & $ \times 1.84 $ & $ \times 2.06 $ \\
 MFs &  $ 0.35 \% $ & $ 0.74 \% $ & $ 2.60 \times 10^{5} $ & $ 0.32 \% $ & $ 0.49 \% $ & $ 7.29 \times 10^{5} $ & $ \times 2.37 $ & $ \times 2.40 $ & $ \times 2.85 $ \\
 BNs &  $ 0.72 \% $ & $ 1.27 \% $ & $ 9.82 \times 10^{4} $ & $ 0.54 \% $ & $ 0.85 \% $ & $ 3.87 \times 10^{5}$ & $ \times 1.38 $ & $ \times 1.37 $ & $ \times 1.51 $ \\
 Pers. BNs & $ 0.32 \% $ & $ 0.60 \% $ & $ 6.95 \times 10^{5} $ & $ 0.29 \% $ & $ 0.53 \% $ & $ 8.29 \times 10^{5} $ & $ \times 2.56 $ & $ \times 2.23 $ & $ \times 3.24 $ \\
 Pers. heat. & $ 0.56 \% $ & $ 0.86 \% $ & $ 3.44 \times 10^{5} $ & $ 0.45 \% $ & $ 0.72 \% $ & $ 4.98 \times 10^{5} $ & $ \times 1.66 $ & $ \times 1.63 $ & $ \times 1.94 $ \\
 ST & $ 0.39 \% $ & $ 0.63 \% $ & $ 4.95 \times 10^{5} $ & $ 0.33 \% $ & $ 0.55 \% $ & $ 7.19 \times 10^{5} $ & $ \times 2.30 $ & $ \times 2.13 $ & $ \times 2.81 $ \\
 All HOS & $ 0.16 \% $ & $ 0.27 \% $ & $ 4.96 \times 10^{6} $ & $ 0.16 \% $ & $ 0.26 \% $ & $ 5.03 \times 10^{6} $ & $ \times 4.82 $ & $ \times 4.41 $ & $ \times 19.65 $ \\

\hline
HOS (non-Gaussian) &  &  &  &  &  &  &  &  \\
 HOM & $ 0.25 \% $ & $ 0.65 \% $ & $ 6.12 \times 10^{5} $ & $ 0.20 \% $ & $ 0.42 \% $ & $ 1.12 \times 10^{6} $ & $ \times 3.72 $ & $ \times 2.76 $ & $ \times 4.37 $ \\
 $\MapMapMap$ &  $ 0.73 \% $ & $ 1.56 \% $ & $ 2.50 \times 10^{5} $ & $ 0.24 \% $ & $ 0.45 \% $ & $ 8.98 \times 10^{5} $ & $ \times 3.17 $ & $ \times 2.59 $ & $ \times 3.51 $ \\
 $\Mapn$ &  $ 0.14 \% $ & $ 0.27 \% $ & $ 2.13 \times 10^{6} $ & $ 0.14 \% $ & $ 0.27 \% $ & $ 2.19 \times 10^{6} $ & $ \times 5.49 $ & $ \times 4.40 $ & $ \times 8.55 $ \\
\hline
\end{tabular} 
\label{tab:forecasts}
\end{table*} 

Having discussed the theoretical modeling and measurement of the WL $\kappa$-PDF, we perform a Fisher forecast for the two parameters $\Omega_{\rm m}$ and $\sigma_8$ using the central region of the WL $\kappa$-PDF at a single scale assuming a \textit{Euclid}-like survey area, shape and source redshift distribution as discussed before. The result shown in the bottom part of Fig.~\ref{fig:PDF_dustgrain_theory} shows reasonably good convergence between the theoretical and numerical results, which reinforces the validity of both sets of forecasts. When adding $w$ to the forecast, the simulated DVs would lead to unrealistically tight constraints compared to the predicted DVs. To avoid an  artificial degeneracy breaking due to numerical noise, we limit the following analysis to two CPs.

\section{Results \done{Nicolas}}
\label{sec:res}

\begin{figure*}
    \centering
    \includegraphics[width=1.0\hsize]{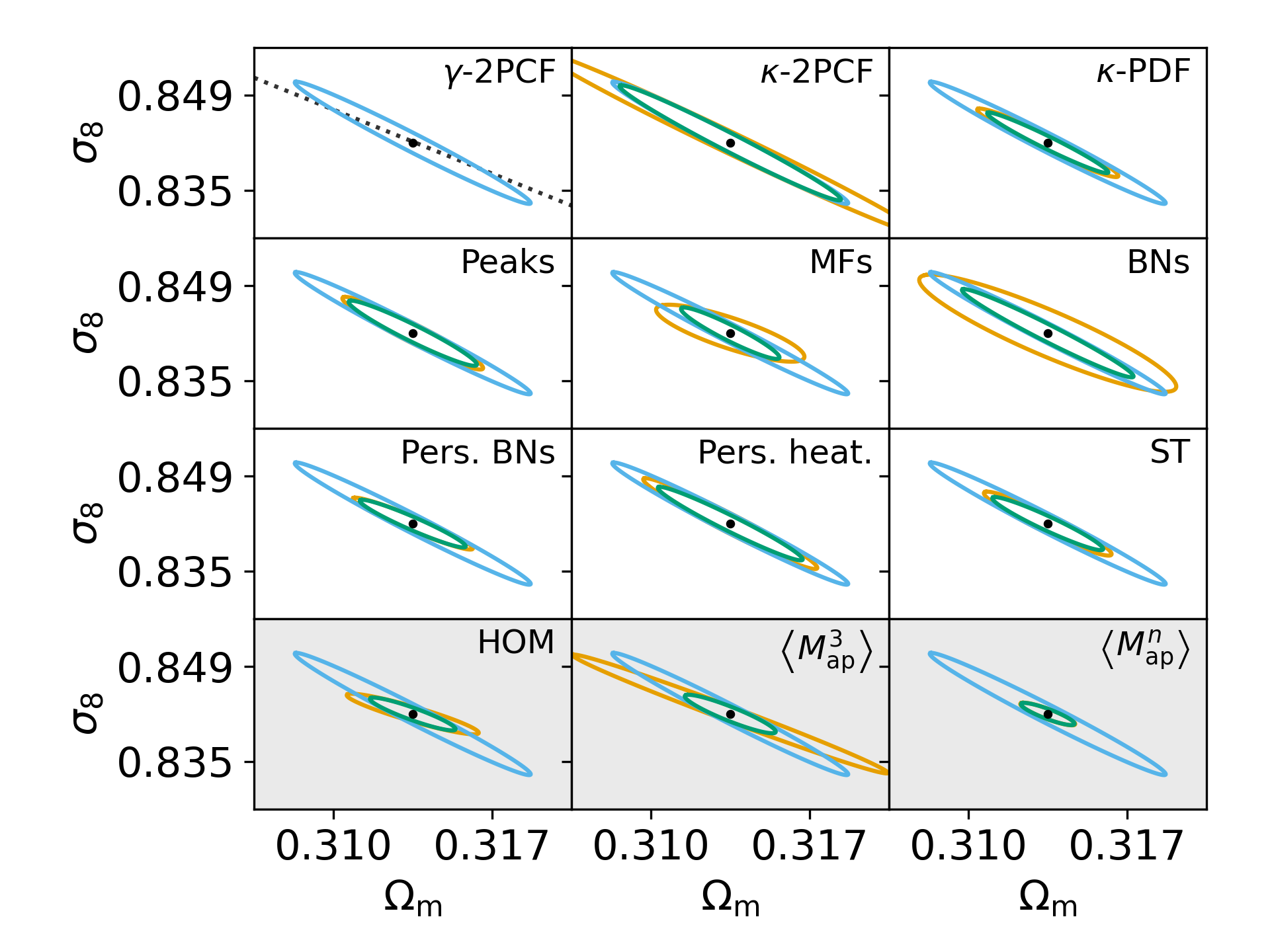}   
    \caption{Individual Fisher forecasts in the $\sigma_8$-$\Omega_{\rm m}$ plane for a nontomographic \Euclid-like survey for the $11$ statistics (orange) and $\gamma$-2PCF (blue), as well as their combination (green). The corresponding marginalized precision on CPs can be found in Table~\ref{tab:forecasts}. The black dashed line in the first quadrant indicates constant $S_8=\sigma_8\sqrt{\Omega_{\rm m}/0.3}$. The bottom row with the gray shaded backgrounds displays probes that are not Gaussian distributed and cannot be robustly interpreted with Fisher forecasts. The abbreviated name of each summary statistic is displayed in the top-right part of each panel: $\gamma$-2PCF and $\kappa$-2PCF for the shear and convergence two-point correlation functions, $\kappa$-PDF for the convergence one-point probability distribution, peaks for aperture mass peak counts, MFs for convergence Minkowski functionals, BNs for convergence Betti numbers, pers. BNs and pers. heat. for aperture mass persistent homology Betti numbers and heatmap, ST for convergence scattering transform coefficients, HOM for higher-order convergence moments, and $\MapMapMap$ and $\Mapn$ for third and $n$-th order aperture mass moments.}
    \label{fig:Fishall}
\end{figure*}

We now discuss the results of the Fisher analysis. We start by considering each probe independently (Sect.~\ref{sec:res;subsec:indiv}), then quantify the gain on CPs from our statistics when combined with the $\gamma$-2PCF alone (Sect.~\ref{sec:res;subsec:2pcf}), and finally investigate the correlations between probes and their combinations (Sect.~\ref{sec:res;subsec:comb}). For all results in this section of the paper, we consider a \Euclid-like nontomographic analysis of $15\,000$ deg$^2$ and vary only $\Omega_{\rm m}$ and $\sigma_8$, as the Fisher derivatives are too noisy for constraining all three parameters (i.e., including $w$) in the nontomographic setup. We also present forecasts for the growth of structure parameter $\Sigma_8 = \sigma_8 \left( \Omega_{\rm m} / 0.3\right)^\alpha$ (equivalent to $S_8$ when $\alpha = 0.5$) in Appendix~\ref{app:resS8} to ease comparison with other analyses; however these results lead to similar conclusions as for $\sigma_8$. The mass maps are smoothed with a $\ang{;2.34;}$ Gaussian,  wavelet or $\Map$ filter except for moments and the $\kappa$-PDF where a $\ang{;4.69;}$ top-hat filter is used, thus probing larger scales. The derivatives are computed with the large variation of CPs in the DUSTGRAIN-\emph{pathfinder} simulations and the covariance from the $924$ SLICS simulations. We also note that these results were cross-checked by running two independent Fisher analysis pipelines that obtained the same output. Finally, we make a distinction between results for Gaussian-distributed DVs, for which the Fisher forecasts can be robustly computed, and results for non-Gaussian DVs, where the Fisher formalism cannot be trusted. The latter are included for consistency of the HOWLS project and because they can be used in future analyses by using a non-Gaussian likelihood.

\subsection{Individual forecasts \done{Nicolas}}
\label{sec:res;subsec:indiv}

Individual Fisher forecasts are displayed in columns $2$ and $3$ of Table~\ref{tab:forecasts} and correspond to the $1\sigma$ marginalized parameter errors given as percentages of the CP fiducial values: $\delta \sigma_8 / \sigma_8$ and $\delta \Omega_{\rm m} / \Omega_{\rm m}$. These forecasts can also be visually appreciated as the orange Fisher ellipses in the $\Omega_{\rm m}$-$\sigma_8$ planes of Fig.~\ref{fig:Fishall}.

If we first focus on second-order statistics, we see that the forecasts are on the order of one percent on both parameters. This is the same order of magnitude as the forecasts from \citet{EuclidVII}. A quantitative comparison is, however, not possible due to large differences in both analyses: we vary only $2$ CPs in a nontomographic setup while they use a minimum of $5$ parameters and a tomographic setup. Additionally, their analysis is based on theoretical predictions, while we are affected by numerical noise because we need to rely on simulations for a fair comparison among HOS. If we compare to analyses more similar to ours, for instance, the cosmo-SLICS based likelihood inference of \citet{Martinet+21a} for $4$ CPs in their nontomographic setup, we also find a good agreement: rescaling the forecasts of their Table 1 to $15\,000$ deg$^2$ gives a $1.9\%$ precision on $\Omega_{\rm m}$ to be compared with the $1.2\%$ of the present analysis. Our Fisher forecasts are tighter because we neglect the degeneracies between our two parameters and the additional two ($w$ and $h$) probed in the aforementioned analysis. Additionally, we know that degeneracies between pairs of CPs, which are treated as simple ellipses in the Fisher formalism, are usually more complex. Looking at the orientation of the blue Fisher ellipse in the top-left quadrant of Fig.~\ref{fig:Fishall}, we also see the usual lensing degeneracy shown by the dashed black line of constant $S_8$. As expected, the $\kappa$-2PCF forecasts are of the same order as the $\gamma$-2PCF, but weaker due to the pixelization of the $\kappa$ map, which destroys signal below the pixel size of $\ang{;0.59;}$.

\begin{figure*}
    \centering
    \begin{tabular}{cl}
    \includegraphics[width=0.48\hsize]{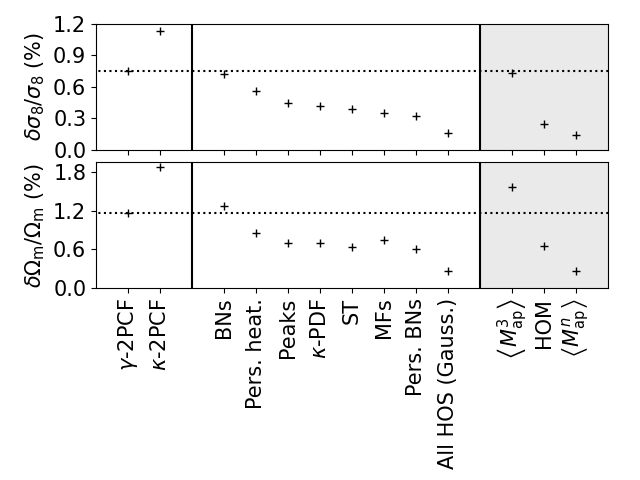}   
    &\includegraphics[width=0.48\hsize]{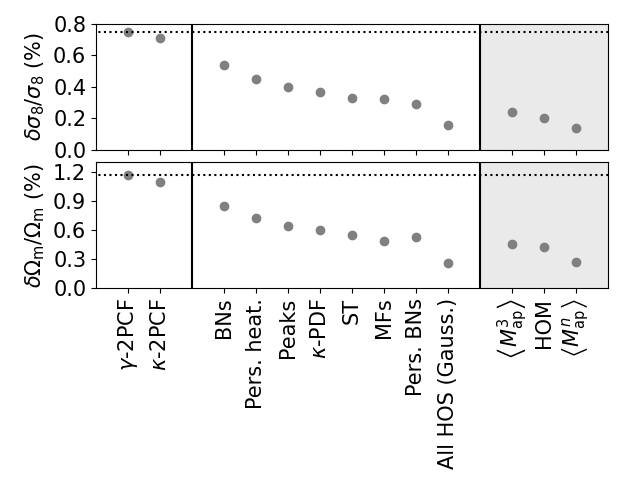} 
        \end{tabular}
    \caption{Fisher forecasts for $\sigma_8$ and $\Omega_{\rm m}$ for a nontomographic \textit{Euclid}-like survey for all statistics. The black crosses (\textit{left}) show the individual forecasts and the gray circles (\textit{right}) the forecasts combined with the $\gamma$-2PCF. Forecasts are ranked within $3$ categories: two-point statistics, higher-order Gaussian statistics, and higher-order non-Gaussian statistics. The latter are displayed on a gray background as they should be confirmed with a non-Gaussian likelihood framework. The corresponding marginalized precision on CPs can be found in Table~\ref{tab:forecasts}.}
    \label{fig:rankedFish}
\end{figure*}

The $\kappa$-PDF forecasts are shown as the orange ellipse in the top-right quadrant of Fig.~\ref{fig:Fishall}. The forecasts in this case are on the order of half a percent and are therefore tighter than those of the $\gamma$-2PCF thanks to the non-Gaussian small-scale information that this statistic is able to probe. The forecasts in this case also show different degeneracy directions. This is also the case for other HOS, with some diversity between DVs ranging from forecasts on the order of a third of a percent to about one percent, and various degeneracy orientations as shown by the different Fisher ellipses of Fig.~\ref{fig:Fishall}. In the left part of Fig.~\ref{fig:rankedFish} we also make an attempt to rank the statistics in terms of constraining power. We see a plateau in this figure where most of the statistics reach similar forecast values, about twice as good as that from the $\gamma$-2PCF. It is therefore hard to identify a single superior statistic, with notably pers. BNs, MFs, ST, $\kappa$-PDF, peaks, and pers. heat. all performing very well.

Finally, we also display the forecasts for the statistics which are not Gaussian distributed at the end of Table~\ref{tab:forecasts} and with a gray shaded background in Fig.~\ref{fig:Fishall} and \ref{fig:rankedFish}: HOM, $\MapMapMap$, and $\Mapn$. These are only displayed for the consistency of the HOWLS challenge but must be interpreted with caution as the Gaussian likelihood assumption is required for the Fisher formalism. There is possible strong constraining power but a refined analysis that goes beyond the simple Fisher method is needed to assess the robustness of these results. We however note that in the case of $\MapMapMap$ our results are consistent with that of the full MCMC implementation of \citet{Heydenreich:2022}.

\subsection{\texorpdfstring{Combination with $\gamma$-2PCF}{Combination with shear-2PCF} \done{Nicolas}}
\label{sec:res;subsec:2pcf}

One striking conclusion of the last section is that almost all of the HOS outperform the $\gamma$-2PCF in terms of constraining power in our analysis. Although this can be in part due to the range of scales available in our simulations, it supports all the recent findings in the HOS literature mentioned in the introduction as well as in Sect.~\ref{sec:hos}. One additional interesting question is the gain that these probes can provide when combined with the standard two-point analysis that is planned for \textit{Euclid}.

This gain is displayed in Table~\ref{tab:forecasts}, where we show the forecasts for each statistic combined with the $\gamma$-2PCF in columns $4$ and $5$ and the gain associated with these combined forecasts with respect to the $\gamma$-2PCF alone in the last two columns. We also show the combined Fisher ellipses in green in the $\Omega_{\rm m}$-$\sigma_8$ planes of Fig.~\ref{fig:Fishall}.

As expected, the $\kappa$-2PCF brings very little additional information (a factor of $1.06$) compared to the $\gamma$-2PCF. In theory there should be no gain at all as $\xi_\kappa$ is equivalent to $\xi_+$ in the absence of systematic B modes. However, the coarser binning of the $\xi_\kappa$ (10 bins between $\ang{;0,6;}<\theta<\ang{;9,23;}$) compared to that of $\xi_+$ (5 bins between $\ang{;0,24;}<\theta<\ang{;8,55;}$) allows the former to probe a small fraction of cosmological information not captured by the latter in the present analysis. Moreover, each of the HOS improves the forecasts by a factor of about $1.5$ to $2.5$ depending on the considered statistic and CP. In some cases, the gain is dominated by the tighter constraints of the HOS, and in others, by the change in the $\Omega_{\rm m}$-$\sigma_8$ degeneracy direction. Most of the time it is due to a combination of the latter two effects.

The ranking order of these combined forecasts, displayed in the right panel of Fig.~\ref{fig:rankedFish}, roughly follows the same order as for the individual forecasts. Here again, most statistics follow a similar trend in the gains provided, with ST, MFs and pers. BNs slightly outperforming other HOS, but closely followed by the $\kappa$-PDF, peaks, and pers. heat..

As a conclusion to this section, we note that the combined forecasts are always around two to three times tighter than that from the $\gamma$-2PCF alone, highlighting that HOS are able to extract the small-scale non-Gaussian information inaccessible to standard two-point estimators.

\subsection{Probe combination \done{Nicolas}}
\label{sec:res;subsec:comb}

\begin{figure*}
    \centering
    \includegraphics[width=1.0\hsize]{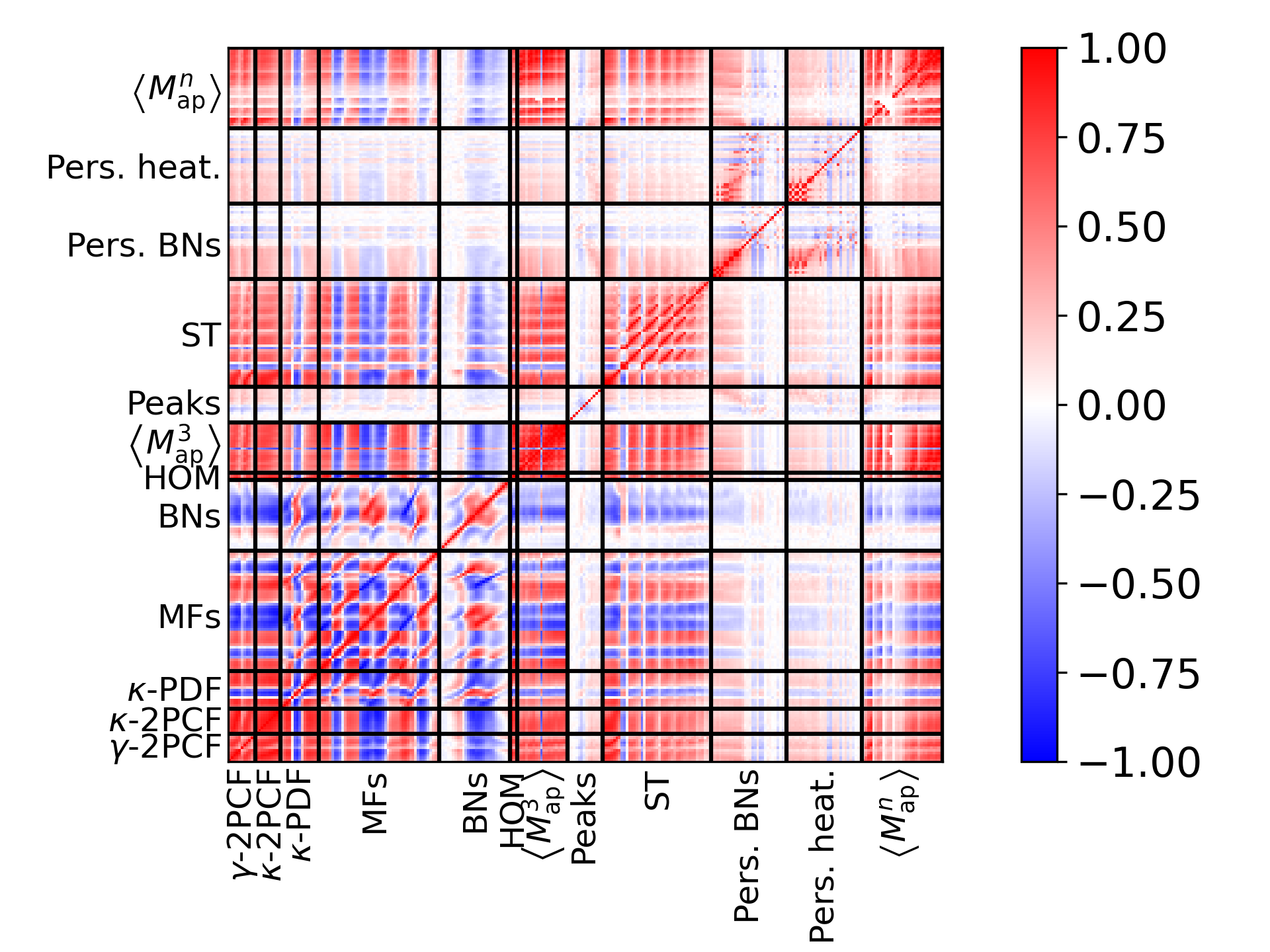}   
    \caption{Correlation matrix for all $12$ statistics tested in our analysis and referenced in Table~\ref{tab:hos}, computed from the same $924$ SLICS simulations. We note the complex correlations between the different statistics due to their expressions as well as to different filtering shapes and scales in the post-processing of the mass maps. The bin sizes of the $\kappa$-PDF, MFs, and BNs DVs have been increased in the figure for visualization purposes.}
    \label{fig:corcovall}
\end{figure*}

\begin{figure}
    \centering
    \includegraphics[width=1.0\hsize]{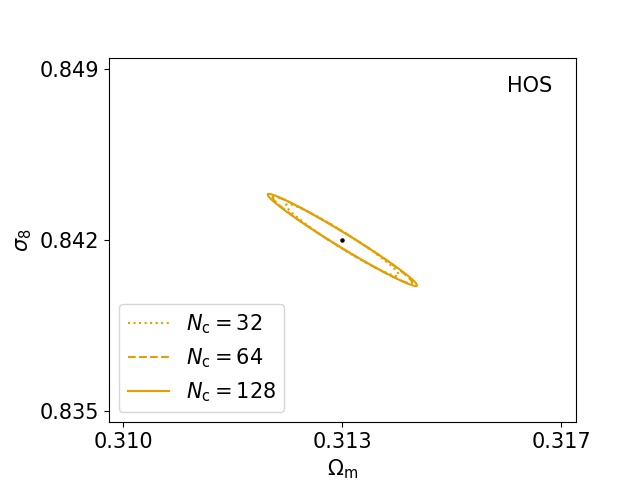} 
    \caption{Same as Fig.~\ref{fig:Fishderiv}, but for all Gaussian HOS together: $\kappa$-PDF, peaks, MFs, BNs, pers. BNs, pers. heat., and ST. Note the change of scales compared to other Fisher contour plots for better visualization.} 
    \label{fig:Fishallcombderiv}
\end{figure}

\begin{figure}
    \centering
    \includegraphics[width=1.0\hsize]{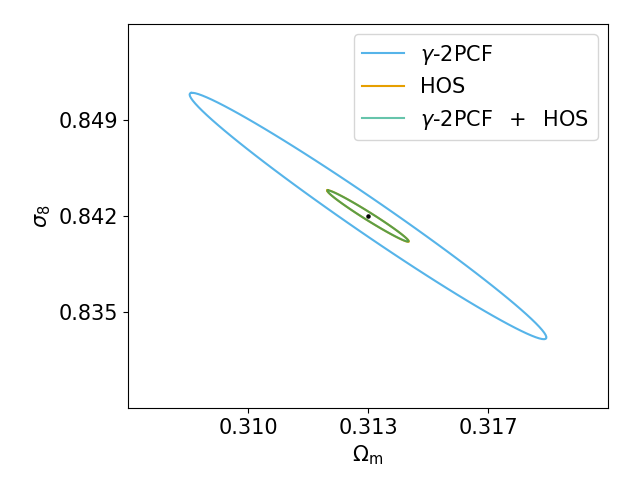} 
    \caption{Same as Fig.~\ref{fig:Fishall}, but for all Gaussian HOS together: $\kappa$-PDF, peaks, MFs, BNs, pers. BNs, pers. heat., and ST. The green and orange ellipses overlap almost exactly.} 
    \label{fig:Fishallcomb}
\end{figure}

Finally, one can wonder whether all these HOS probe independent information or if their combination can extract even more of the non-Gaussian small-scale information. To answer this question, we first consider the correlation matrix of the concatenated DV made of all individual DVs. This has been computed from the $924$ independent SLICS simulations and is shown in Fig.~\ref{fig:corcovall}. We see strong correlations between the different statistics with amplitude affected by the filtering of the $\kappa$ map or $\gamma$ field. In addition, we can identify patterns that are common to the probes that are of the same nature: for instance, the $\gamma$-2PCF and $\kappa$-2PCF, the HOM and the $\MapMapMap$, or the peaks and the $\kappa$-PDF. All these correlations suggest that one could use a data compression scheme to create a DV that optimally combines the information from these different statistics. This is, however, left for a future article in the HOWLS series.

Instead, we simply use the concatenated DV to compute the Fisher forecasts here. This increases the noise compared to the aforementioned optimal setting but the large number of realizations used for the covariance matrix still allows us to derive robust forecasts. These results are also checked against numerical noise as for the individual probes, finding converged forecasts when using $64$ or $128$ DUSTGRAIN-\emph{pathfinder} realizations, as shown in Fig.~\ref{fig:Fishallcombderiv}. In more detail, we combine the $\kappa$-PDF, peaks, MFs, BNs, pers. BNs, pers. heat., and ST and perform a comparison with and without the $\gamma$-2PCF in Fig.~\ref{fig:Fishallcomb}. This DV consists of $511$ elements, that is about half the number of SLICS realizations and gives a \citet{Sellentin+17} correction factor of $0.461$. The corresponding forecast values are given in Table~\ref{tab:forecasts} and ranked compared to the individual probes in Fig.~\ref{fig:rankedFish}. We find an improvement by a factor of four to five in that case compared to the $\gamma$-2PCF alone. This improvement is mainly driven by the combination of forecasts with slight different degeneracies, which manage to partially break the classical lensing degeneracy. This was already suggested in several analyses showing that HOS present different degeneracies to the $\gamma$-2PCF, but this is the first time that so many statistics have been combined to further lift the lensing degeneracy. Furthermore, the addition of the $\gamma$-2PCF to this combined DV only marginally improves the forecast precision, suggesting that the information contained in the $\gamma$-2PCF has also been captured by the combined HOS with the scales considered here.

\section{Conclusion \done{Nicolas}}
\label{sec:ccl}

HOWLS is a collaborative effort to explore the use of weak-lensing HOS for the interpretation of {\it Euclid} data. With this aim, we have compared two two-point statistics and ten HOS from the same set of {\it Euclid}-like mocks: the shear two-point correlation functions ($\gamma$-2PCF), the convergence two-point correlation function ($\kappa$-2PCF), the convergence one-point probability distribution ($\kappa$-PDF), convergence Minkowski functionals (MFs), convergence Betti numbers (BNs), aperture mass peak counts (peaks), higher-order convergence moments (HOM), third order aperture mass moments ($\smash{\MapMapMap}$), $n$-th order aperture mass moments ($\smash{\Mapn}$), aperture mass persistent homology Betti numbers (pers. BNs), aperture mass persistent homology heatmap (pers. heat.), and convergence scattering transform coefficients (ST). Our mocks are based on the DUSTGRAIN-\emph{pathfinder} N-body simulations, consisting of $12$ simulations varying $\Omega_{\rm m}$, $\sigma_8$, and $w$, with $128$ light cones for each of these plus one simulation at our fiducial cosmology with $256$ light cones for covariance estimation. We also used the $924$ independent SLICS simulations to improve the accuracy of our covariance matrix. In total, we generated $2460$ mocks, which are all representative of the {\it Euclid} survey in terms of galaxy density, redshift distribution, and shape noise. For the fiducial analysis though we used only $1436$, focusing on the four DUSTGRAIN-\emph{pathfinder} simulations with large variation of $\Omega_{\rm m}$ and $\sigma_8$, and the SLICS set for the covariance. We then built $\kappa$ maps for all mocks with the internal {\it Euclid} mass mapping pipeline and computed each statistic from them, except for the $\gamma$-2PCF and the aperture masses, which were computed at the ellipticity catalog level. We performed a Fisher analysis to forecast the constraining power of all these statistics for a $15\,000$ deg$^2$ {\it Euclid}-like survey. The generated mass maps, DVs, and inverse Fisher matrices are being publicly released\footnote{\url{https://archive.lam.fr/GECO/HOWLS}} with the paper to ensure reproducibility and to allow for a fair comparison with future new HOS. 

We applied more than twice as many HOS as in any other publication to date, offering the first comprehensive overview of almost all WL statistics studied in the literature today. We found that combining any of these statistics with the standard $\gamma$-2PCF improves the precision of the individual forecasts on $\Omega_{\rm m}$ and $\sigma_8$ by a factor of $\sim 2$ to $3$, highlighting the ability of HOS to probe non-Gaussian small-scale information missed by two-point estimators. Combining these HOS further increases the gain to a factor of $4.5$ compared to the $\gamma$-2PCF thanks to the different degeneracy direction of each statistic in the $\Omega_{\rm m}$-$\sigma_8$ plane, which helps lift the classical lensing degeneracy.

Finally, while performing our analysis, we identified several points that require further investigation and will be addressed in future publications of the HOWLS series. First, we found that numerical noise in the Fisher derivatives could artificially break degeneracies between parameters, which prevented us from measuring robust forecasts for $w$ and from using simulations with very small CP variations. Although this might be solved by a tomographic approach in the specific case of $w$, it also encourages more independent N-body simulations to be run at each cosmology, and not only at the fiducial one used for covariance estimates. We note that this could also be linked to the Fisher formalism and that this difficulty could be less of a problem in approaches where degeneracies are better explored in a CP hypercube. Second, we note that some statistics do not follow a multivariate Gaussian distribution, namely the HOM, $\MapMapMap$, and $\Mapn$. The Fisher formalism breaks down for these statistics. They were included in the paper for consistency but not discussed on equal footing with other more robust results. This issue will be solved when moving from the Fisher formalism to an emulator-based model sampling the CP hypercube, which is a necessary step for measuring cosmological constraints from HOS in observations. Third, we calculated large correlations with complex patterns between all statistics. Although in a nontomographic setup it was possible to keep a large correlated DV in the joint analysis, this will no longer be the case when including tomography, requiring the development of a robust DV compression scheme. Finally, no systematic effects beyond shape noise were included in the present paper. Working with mass maps will notably introduce some new systematic effects related to masks and the $\kappa$ reconstruction, and different HOS will have different responses to astrophysical biases such as intrinsic alignments and baryons that are particularly important on small scales \citep[e.g., ][]{Semboloni+13}. Besides the impressive gain in precision that we found, there is therefore also the hope that adding HOS to the $\gamma$-2PCF might help break degeneracies between systematics and CPs thanks to the different nature of these probes \citep[e.g.,][]{Patton2017,Pyne+21}.

\begin{acknowledgements}

NM acknowledges support from a fellowship of the Centre National d’Etudes Spatiales (CNES). CG acknowledges the support from the grant PRIN-MIUR 2017 WSCC32 ZOOMING, from the grant ASI n.2018-23-HH.0, from the Italian National Institute of Astrophysics under the grant "Bando PrIN 2019", PI: Viola Allevato, from the HPC-Europa3 Transnational Access programme HPC17VDILO,  from INAF theory Grant 2022: Illuminating Dark Matter using Weak Lensing by Cluster Satellites, PI: Carlo Giocoli. IT acknowledges support from the Fundacao para a Ciencia e a Tecnologia (FCT) through the projects UIDB/04434/2020, UIDP/04434/2020 and the Investigador FCT Contract No. IF/01518/2014 and POCH/FSE (EC).

\AckEC \\~\\

\textit{Authors’ contributions}: All authors have significantly contributed to this publication and/or made multiyear essential contributions to the \textit{Euclid} project which made this work possible. The paper-specific main contributions of the lead authors are as follows: HOWLS project management: V. Cardone, C. Giocoli, N. Martinet, S. Pires, and I. Tereno; DUSTGRAIN-\emph{pathfinder} simulations: M. Baldi and C. Giocoli; SLICS simulations: J. Harnois-D\'eraps; mocks and convergence maps: S. Pires and N. Martinet; Fisher analysis and paper writing coordination: V. Cardone and N. Martinet; theoretical predictions: A. Barthelemy, A. Boyle, V. Cardone, S. Heydenreich, L. Linke, L. Porth, and C. Uhlemann; $\gamma$-2PCF: N. Martinet; $\kappa$-2PCF, MFs, and BNs: S. Vinciguerra and V. Cardone; $\kappa$-PDF: A. Barthelemy, A. Boyle, S. Codis, and C. Uhlemann; peaks: S. Pires; pers. BNs and pers. heat.: S. Heydenreich and P. Burger; ST: S. Cheng; HOM: A. Boyle, A. Barthelemy, S. Codis, V. Cardone, C. Uhlemann, and S. Vinciguerra; $\MapMapMap$: L. Linke, S. Heydenreich, and P. Burger; $\Mapn$: L. Porth; earlier development of HOWLS: V. Ajani, M. Kilbinger, F. Lanusse, C. Llinares, C. Parroni, A. Peel, and M. Vicinanza.

\end{acknowledgements}

\bibliographystyle{aa}
\bibliography{HOWLS_KPpaper1}

\appendix

\section{\texorpdfstring{Forecasts for $\Sigma_8$}{Forecasts for S8}}
\label{app:resS8}

In order to ease comparison with the existing literature we also produce forecasts in the $\Sigma_8$-$\Omega_{\rm m}$ plane in addition to the $\sigma_8$-$\Omega_{\rm m}$ in the rest of the paper. We use the general definition of the growth of structure parameter $\Sigma_8 = \sigma_8 \left( \Omega_{\rm m}/0.3 \right)^\alpha$ and choose the optimal $\alpha$ value for the $\gamma$-2PCF, that is the one that coincides with the semi-minor axis of the corresponding Fisher ellipse. We find $\alpha=0.63$ slightly different from the $\alpha=0.5$ value of $S_8$ as the lensing degeneracy is not fully captured by the simplistic Fisher formalism. We compute the transformation matrix \citep[e.g.,][]{Coe09} between the two sets of parameters to derive the Fisher matrix for $\Sigma_8$-$\Omega_{\rm m}$, assuming that $\sigma_8$ depends on $\Omega_{\rm m}$ and $\Sigma_8$ through $\sigma_8 = \Sigma_8 \left(0.3/\Omega_{\rm m}\right)^\alpha$ and that $\Omega_{\rm m}$ is independent:
\begin{equation}
  \tens{F}(\Omega_{\rm m},\Sigma_8) =  \tens{M}^\mathrm{T} \, \tens{F}(\Omega_{\rm m},\sigma_8) \, \tens{M} \; ,
\end{equation}
with
\begin{equation}
\tens{M} =
    \begin{pmatrix}
    \frac{\partial \Omega_{\rm m}}{\partial \Omega_{\rm m}} & \frac{\partial \Omega_{\rm m}}{\partial \Sigma_8} \\
    \frac{\partial \sigma_8}{\partial \Omega_{\rm m}} & \frac{\partial \sigma_8}{\partial \Sigma_8} 
  \end{pmatrix}
  =
  \begin{pmatrix}
    1 & 0 \\
    - \alpha \frac{\sigma_8}{\Omega_{\rm m}} &  \left(\frac{0.3}{\Omega_{\rm m}}\right)^\alpha
  \end{pmatrix} \; .
\end{equation}
In the above equations, $\Omega_{\rm m}$ and $\sigma_8$ are evaluated at the fiducial cosmology of the DUSTGRAIN-\emph{pathfinder} simulations (see Table~\ref{tab:allsim}).

\begin{figure*}
    \centering
    \includegraphics[width=1.0\hsize]{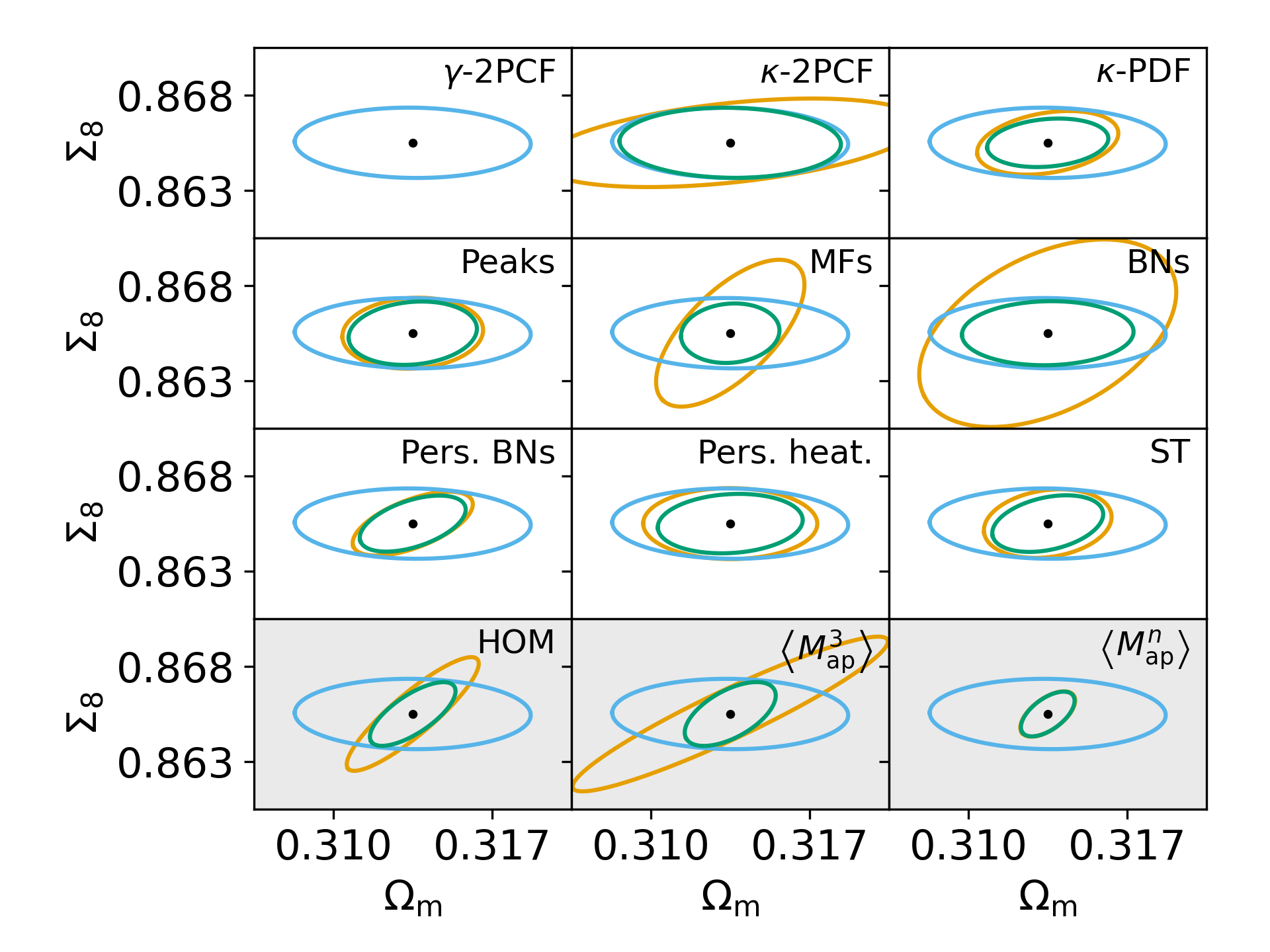} 
    \caption{Same as Fig.~\ref{fig:Fishall}, but for ($\Omega_{\rm m}$, $\Sigma_8$) instead of ($\Omega_{\rm m}$, $\sigma_8$), with $\Sigma_8 = \sigma_8 \left( \Omega_{\rm m}/0.3 \right)^\alpha$ and $\alpha=0.63$. Individual Fisher forecasts in the $\Sigma_8$-$\Omega_{\rm m}$ plane for a nontomographic \Euclid-like survey for the $11$ statistics (orange) and $\gamma$-2PCF (blue), as well as their combination (green). The corresponding marginalized precision on CPs can be found in Table~\ref{tab:forecasts_S8}. The bottom row with the gray shaded backgrounds displays probes that are not Gaussian distributed and cannot be robustly interpreted with Fisher forecasts. The abbreviated name of each summary statistic is displayed in the top-right part of each panel: $\gamma$-2PCF and $\kappa$-2PCF for the shear and convergence two-point correlation functions, $\kappa$-PDF for the convergence one-point probability distribution, peaks for aperture mass peak counts, MFs for convergence Minkowski functionals, BNs for convergence Betti numbers, pers. BNs and pers. heat. for aperture mass persistent homology Betti numbers and heatmap, ST for convergence scattering transform coefficients, HOM for higher-order convergence moments, and $\MapMapMap$ and $\Mapn$ for third and $n$-th order aperture mass moments.}
    \label{fig:Fishall_S8}
\end{figure*}

The results in the new set of parameters are presented in Fig.~\ref{fig:Fishall_S8} and Table~\ref{tab:forecasts_S8} which are respectively analogous to Fig.~\ref{fig:Fishall} and Table~\ref{tab:forecasts}. The transformation of CPs implies reading Fig.~\ref{fig:Fishall} along the semi-major axis of the $\gamma$-2PCF Fisher ellipse (we note that for $S_8$ that would be equivalent to reading this figure along the dashed black line of the top left quadrant). All ellipses appear rotated in the new CPs plane. Hence the gain on the FoM is preserved compared to the original case. As expected as well the forecasts of $\Sigma_8$ are tighter than on the other two parameters. The general ranking of the statistics is the same whether one considers $\Sigma_8$-$\Omega_{\rm m}$ or $\sigma_8$-$\Omega_{\rm m}$. However, the gain from combining a given HOS with the $\gamma$-2PCF is less important for $\Sigma_8$, ranging from a factor $1.1$ to $1.45$ improvement in precision. This is because we probe the direction perpendicular to the lensing degeneracy while HOS are complementary to the $\gamma$-2PCF as they present different degeneracy directions. The FoM, which quantifies the improved precision for the set of probed parameters remains the same whether we consider $\Sigma_8$-$\Omega_{\rm m}$ or $\sigma_8$-$\Omega_{\rm m}$.

\begin{table*} 
\caption[]{Same as Table.~\ref{tab:forecasts}, but for ($\Omega_{\rm m}$, $\Sigma_8$) instead of ($\Omega_{\rm m}$, $\sigma_8$), with $\Sigma_8 = \sigma_8 \left( \Omega_{\rm m}/0.3 \right)^\alpha$ and $\alpha=0.63$. Fisher forecasts for a \textit{Euclid}-like nontomographic analysis. The precision on $\Sigma_8$ and $\Omega_{\rm m}$ is given as a percentage of the fiducial values for the probes taken individually and when combined with the $\gamma$-2PCF. We also report the figure of merit (FoM) as defined in the Dark Energy Task Force \citep{Albrecht+06}. The last columns display the expected gain over the $\gamma$-2PCF analysis for the scales available in the DUSTGRAIN-\emph{pathfinder} simulations. Results for non-Gaussian statistics are also presented as part of the HOWLS project; we stress, however, that a non-Gaussian likelihood would be necessary to assess the robustness of these three probes.}
\centering 
\begin{tabular}{lccccccccc} 
\hline 
\hline
Statistics & \multicolumn{3}{c}{individual
} & \multicolumn{3}{c}{added $\gamma$-2PCF} & \multicolumn{3}{c}{Gain over $\gamma$-2PCF} \\
& $\delta \Sigma_8 / \Sigma_8$ & $\delta \Omega_{\rm m} / \Omega_{\rm m}$ & FoM & $\delta \Sigma_8 / \Sigma_8$ & $\delta \Omega_{\rm m} / \Omega_{\rm m}$ & FoM & $\delta \Sigma_8$ & $\delta \Omega_{\rm m}$ & FoM \\

\hline

2nd order statistics &  &  &  &  &  &  &  & \\
 $\gamma$-2PCF & $ 0.13 \% $ & $ 1.17 \% $ & $ 2.49 \times 10^{5} $ & $ - $ & $ - $ & $ - $ & $ - $ & $ - $ & $ - $ \\
 $\kappa$-2PCF & $ 0.16 \% $ & $ 1.88 \% $ & $ 1.37 \times 10^{5} $ & $ 0.13 \% $ & $ 1.10 \% $ & $ 2.67 \times 10^{5} $ & $ \times 1.00 $ & $ \times 1.07 $ & $ \times 1.07 $ \\

\hline
HOS (Gaussian) &  &  &  &  &  &  &  &  \\
 $\kappa$-PDF & $ 0.12 \% $ & $ 0.70 \% $ & $ 4.82 \times 10^{5} $ & $ 0.09 \% $ & $ 0.60 \% $ & $ 7.16 \times 10^{5} $ & $ \times 1.45 $ & $ \times 1.95 $ & $ \times 2.87 $ \\
 Peaks & $ 0.13 \% $ & $ 0.70 \% $ & $ 4.16 \times 10^{5} $ & $ 0.11 \% $ & $ 0.64 \% $ & $ 5.14 \times 10^{5} $ & $ \times 1.11 $ & $ \times 1.84 $ & $ \times 2.06 $ \\
 MFs &  $ 0.27 \% $ & $ 0.74 \% $ & $ 2.53 \times 10^{5} $ & $ 0.11 \% $ & $ 0.49 \% $ & $ 7.09 \times 10^{5} $ & $ \times 1.18 $ & $ \times 2.40 $ & $ \times 2.84 $ \\
 BNs &  $ 0.34 \% $ & $ 1.27 \% $ & $ 9.55 \times 10^{4} $ & $ 0.12 \% $ & $ 0.85 \% $ & $ 3.77 \times 10^{5} $ & $ \times 1.10 $ & $ \times 1.38 $ & $ \times 1.51 $ \\
 Pers. BNs & $ 0.12 \% $ & $ 0.60 \% $ & $ 6.76 \times 10^{5} $ & $ 0.10 \% $ & $ 0.53 \% $ & $ 8.07 \times 10^{5} $ & $ \times 1.25 $ & $ \times 2.23 $ & $ \times 3.23 $ \\
 Pers. heat. & $ 0.13 \% $ & $ 0.86 \% $ & $ 3.35 \times 10^{5} $ & $ 0.11 \% $ & $ 0.72 \% $ & $ 4.84 \times 10^{5} $ & $ \times 1.18 $ & $ \times 1.63 $ & $ \times 1.94 $ \\
 ST & $ 0.12 \% $ & $ 0.63 \% $ & $ 4.82 \times 10^{5} $ & $ 0.10 \% $ & $ 0.55 \% $ & $ 6.99 \times 10^{5} $ & $ \times 1.24 $ & $ \times 2.13 $ & $ \times 2.80 $ \\
 All HOS & $ 0.03 \% $ & $ 0.27 \% $ & $ 4.84 \times 10^{6} $ & $ 0.03 \% $ & $ 0.27 \% $ & $ 4.89 \times 10^{6} $ & $ \times 3.99 $ & $ \times 4.41 $ & $ \times 19.62 $ \\

\hline
HOS (non-Gaussian) &  &  &  &  &  &  &  &  \\
 HOM & $ 0.21 \% $ & $ 0.65 \% $ & $ 5.95 \times 10^{5} $ & $ 0.12 \% $ & $ 0.42 \% $ & $ 1.09 \times 10^{6} $ & $ \times 1.10 $ & $ \times 2.76 $ & $ \times 4.36 $ \\
 $\MapMapMap$ &  $ 0.28 \% $ & $ 1.56 \% $ & $ 2.43 \times 10^{5} $ & $ 0.12 \% $ & $ 0.45 \% $ & $ 8.73 \times 10^{5} $ & $ \times 1.10 $ & $ \times 2.59 $ & $ \times 3.50 $ \\
 $\Mapn$ &  $ 0.08 \% $ & $ 0.27 \% $ & $ 2.08 \times 10^{6} $ & $ 0.08 \% $ & $ 0.27 \% $ & $ 2.13 \times 10^{6} $ & $ \times 1.57 $ & $ \times 4.40 $ & $ \times 8.54 $ \\

\hline
\end{tabular} 
\label{tab:forecasts_S8}
\end{table*}

\end{document}